\documentclass[11pt]{elsarticle}

\usepackage{amssymb}
\usepackage{amsmath}
\usepackage{graphicx}
\usepackage{dcolumn}
\usepackage{bm}
\usepackage{psfrag}
\usepackage{tikz}
\usepackage{pgfplots}
\usepackage{hyperref}
\allowdisplaybreaks
\pgfmathdeclarefunction{gauss}{2}{%
  \pgfmathparse{1/(#2*sqrt(2*pi))*exp(-((x-#1)^2)/(2*#2^2))}%
}

\begin{document}

\begin{frontmatter}



\title{A kinetic scheme with variable velocities and relative entropy}


\author[1]{Shashi Shekhar Roy\corref{cor1}%
\fnref{fn1}}
\author[2]{S. V. Raghurama Rao\fnref{fn2}}
\cortext[cor1]{Corresponding author}
\fntext[fn1]{E-mail address: shashi@iisc.ac.in, shashisroy@gmail.com}
\fntext[fn2]{E-mail address: raghu@iisc.ac.in}
\affiliation[1]{organization={Research Scholar, Department of Aerospace Engineering, Indian Institute of Science},
            city={Bangalore},
            country={India}}
						
\affiliation[2]{organization={Department of Aerospace Engineering, Indian Institute of Science},
            city={Bangalore},
            country={India}}

\begin{abstract}
A new kinetic model is proposed where the equilibrium distribution with bounded support has a range of velocities about two average velocities in 1D. In 2D, the equilibrium distribution function has a range of velocities about four average velocities, one in each quadrant. In the associated finite volume scheme, the average velocities are used to enforce the Rankine-Hugoniot jump conditions for the numerical diffusion at cell-interfaces, thereby capturing steady discontinuities exactly.  The variable range of velocities is used to provide additional diffusion in smooth regions. Further, a novel kinetic theory based expression for relative entropy is presented which, along with an additional criterion, is used to identify expansions and smooth flow regions. Appropriate flow tangency and far-field boundary conditions are formulated for the proposed kinetic model. Several benchmark 1D and 2D compressible flow test cases are solved to demonstrate the efficacy of the proposed solver.  

\end{abstract}

%


\end{frontmatter}



\section{Introduction} 
Numerical methods to solve Euler equations and other nonlinear hyperbolic problems constitute an area of intense ongoing research.  Of the several approaches introduced for spatial discretization, {\em upwinding} has emerged as a popular technique to obtain stability.  Upwind schemes use stencils biased along the direction of incoming wave(s). The upwind schemes can be categorized as (exact and approximate) Riemann solvers, flux vector splitting methods, kinetic/Boltzmann schemes and relaxation schemes.  Of these, the approximate Riemann solvers are quite popular, due to their low numerical diffusion.  However, these schemes suffer from many drawbacks like admitting entropy-violating solutions, carbuncle phenomenon, kinked Mach stems, instabilities due to odd even decoupling, etc. \cite{quirk1997contribution}.  Another undesirable feature of these schemes is the strong dependence on the underlying eigen-structure.

Kinetic theory based schemes provide elegant alternatives to traditional Riemann solvers. The governing equation at the kinetic level is the Boltzmann equation. The macroscopic equations can be retrieved from the Boltzmann equation by taking suitable moments. One of the major advantages that kinetic schemes offer is the linearity of the advection term in Boltzmann equation, which simplifies upwinding. The first Boltzmann scheme was developed by Chu \cite{chu1965kinetic}, who used a finite difference method to solve the Boltzmann-BGK equation. Sanders and Prendergast \cite{sanders1974possible} developed the {\em Beam scheme}, in which the Maxwellian equilibrium distribution is replaced by a set of weighted Dirac-delta functions, named beams. Pullin \cite{pullin1980direct} introduced the Equilibrium Flux Method (EFM) in which fluxes at the interface between cells are calculated based on half Maxwellians.  The Kinetic Numerical Method (KNM) of Reitz \cite{reitz1981one}, based on an operator splitting involving advection and collision steps, is based on tracing the foot of the characteristic in the convection step in Boltzmann equation with appropriate velocity discretization.  Deshpande introduced \cite{deshpande1986second} a similar kinetic scheme, obtaining second order accuracy by using Chapman-Enskog distribution function to provide an anti-diffusive correction.  Deshpande \cite{deshpande1986kinetic} and Mandal \& Deshpande \cite{mandal1994kinetic} introduced a Kinetic Flux Vector Splitting (KFVS) scheme with the upwinding based on molecular velocities. The final expressions for the split fluxes in KFVS method are the same as in the Equilibrium Flux Method, though the approaches are different. Kaniel \cite{kaniel1988kinetic} and Perthame \cite{perthame1990boltzmann} introduced kinetic schemes based on equilibrium distributions with bounded support.  Prendergast and Kun Xu \cite{prendergast1993numerical} utilized an approximate local solution of the Boltzmann-BGK equation, while using a pressure sensor to detect shocks, thereby introducing an unsplit kinetic scheme. Raghurama Rao and Deshpande \cite{raghurama1995peculiar} introduced the Peculiar Velocity based upwind (PVU) method, replacing the molecular velocity by the sum of fluid velocity ($u$) and peculiar velocity ($c= v-u$), leading to a convection-pressure splitting based upwind method.  While there are more kinetic schemes in the literature, a common drawback they share is their inability to capture steady discontinuities exactly, a feat shared by some of the popular macroscopic upwind schemes.  

Another interesting line of research is to introduce kinetic schemes based on discrete velocities.  Natalini \cite{natalini1998discrete} and  Aregba-Driollet and Natalini \cite{aregba2000discrete} developed kinetic schemes starting from a discrete velocity Boltzmann equation with a BGK model.  In such a discrete velocity formulation, the equilibrium distribution functions are not Maxwellians but simple linear combinations of conserved variable vectors and flux vectors. One inherent advantage that the discrete velocity based schemes have over the continuous velocity based kinetic schemes is that the complex integrals are replaced by summations while taking moments. For the two-velocity model with velocities $\lambda$ and $-\lambda$ in 1D, the discrete velocity model is equivalent to the relaxation model of Jin and Xin \cite{jin1995relaxation}. Bouchut \cite{bouchut1999construction} introduced BGK models with a family of kinetic entropies for a given system of hyperbolic conservation laws.  Shrinath {\em et al.} \cite{shrinath2023kinetic} introduced the Kinetic Flux Difference Splitting Scheme based on discrete velocity Boltzmann equations, utilizing the two discrete velocities to satisfy the R-H conditions, while the third velocity is used to provide additional diffusion in smooth regions.  For this purpose, they utilized the Kullback-Leibler divergence, in the molecular velocity framework, to distinguish among different regions.

In the present work, we introduce a new kinetic formulation based on variable velocities, based on compactly supported distributions.  Each set of variable velocities is centered on an average velocity.  While the average velocities are utilized to enforce Rankine-Hugoniot jump conditions in the discretization, thereby capturing grid-aligned steady discontinuities exactly, the variable range of velocities is utilized to add additional numerical diffusion in the smooth regions.  For this purpose, we introduce a novel discrete velocity version of relative entropy.  The advantage that our formulation offers over the earlier compactly supported distribution functions is that in our model, both the average velocities and the range of velocities are flexible.  Further, new flow tangency and far-field boundary conditions are formulated for our kinetic model and are utilized in the numerical simulations.  
 
\section{Gas-Kinetic Theory}
The kinetic theory of gases is governed by the Boltzmann equation, given by 
\begin{equation}
\frac{\partial f}{\partial t}+ \textbf{v} \cdot\frac{\partial f}{\partial \textsl{\textbf{x}}}= Q(f)
\label{eq:2_1}
\end{equation}
Here, $f(t,\textbf{x}, \textbf{v})$ is the velocity distribution function, $\textbf{v}$ is molecular velocity, 
and $Q(f)$ is the collision term. The L.H.S. represents the rate of change of $f$ due to molecular motion of particles. The R.H.S. represents the rate of change of $f$ due to binary collisions among particles. The advection term is linear; the non-linearity is present in the collision term. The collision term drives the distribution function $f$ to equilibrium, vanishing at the limit. The equilibrium distribution given by kinetic theory of gases is the Maxwell-Boltzmann distribution function
\begin{equation}
f^{eq} = f_{Maxwell}= \frac{\rho}{I_{0}} \left(\frac{\beta}{\pi}\right)^{N/2} exp\left(-\beta|\textbf{v}-\textbf{u}|^{2}\right) exp\left(-I/I_{0}\right) 
\label{eq:2_2}
\end{equation}
where $\beta$= $\frac{1}{2RT}$, $\textbf{u}(t,\textbf{x})$ is the macroscopic velocity, $N$ is the translational degrees of freedom, $I$ is the internal energy variable corresponding to non-translational degrees of freedom, $I_{0}~=~\frac{2- N(\gamma -1)}{2(\gamma -1)}RT$ and $\gamma$= $\frac{c_{p}}{c_{v}}$. The combined mass, momentum and total energy of particles is conserved during collisions. Thus, 1, $\textbf{v}$ and $I+ \frac{|\textbf{v}|^{2}}{2}$ are the collisional invariants. Multiplying the Boltzmann equation with the moment vector $\bm{\Psi}= \left[1, v_{1}, ..,v_{N}, I+ \frac{|\textbf{v}|^{2}}{2}\right]^{T}$ and integrating w.r.t. $\textbf{v}$ and $I$, {\em i.e.}, taking moments gives us the macroscopic conservation laws of mass, momentum and energy.\\\\
Utilizing the popular simplification to the collision term, the BGK collision model \cite{bhatnagar1954model}, equation (\eqref{eq:2_1}) becomes 
\begin{equation}
\frac{\partial f}{\partial t}+ \textbf{v} \cdot \frac{\partial f}{\partial \textsl{\textbf{x}}}= 
- \frac{1}{\epsilon} \left[ f - f^{eq} \right] 
\label{BGK_equation}
\end{equation}
where $\epsilon$ is the relaxation time.  Then, using operator splitting for separating advection and collision terms and further using instantaneous relaxation to equilibrium in the collision step ($\epsilon \rightarrow 0$, {\em i.e.}, $f= f^{eq}$), we can write the moments of the Boltzmann equation as 
\begin{equation}
\int_{\mathbb{R}^{N}}d\textbf{v} \int_{\mathbb{R}^{+}}dI \ \bm{\Psi}\left(\frac{\partial f}{\partial t}+ \frac{\partial (v_{i}f)}{\partial x_{i}}= 0,f= f^{eq}\right)
\label{eq:2_3}
\end{equation}
These moments in Equation \eqref{eq:2_3} form an elegant way of representing the inviscid Euler equations at the macroscopic level, given by 
 \begin{equation}
\frac{\partial \textbf{U}}{\partial t}+ \frac{\partial \textbf{G}_{i}}{\partial x_{i}}= 0
\label{eq:2_4}
\end{equation}
with
\begin{equation}
\textbf{U}= \begin{bmatrix} \rho \\ \rho u_{j} \\ \rho E \end{bmatrix}, \textbf{G}_{i}= \begin{bmatrix} \rho u_{i} \\ \rho u_{i}u_{j}+ p \delta_{ij} \\ (\rho E+ p)u_{i} \end{bmatrix}, E= e+ \frac{u^{2}_{i}}{2}
\label{eq:2_5}
\end{equation}
Here, the moment relations are
\begin{subequations}
	\label{eq:2_6}
 \begin{equation}
    \int_{\mathbb{R}^{N}}d\textbf{v} \int_{\mathbb{R}^{+}}dI \ \bm{\Psi} f= \int_{\mathbb{R}^{N}}d\textbf{v} \int_{\mathbb{R}^{+}}dI \ \bm{\Psi} f^{eq}=  \textbf{U}
 \end{equation}
 \begin{equation}
     \int_{\mathbb{R}^{N}}v_{i}d\textbf{v} \int_{\mathbb{R}^{+}}dI \ \bm{\Psi} f^{eq}= \textbf{G}_{i}
 \end{equation}
 \end{subequations}
	Multiplying the Boltzmann equation by $\ln f$ and taking its moment, we get the kinetic entropy inequality, {\em i.e.}, the H-theorem
	\begin{equation}
	\frac{\partial H}{\partial t}+ \frac{\partial H_{v,i}}{\partial x_{i}} \leq 0
	\label{eq:2_7}
	\end{equation}
	with
	\begin{subequations}
	\label{eq:2_8}
	\begin{equation}
	\int_{\mathbb{R}^{N}}d\textbf{v} \int_{\mathbb{R}^{+}}dI \ f \ln f= H,
	\end{equation}
	\begin{equation}
	\int_{\mathbb{R}^{N}}v_{i}d\textbf{v} \int_{\mathbb{R}^{+}}dI \ f \ln f= H_{v,i},
	\end{equation}
	\begin{equation}
	\int_{\mathbb{R}^{N}}d\textbf{v} \int_{\mathbb{R}^{+}}dI \ Q(f) \ln f \leq 0
	\end{equation}
	\end{subequations}
In the next section, we introduce our new compactly supported distribution in 1-D with variable velocities, centered around two averaged velocities. 	
\section{Equilibrium distribution in 1D}
\begin{figure}[h!] 
\centering
\begin{tikzpicture}
\small\begin{axis}
 [every axis plot post/.append style={
  mark=none,domain=-2.0:2.0,samples=50,smooth},
  xmin=-2.5,xmax=2.5,ymin=0.0,ymax=1.0,
	ylabel={$\hat{f}^{eq}_{i}$},xlabel={v},
	axis x line=bottom, 
  axis y line=center,
	xtick={-0.9,0.9},
	xticklabels={\color{red} $-\lambda_{i}$, \color{red}  $\lambda_{i}$},  
	yticklabel={\empty},
	tick label style={major tick length=0pt}]
	\draw[red] (axis cs:0.65,0) rectangle (axis cs:1.15,0.8);
\draw[red] (axis cs:-0.65,0) rectangle (axis cs:-1.15,0.6);
\draw[black,dashed] (axis cs:0.9,0) -- (axis cs:0.9,0.8)node[anchor=south]{};
\draw[black,dashed] (axis cs:-0.9,0) -- (axis cs:-0.9,0.6)node[anchor=south]{};
\draw[thick,stealth-stealth] (axis cs:0.65,0.82) -- (axis cs:1.15,0.82) node[anchor=south east] {\color{red} $2\delta \lambda_{i}$};
\draw[thick,stealth-stealth] (axis cs:-1.15,0.62) -- (axis cs:-0.65,0.62) node[anchor=south east] {\color{red} $2\delta \lambda_{i}$};
\node[] at (axis cs:1.32,0.8) {\color{red} $f^{eq}_{+,i}$};
\node[] at (axis cs:-0.47,0.6) {\color{red} $f^{eq}_{-,i}$};
\node[] at (axis cs:0.22,0.50) {\color{blue}$\hat{f}^{eq}_{i, Maxwellian}$}; 
  \addplot {gauss(0.2,0.7)}; 
\node[] at (axis cs:-2.3,0.02) {-$\infty$}; 
\node[] at (axis cs:2.3,0.02) {$\infty$};
 
\end{axis}
\end{tikzpicture}
\caption{Equilibrium distribution in 1D: Maxwellian replaced by compactly supported variable velocity distributions}
\label{fig:1}
\end{figure}
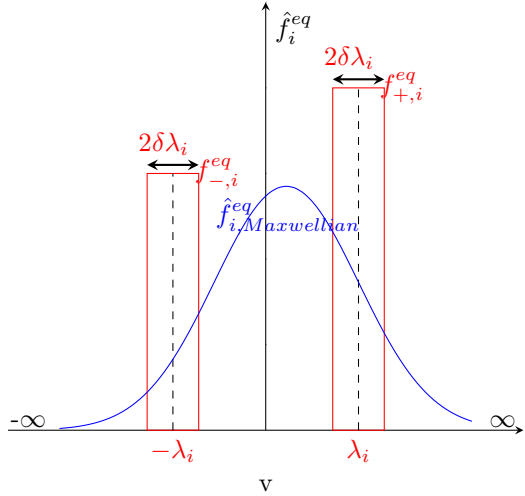

In this research work, we focus only on modeling in the velocity ($v$) space. Therefore, we first introduce a truncated equilibrium distribution function $\breve{f}^{eq}$, by first integrating w.r.t. to the internal energy variable $I$ as follows. 
\begin{equation}
	\breve{f}^{eq} = \int^{\infty}_{0} \ f^{eq} dI
\label{eq:3_2}
\end{equation} 
Then, the moments in 1-D become 
\begin{equation}
    U_{i} = \int_{-\infty}^{\infty} dv \ \Psi_{i} \breve{f}^{eq} \ \textrm{and} \ 
    G_{i} = \int_{-\infty}^{\infty} dv \ v \Psi_{i} \breve{f}^{eq} 
\end{equation}  
Now, defining 
\begin{equation}
    \hat{f}^{eq}_{i} = \Psi_{i} \breve{f}^{eq} 
\end{equation} 
the moment relations take the form 
\begin{equation}
    U_{i} = \int_{-\infty}^{\infty} dv \ \hat{f}^{eq}_{i} = \langle \hat{f}^{eq}_{i} \rangle \ \textrm{and} \ 
    G_{i} = \int_{-\infty}^{\infty} dv \ v \hat{f}^{eq}_{i} = \langle v \hat{f}^{eq}_{i} \rangle
\end{equation} 
The basic strategy of our model is to replace the 1-D Maxwellians $\hat{f}^{eq}_{i, Maxwellian}$ (represented by a Gaussian curve in the space of molecular velocities) by a set of two rectangular distributions, with control over the the locations as well as widths of these rectangular distributions, such that the moment relations still hold.  The rectangular distributions satisfy the requirement of distributions with compact support, thus leading to simplifications.  The control over the locations of these distributions will later help us in enforcing Rankine-Hugoniot conditions in the discretization process.  The control over the widths of these distributions will help us in adding additional numerical diffusion wherever required, {\em e.g.}, in expansions which may contain sonic points and thus need additional diffusion to avoid nonphysical solutions.  The new set of compactly supported distributions is illustrated in Figure \ref{fig:1} for the 1-D case. 
We introduce the compactly supported rectangular distributions with flexible velocities on both sides of $v=0$ for convenience.  In the limit of vanishing support, these new distributions degenerate to Dirac-delta distributions, leading to the discrete velocities $\lambda_{i}$ and $-\lambda_{i}$.    
For our model, $\hat{f}^{eq}_{i}$ (shown in Figure \ref{fig:1} ) is defined as
\begin{equation}
	\hat{f}^{eq}_{i} = \left\{\begin{array}{cc}
f^{eq}_{+,i},& \lambda_{i}-\delta \lambda_{i}\leq v \leq\lambda_{i}+\delta\lambda_{i},\\
f^{eq}_{-,i},& -\lambda_{i}-\delta\lambda_{i}\leq v\leq-\lambda_{i}+\delta \lambda_{i},\\
0,&  \text{otherwise}\end{array} \right\}
	\label{eq:3_3}
	\end{equation}
Thus, $\hat{f}^{eq}_{i}$ is zero everywhere except in the velocity ranges $[-\lambda_{i}-\delta \lambda_{i}, -\lambda_{i}+\delta\lambda_{i}]$ and  $[\lambda_{i}-\delta \lambda_{i}, \lambda_{i}+\delta\lambda_{i}]$. The Figure \ref{fig:1} corresponds to the first component of $\textbf{U}$.   
Now, let us recall the moment relations in 1D. 
	\begin{subequations}
	\label{eq:3_1}
	\begin{equation}
	U_{i} = \int^{\infty}_{-\infty}dv \ \hat{f}^{eq}_{i}= \left\langle \hat{f}^{eq}_{i} \right\rangle 
	\end{equation}
	\begin{equation}
	G_{i} = \int^{\infty}_{-\infty}vdv \ \hat{f}^{eq}_{i}= \left\langle v \hat{f}^{eq}_{i} \right\rangle 
	\end{equation}
	\end{subequations}
Here, $U_{i}$ and $G_{i}$ are the $i^{th}$ conserved variable and inviscid flux terms, respectively. Substituting (\ref{eq:3_3}) in the moment relations \eqref{eq:3_1}, we obtain  
        \begin{subequations}
        \label{distributions-derivation} 
        \begin{equation} 
        2 \delta \lambda_{i}  f^{eq}_{+, i} + 2 \delta \lambda_{i}f^{eq}_{-, i} \ = f^{eq}_{1i}+ f^{eq}_{2i}= U_{i} 
        \label{dist-deriv1} 
        \end{equation} 
        \begin{equation} 
         \lambda{_i} \left( 2 \delta \lambda_{i}  f^{eq}_{+, i} - 2 \delta \lambda_{i}f^{eq}_{-, i} \right)= \lambda_{i} \left(  f^{eq}_{1i} - f^{eq}_{2i} \right) = G_{i} 
        \label{dist-deriv2} 
        \end{equation} 
        \end{subequations}
where $f^{eq}_{1i}= 2 \delta \lambda_{i}  f^{eq}_{+, i}$ and $f^{eq}_{2i}= 2 \delta \lambda_{i}  f^{eq}_{-, i}$ are areas under respective rectangles. Solving the above for $f^{eq}_{1i}$ and $f^{eq}_{2i}$, we obtain 
	\begin{equation}
	f^{eq}_{1i}= \frac{U_{i}}{2}+\frac{G_{i}}{2\lambda_{i}}, f^{eq}_{2i}= \frac{U_{i}}{2}-\frac{G_{i}}{2\lambda_{i}}
	\label{eq:3_4}
	\end{equation}
	Further, computing the $v^{2}$ moment, we get
	\begin{subequations}
	\label{eq:3_5}
	\begin{eqnarray}
	\left\langle v^{2} \hat{f}^{eq}_{i} \right\rangle &&= (\lambda_{i}^{2}+ \frac{(\delta \lambda_{i})^{2}}{3})2(\delta \lambda_{i})(f^{eq}_{+,i}+ f^{eq}_{-,i})\\
 &&= (\lambda_{i}^{2}+ \frac{(\delta \lambda_{i})^{2}}{3})(f^{eq}_{1i}+ f^{eq}_{2i})\\
	&&=  (\lambda^{2}_{i}+ \frac{(\delta \lambda_{i})^{2}}{3}) U_{i}\text{ (from }\eqref{dist-deriv1})\\
    &&= \widetilde{\lambda}^{2}_{i} U_{i}
	\end{eqnarray}
	\end{subequations}
	We note that as $\delta \lambda_{i}$ changes, the zeroth and first moment relations in Equation \eqref{distributions-derivation} remain the same (in terms of $f^{eq}_{1i}$ and $f^{eq}_{2i}$). However, the second moment (Equation \eqref{eq:3_5}) changes. Thus, for our model, we preserve all three moments. To understand the significance of $v^{2}$ moment and the width $\delta \lambda_{i}$, we perform a Chapman-Enskog type analysis for a first order approximation to $\textbf{f}_{i}$ (\ref{appendix:a1}). We observe that the  $\left\langle v^{2} \hat{f}^{eq}_{i} \right\rangle$ moment acts as a viscous term. Thus a non zero $\delta\lambda_{i}$ adds to viscosity.
	
	\subsection{\texorpdfstring{Limiting case: $\delta\lambda_{i}\rightarrow$ 0}{Limiting case: ??i ? 0}}
	In the limiting case of $\delta\lambda_{i}\rightarrow$ 0, the equilibrium distribution function reduces to Dirac-delta distributions at the discrete velocities $\pm \lambda_{i}$. That is, in this limiting case, 
	\begin{equation}
	\hat{f}^{eq}_{i}= f^{eq}_{1i}\delta(v- \lambda_{i})+ f^{eq}_{2i}\delta(v+ \lambda_{i})
	\label{eq:3_6}
	\end{equation}
	The $\left\langle \hat{f}^{eq}_{i} \right\rangle$ and $\left\langle v \hat{f}^{eq}_{i} \right\rangle$ moment relations then give us:
	\begin{equation}
	f^{eq}_{1i}=\frac{U_{i}}{2}+ \frac{G_{i}}{2\lambda_{i}},  f^{eq}_{2i}=\frac{U_{i}}{2}- \frac{G_{i}}{2\lambda_{i}}
	\label{eq:3_7}
	\end{equation}
	The $\left\langle v^{2} \hat{f}^{eq}_{i} \right\rangle$ moment in this limiting case is
	\begin{equation}
	\left\langle v^{2} \hat{f}^{eq}_{i} \right\rangle= \lambda_{i}^{2}(f^{eq}_{1i}+ f^{eq}_{2i})= \lambda_{i}^{2}U_{i}
	\label{eq:3_8}
  \end{equation}

\section{Kinetic scheme for 1D Euler equations}
  To simplify our numerical method, we define an equivalent {\em Flexible Velocity Boltzmann Equation} such that the zeroth, first and second moment relations for the equilibrium distribution function given by \eqref{eq:3_1} and \eqref{eq:3_5} are satisfied. That equation is
\begin{equation}
\frac{\partial \textbf{f}_{i}}{\partial t}+ \frac{\partial (\widetilde{\Lambda}_{i}\textbf{f}_{i})}{\partial x}= 
- \frac{1}{\epsilon} \left[\textbf{f}_{i}- \widetilde{\textbf{f}}_{i}^{eq}\right] 
\label{eq:4_1}
\end{equation}
Here,
\begin{subequations}
\label{eq:4_2}
\begin{equation}
\widetilde{\textbf{f}}_{i}^{eq}= \begin{bmatrix} \widetilde{f}^{eq}_{1i} \\ \widetilde{f}^{eq}_{2i} \end{bmatrix}, \widetilde{\Lambda}_{i}= \begin{bmatrix} \widetilde{\lambda}_{i} & 0\\ 0 & -\widetilde{\lambda}_{i} \end{bmatrix},
\end{equation}
\begin{equation}
\widetilde{f}^{eq}_{1i}=\frac{U_{i}}{2}+ \frac{G_{i}}{2\widetilde{\lambda}_{i}}, \ \widetilde{f}^{eq}_{2i}=\frac{U_{i}}{2}- \frac{G_{i}}{2\widetilde{\lambda}_{i}}; \ \widetilde{\lambda}_{i} = \sqrt{\lambda^{2}_{i}+ \frac{(\delta \lambda_{i})^{2}}{3}}
\end{equation}
\end{subequations}
Now, given the row vector $\textbf{P}_{i}= \begin{bmatrix}1 & 1\end{bmatrix}$, the different moments of $\widetilde{\textbf{f}}_{i}^{eq}$ are
\begin{subequations}
\label{eq:4_3}
\begin{equation}
\textbf{P}_{i} \widetilde{\textbf{f}}_{i}^{eq}= \widetilde{f}^{eq}_{1i}+ \widetilde{f}^{eq}_{2i} = U_{i}
\end{equation}
\begin{equation}
\textbf{P}_{i}\widetilde{\Lambda}_{i} \widetilde{\textbf{f}}_{i}^{eq}= \widetilde{\lambda}_{i} \widetilde{f}^{eq}_{1i}-\widetilde{\lambda}_{i} \widetilde{f}^{eq}_{2i} = G_{i}
\end{equation}
\begin{equation}
\textbf{P}_{i}\widetilde{\Lambda}_{i}^{2} \widetilde{\textbf{f}}_{i}^{eq}= \widetilde{\lambda}^{2}_{i} (\widetilde{f}^{eq}_{1i}+ \widetilde{f}^{eq}_{2i}) =\widetilde{\lambda}^{2}_{i} U_{i} = (\lambda^{2}_{i}+ \frac{(\delta \lambda_{i})^{2}}{3})U_{i}
\end{equation}
\end{subequations}
Thus, all three moment relations are satisfied. We work in a finite volume framework and solve the conservation form of Boltzmann equations \eqref{eq:4_1} in $j^{th}$ cell. To solve the Boltzmann equations, we use operator splitting strategy. At the end of the $n^{th}$ time step, we allow the distribution function to relax (relaxation step) instantaneously to the equilibrium distribution function. Next, in the advection step we solve the advective part of Boltzmann equation numerically to get the distribution function for $(n+1)^{th}$ time step. Thus,\\\\
Relaxation step:\hspace{1cm} Instantaneous, {\em i.e.}, $\epsilon \rightarrow$ 0. Thus,
\begin{equation}
(\textbf{f}_{i})^{n}_{j}=  (\widetilde{\textbf{f}}_{i}^{eq})^{n}_{j}
\label{eq:4_4}
\end{equation}
Advection step:\hspace{1cm} The advective part of Boltzmann equation is
\begin{equation}
\frac{\partial \textbf{f}_{i}}{\partial t}+ \frac{\partial \textbf{h}_{i}}{\partial x}= 0; \textbf{h}_{i}= \widetilde{\Lambda}_{i}\widetilde{\textbf{f}}_{i}^{eq}
\label{eq:4_5}
\end{equation}
Rewriting the advection equation in integral form for $j^{th}$ cell,
\begin{equation}
\frac{d (\textbf{f}_{i})_{j}}{dt} = -(R_{i})^{n}_{j}= -\frac{1}{\Delta x}\left[ (\textbf{h}_{i})_{j+1/2}^{n}- (\textbf{h}_{i})_{j-1/2}^{n}\right]
\label{eq:4_6}
\end{equation}
The numerical flux is split using Courant splitting as follows
\begin{eqnarray}
(\textbf{h}_{i})_{j+1/2} &&= (\textbf{h}_{i}^{+})_{L}+(\textbf{h}_{i}^{-})_{R}\nonumber\\
&&= (\widetilde{\Lambda}_{i}^{+}\widetilde{\textbf{f}}_{i}^{eq})_{L}+ (\widetilde{\Lambda}_{i}^{-}\widetilde{\textbf{f}}_{i}^{eq})_{R}\text{, where } \widetilde{\Lambda}_{i}^{\pm}= \frac{1}{2}\left(\widetilde{\Lambda}_{i}\pm |\widetilde{\Lambda}_{i}|\right)\nonumber\\
&&=\frac{1}{2}\left\{(\textbf{h}_{i})_{L}+ (\textbf{h}_{i})_{R}\right\}-\frac{1}{2}\left\{(\Delta \textbf{h}_{i}^{+})_{j+1/2}- (\Delta \textbf{h}_{i}^{-})_{j+1/2}\right\};\nonumber\\
&&(\Delta \textbf{h}_{i}^{\pm})_{j+ 1/2} = \left[\widetilde{\Lambda}_{i}^{\pm}\left\{(\widetilde{\textbf{f}}_{i}^{eq})_{R}- (\widetilde{\textbf{f}}_{i}^{eq})_{L}\right\}\right]_{j+ 1/2}
\label{eq:4_7}
\end{eqnarray}
For 1st order overall accuracy, spatially we assume piecewise constant approximation for $\widetilde{\textbf{f}}_{i}^{eq}$ in each cell. Thus, $(\widetilde{\textbf{f}}_{i}^{eq})_{L}= (\widetilde{\textbf{f}}_{i}^{eq})_{j}$, $(\widetilde{\textbf{f}}_{i}^{eq})_{R}= (\widetilde{\textbf{f}}_{i}^{eq})_{j+1}$. Temporal derivative is approximated using Euler method as $ \frac{d (\textbf{f}_{i})_{j}}{dt}= \frac{(\textbf{f}_{i})_{j}^{n+1}- (\widetilde{\textbf{f}}_{i}^{eq})_{j}^{n}}{\Delta t}$.

To obtain the macroscopic update formula for the $i^{th}$ conserved variable $U_{i}$ in the $j^{th}$ cell, we take moment of \eqref{eq:4_6} by multiplying it by $\textbf{P}_{i}$. This gives us 
\begin{equation}
\frac{d (U_{i})_{j}}{dt} =  -\frac{1}{\Delta x}\left[ (G_{i})_{j+1/2}^{n}- (G_{i})_{j-1/2}^{n}\right]
\label{eq:4_7a}
\end{equation}
where the macroscopic flux at the interface is given by
\begin{eqnarray}
(G_{i})_{j+1/2} =&& \textbf{P}_{i} (\textbf{h}_{i})_{j+1/2}\nonumber\\
=&&\frac{1}{2}\left\{(G_{i})_{L}+ (G_{i})_{R}\right\} -\frac{(\widetilde{\lambda}_{i})_{j+1/2}}{2}\left\{(U_{i})_{R}- (U_{i})_{L}\right\}\nonumber\\
=&&\frac{1}{2}\left\{(G_{i})_{L}+ (G_{i})_{R}\right\}-\frac{1}{2}\left\{(\Delta G_{i}^{+})_{j+1/2}- (\Delta G_{i}^{-})_{j+1/2}\right\};\nonumber\\
&&(\Delta G_{i}^{\pm})_{j+ 1/2} = \left[{\pm}\frac{\widetilde{\lambda}_{i}}{2}\left\{(U_{i})_{R}- (U_{i})_{L}\right\}\right]_{j+ 1/2}
\label{eq:4_7b}
\end{eqnarray}
  
\subsection{Stability of Kinetic model}
We consider Bouchut's criterion \cite{bouchut1999construction} for existence of kinetic entropies. It requires the eigenvalues of derivatives of the equilibrium distributions, {\em i.e.}, $\widetilde{\textbf{f}}^{eq}_{1}(=\widetilde{f}^{eq}_{1i})$ and  $\widetilde{\textbf{f}}^{eq}_{2}(=\widetilde{f}^{eq}_{2i})$, to be real and non negative. These derivatives are:
		\begin{subequations}
		\label{eq:4_8}
		\begin{equation}
		\frac{\partial \widetilde{\textbf{f}}^{eq}_{1}}{\partial \textbf{U}}= \frac{1}{2}\begin{bmatrix}1 &0 &0\\ 0& 1 &0 \\ 0& 0& 1\end{bmatrix}+ \frac{1}{2}\begin{bmatrix}\frac{1}{\widetilde{\lambda}_{1}} &0 &0\\ 0& \frac{1}{\widetilde{\lambda}_{2}} &0 \\ 0& 0& \frac{1}{\widetilde{\lambda}_{3}}\end{bmatrix}\frac{\partial \textbf{G}}{\partial \textbf{U}}
		\end{equation}
		\begin{equation}
		\frac{\partial \widetilde{\textbf{f}}^{eq}_{2}}{\partial \textbf{U}}= \frac{1}{2}\begin{bmatrix}1 &0 &0\\ 0& 1 &0 \\ 0& 0& 1\end{bmatrix}- \frac{1}{2}\begin{bmatrix}\frac{1}{\widetilde{\lambda}_{1}} &0 &0\\ 0& \frac{1}{\widetilde{\lambda}_{2}} &0 \\ 0& 0& \frac{1}{\widetilde{\lambda}_{3}}\end{bmatrix}\frac{\partial \textbf{G}}{\partial \textbf{U}}
		\end{equation}
	\end{subequations}
	Then, as per Bouchut's criterion 
	\begin{equation}
	\sigma\left(\frac{\partial \widetilde{\textbf{f}}^{eq}_{1,2}}{\partial \textbf{U}}\right) \subset\left[0,\infty\right)
	\label{eq:4_9}
	\end{equation}
	Here $\sigma$ represents the spectral radius.  For our scheme, we take $\lambda_{i}=\lambda$ and $\delta \lambda_{i}= \delta \lambda$ (thus $\widetilde{\lambda}_{i}=\widetilde{\lambda}$, {\em i.e.}, numerical diffusion is a scalar). Then, \eqref{eq:4_8} and \eqref{eq:4_9} lead to:
	\begin{equation}
	\widetilde{\lambda} \geq max\left(\left|u-a\right|, \left|u\right|, \left|u+a\right|\right)
	\label{eq:4_10}
	\end{equation}

\subsection{\texorpdfstring{Fixing $\lambda$}{Fixing ?}}
We take $\lambda_{i}=\lambda$ and $\delta \lambda_{i}= \delta \lambda$ (thus $\widetilde{\lambda}_{i}=\widetilde{\lambda}$). $\lambda$ is defined such that it satisfies the Rankine-Hugoniot jump conditions at a steady discontinuity. First we define the following speeds at the cell interface:
\begin{subequations}
\label{eq:4_15}
\begin{equation}
\lambda_{s}= min_{i}(\frac{\left|\Delta G_{i}\right|}{\left|\Delta U_{i}\right|+ \epsilon_{0}}),\text{ }\Delta = ()_{R}- ()_{L} 
\end{equation}
\begin{equation}
\lambda_{min}= max\left\{ min(|u-a|,|u|,|u+a|)_{L}, min(|u-a|,|u|,|u+a|)_{R}\right\}
\end{equation}
\end{subequations}
Then, we define $\lambda$ as follows
\begin{equation}
\lambda=\left\{\begin{array}{l}
\lambda_{min},\text{ if }\left\|\Delta \textbf{U}\right\| \leq \epsilon_{1}\text{ (uniform flow)}\\
\lambda_{s},\text{ if }\left\|\Delta \textbf{U}\right\| > \epsilon_{1}, \left\|\Delta \textbf{G}\right\| \leq \epsilon_{2}\text{ (steady discontinuity)}\\
max(\lambda_{min}, \lambda_{s}),\text{ otherwise} 
\end{array}\right\}
\label{eq:4_16}
\end{equation}
Here, we use $\lambda_{min}$ to provide a lower limit to $\lambda$ everywhere except at a steady discontinuity. We take $\epsilon_{1}= 10^{-5}$, $\epsilon_{2}= 10^{-8}$ and $\epsilon_{0}= 10^{-14}$ for all the test cases.
\subsection{\texorpdfstring{Fixing $\delta\lambda$}{Fixing ??}}
$\lambda$ as defined in the previous subsection is still a very low numerical coefficient of diffusion, which can lead to formation of entropy-violating expansion shocks when expansive sonic points are present. Thus we augment the coefficient of diffusion by having a non-zero $\delta\lambda$  in smooth regions, which include expansions. In smooth regions, we set $\delta\lambda$  based on equation \eqref{eq:4_10}, as follows.  
\begin{equation}
\widetilde{\lambda}^{2}= (\lambda^{2}+ \frac{(\delta \lambda)^{2}}{3}) \geq \left[max\left(\left|u-a\right|, \left|u\right|, \left|u+a\right|\right)\right]^{2}
\label{eq:4_17}
\end{equation}
Thus,
\begin{subequations}
\label{eq:4_18}
\begin{equation}
\delta \lambda= \left\{\begin{array}{l}
\sqrt{3(\lambda^{2}_{max}- \lambda^{2})},\text{ in smooth regions}\\
0,\text{ otherwise}
\end{array}\right\}
\end{equation}
\begin{equation}
\lambda_{max}= max\left\{ max(|u-a|,|u|,|u+a|)_{L}, max(|u-a|,|u|,|u+a|)_{R}\right\}
\end{equation}
\end{subequations}
Thus, in smooth regions, the numerical diffusion in our scheme changes to that of the Rusanov or Local Lax-Friedrich (LLF) scheme.  
\subsection{Relative entropy}
\begin{figure}
\includegraphics[width=14cm]{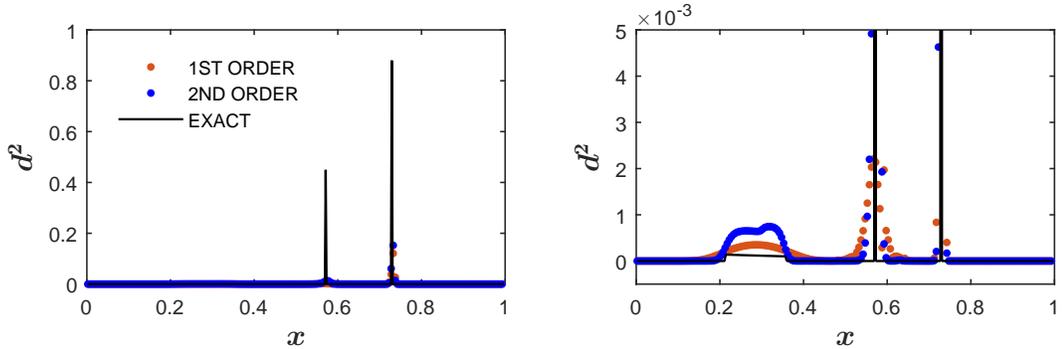}
\caption{a) Relative entropy $d^{2}$ for Sod's shock tube problem at t=0.2, b) Zoomed figure} 
\label{fig:2}
\end{figure}

Now, the next problem is identifying the smooth flow regions in contrast to the discontinuities (high gradient regions). For this purpose, we introduce a novel formulation for the relative entropy, $d^{2}$(first presented at \cite{shashi2022entropy}).  First we introduce a {\em kinetic entropy variable}, $\omega$ as 
\begin{equation}
    \omega = \displaystyle \frac{\partial \hat{H}}{\partial f^{eq}} 
\end{equation}  
where $\hat{H}$ is the kinetic entropy.  For the classical case, it refers to \eqref{eq:2_8} without the moments 
($\hat{H} = f^{eq} \ln f^{eq}$).  We now define the {\em relative entropy} as the {\em kinetic entropy distance}, given by  
\begin{equation}
    d^{2} = \langle \Delta \omega \cdot \Delta f^{eq} \rangle ; \ \Delta=()_{R} - ()_{L} 
\end{equation} 
where $\left\langle \right\rangle$ refers to taking moments.  For the classical case with continuous molecular velocity, from \eqref{eq:2_8}, we have 
\begin{equation}
    \omega = 1+ \ln f^{eq}  
\end{equation} 
and thus we recover 
\begin{equation}
d^{2}= \left\langle \Delta \left(lnf^{eq}\right) \cdot \Delta f^{eq} \right\rangle = \left\langle ln\frac{f^{eq}_{R}}{f^{eq}_{L}}(f^{eq}_{R}- f^{eq}_{L})\right\rangle
\label{eq:4_20}
\end{equation}
which is the Kullback-Leibler divergence \cite{kullback1951information}.  However, as our kinetic model is closer to the discrete Boltzmann system, we use Bouchut's form of kinetic entropy function \cite{bouchut1999construction}, while noting that Bouchut's stability criterion is satisfied in smooth regions, ensuring existence of kinetic entropies.  To begin, we rewrite our equilibrium distribution function as
\begin{eqnarray}
\widetilde{\textbf{f}}_{i}^{eq}&&=\begin{bmatrix} \widetilde{\textbf{f}}^{eq}_{1i} \\ \widetilde{\textbf{f}}^{eq}_{2i} \end{bmatrix}= \begin{bmatrix} \frac{U_{i}}{2}+ \frac{G_{i}}{2\widetilde{\lambda}}\\ \frac{U_{i}}{2}- \frac{G_{i}}{2\widetilde{\lambda}}\end{bmatrix} = \begin{bmatrix} \frac{1}{2} \\ \frac{1}{2} \end{bmatrix}U_{i} + \begin{bmatrix} \frac{1}{2\widetilde{\lambda}} \\ -\frac{1}{2\widetilde{\lambda}} \end{bmatrix}G_{i}
\label{eq:4_21a}
\end{eqnarray}
or,
\begin{eqnarray}
\widetilde{\textbf{f}}^{eq}&&= \begin{bmatrix} \widetilde{f}^{eq}_{11} \\ \widetilde{f}^{eq}_{21} \\  \widetilde{f}^{eq}_{12} \\ \widetilde{f}^{eq}_{22} \\  \widetilde{f}^{eq}_{13} \\ \widetilde{f}^{eq}_{23} \\ \end{bmatrix}= \frac{1}{2}\begin{bmatrix} 1&0&0\\ 1&0&0\\ 0&1&0 \\ 0&1&0 \\ 0&0&1\\ 0&0&1 \end{bmatrix}\textbf{U}+ \frac{1}{2\widetilde{\lambda}}\begin{bmatrix} 1 &0 &0\\ -1 &0 &0\\ 0& 1 &0 \\ 0& -1 &0 \\ 0& 0& 1\\ 0& 0& -1 \end{bmatrix}\textbf{G}\nonumber\\
&&= \boldsymbol{\alpha}_{0}\textbf{U}+ \boldsymbol{\alpha}_{1}\textbf{G}
\label{eq:4_21}
\end{eqnarray}
Then, the kinetic entropy $\hat{H}$ is given by
\begin{equation}
\hat{H} = \boldsymbol{\alpha}_{0}\eta+ \boldsymbol{\alpha}_{1} \psi
\label{eq:4_22}
\end{equation}
where ($\eta, \psi$) are the macroscopic entropy - entropy flux pairs. For Euler equations, $\eta= \rho s$ and $\psi= \rho u s$, where $s$= $c_{v} \ln \frac{p}{\rho^{\gamma}} + \textrm{constant}$. Now,
\begin{subequations}
\label{eq:4_23}
\begin{equation}
\left\langle \widetilde{\textbf{f}}^{eq} \right\rangle= \textbf{P}\widetilde{\textbf{f}}^{eq}=\textbf{U} \Rightarrow \textbf{P}\boldsymbol{\alpha}_{0}= I, \textbf{P}\boldsymbol{\alpha}_{1}= 0
\end{equation}
\begin{equation}
\left\langle \widetilde{\Lambda}\widetilde{\textbf{f}}^{eq} \right\rangle= \textbf{P}\widetilde{\Lambda}\widetilde{\textbf{f}}^{eq}=\textbf{G} \Rightarrow \textbf{P}\widetilde{\Lambda}\boldsymbol{\alpha}_{0}= 0, \textbf{P}\widetilde{\Lambda}\boldsymbol{\alpha}_{1}= I
\end{equation}
\end{subequations}
Therefore,
\begin{subequations}
\label{eq:4_24}
\begin{equation}
\left\langle \hat{H} \right\rangle= \textbf{P}(\boldsymbol{\alpha}_{0}\eta+ \boldsymbol{\alpha}_{1} \psi)= \eta
\end{equation}
\begin{equation}
\left\langle \widetilde{\Lambda} \hat{H} \right\rangle= \textbf{P}\widetilde{\Lambda}(\boldsymbol{\alpha}_{0} \eta+ \boldsymbol{\alpha}_{1} \psi)= \psi
\end{equation}
\end{subequations}
Now the relative entropy is 
\begin{eqnarray}
d^{2}&&= \langle \Delta \omega \cdot \Delta \widetilde{\textbf{f}}^{eq} \rangle = 
\left\langle \Delta \left\{ \frac{\partial \hat{H} (\widetilde{\textbf{f}}^{eq})}{\partial \widetilde{\textbf{f}}^{eq}} \right\} \cdot \Delta \widetilde{\textbf{f}}^{eq}\right\rangle \nonumber\\
&&= \left\langle \Delta \left\{ \boldsymbol{\alpha}_{0} \frac{\partial \eta}{\partial \widetilde{\textbf{f}}^{eq}}+ \boldsymbol{\alpha}_{1} \frac{\partial \psi}{\partial \widetilde{\textbf{f}}^{eq}} \right\} \cdot  \Delta \widetilde{\textbf{f}}^{eq} \right\rangle \nonumber\\ 
&&= \left\langle \left\{ \boldsymbol{\alpha}_{0} \Delta \left( \frac{\partial \eta}{\partial \widetilde{\textbf{f}}^{eq}} \right) + \boldsymbol{\alpha}_{1} \Delta \left( \frac{\partial \psi}{\partial \widetilde{\textbf{f}}^{eq}} \right) \right\} \cdot \Delta\widetilde{\textbf{f}}^{eq} \right\rangle\text{ (since $\lambda, \delta \lambda$= f(L,R))} \nonumber\\
&&=\textbf{P}\boldsymbol{\alpha}_{0} \Delta \left( \frac{\partial \eta}{\partial \widetilde{\textbf{f}}^{eq}} \right) \cdot \Delta\widetilde{\textbf{f}}^{eq}+ \textbf{P}\boldsymbol{\alpha}_{1} \Delta \left( \frac{\partial \psi}{\partial \widetilde{\textbf{f}}^{eq}} \right) \cdot \Delta\widetilde{\textbf{f}}^{eq} \nonumber\\
&&=\Delta \left( \frac{\partial \eta}{\partial \widetilde{\textbf{f}}^{eq}} \right) \cdot \Delta\widetilde{\textbf{f}}^{eq} \nonumber\\
&&=\Delta\begin{bmatrix}\frac{\partial \eta}{\partial U_{1}}\frac{\partial U_{1}}{\partial \widetilde{f}^{eq}_{11}}+ \frac{\partial \eta}{\partial U_{2}}\frac{\partial U_{2}}{\partial \widetilde{f}^{eq}_{11}}+\frac{\partial \eta}{\partial U_{3}}\frac{\partial U_{3}}{\partial \widetilde{f}^{eq}_{11}}\\
\frac{\partial \eta}{\partial U_{1}}\frac{\partial U_{1}}{\partial \widetilde{f}^{eq}_{21}}+ \frac{\partial \eta}{\partial U_{2}}\frac{\partial U_{2}}{\partial \widetilde{f}^{eq}_{21}}+\frac{\partial \eta}{\partial U_{3}}\frac{\partial U_{3}}{\partial \widetilde{f}^{eq}_{21}}\\
\frac{\partial \eta}{\partial U_{1}}\frac{\partial U_{1}}{\partial \widetilde{f}^{eq}_{12}}+ \frac{\partial \eta}{\partial U_{2}}\frac{\partial U_{2}}{\partial \widetilde{f}^{eq}_{12}}+\frac{\partial \eta}{\partial U_{3}}\frac{\partial U_{3}}{\partial \widetilde{f}^{eq}_{12}}\\
\frac{\partial \eta}{\partial U_{1}}\frac{\partial U_{1}}{\partial \widetilde{f}^{eq}_{22}}+ \frac{\partial \eta}{\partial U_{2}}\frac{\partial U_{2}}{\partial \widetilde{f}^{eq}_{22}}+\frac{\partial \eta}{\partial U_{3}}\frac{\partial U_{3}}{\partial \widetilde{f}^{eq}_{22}}\\
\frac{\partial \eta}{\partial U_{1}}\frac{\partial U_{1}}{\partial \widetilde{f}^{eq}_{13}}+ \frac{\partial \eta}{\partial U_{2}}\frac{\partial U_{2}}{\partial \widetilde{f}^{eq}_{13}}+\frac{\partial \eta}{\partial U_{3}}\frac{\partial U_{3}}{\partial \widetilde{f}^{eq}_{13}}\\
\frac{\partial \eta}{\partial U_{1}}\frac{\partial U_{1}}{\partial \widetilde{f}^{eq}_{23}}+ \frac{\partial \eta}{\partial U_{2}}\frac{\partial U_{2}}{\partial \widetilde{f}^{eq}_{23}}+\frac{\partial \eta}{\partial U_{3}}\frac{\partial U_{3}}{\partial \widetilde{f}^{eq}_{23}}\end{bmatrix} \cdot \Delta\begin{bmatrix} \widetilde{f}^{eq}_{11} \\ \widetilde{f}^{eq}_{21} \\ \widetilde{f}^{eq}_{12} \\ \widetilde{f}^{eq}_{22} \\ \widetilde{f}^{eq}_{13} \\ \widetilde{f}^{eq}_{23} \end{bmatrix} \\
d^{2} &&=\Delta\begin{bmatrix}\frac{\partial \eta}{\partial U_{1}} \\\frac{\partial \eta}{\partial U_{1}} \\
\frac{\partial \eta}{\partial U_{2}} \\\frac{\partial \eta}{\partial U_{2}} \\ \frac{\partial \eta}{\partial U_{3}} \\\frac{\partial \eta}{\partial U_{3}} \end{bmatrix} \cdot \Delta\begin{bmatrix} \widetilde{f}^{eq}_{11} \\ \widetilde{f}^{eq}_{21} \\ \widetilde{f}^{eq}_{12} \\ \widetilde{f}^{eq}_{22} \\ \widetilde{f}^{eq}_{13} \\ \widetilde{f}^{eq}_{23} \end{bmatrix} \nonumber\\
&&=\Delta \frac{\partial \eta}{\partial U_{1}} \Delta( \widetilde{f}^{eq}_{11}+\widetilde{f}^{eq}_{21}) +\Delta\frac{\partial \eta}{\partial U_{2}} \Delta( \widetilde{f}^{eq}_{12}+\widetilde{f}^{eq}_{22}) +\Delta\frac{\partial \eta}{\partial U_{3}} \Delta(\widetilde{f}^{eq}_{13}+\widetilde{f}^{eq}_{23})\nonumber\\
&&=  \Delta \frac{\partial \eta}{\partial U_{1}} \Delta U_{1}+\Delta \frac{\partial \eta}{\partial U_{2}} \Delta U_{2}+\Delta \frac{\partial \eta}{\partial U_{3}} \Delta U_{3}\nonumber\\
d^{2} &&= \left(\Delta \frac{\partial \eta}{\partial \textbf{U}}\right)^{T} \cdot \Delta \textbf{U}\\
&&=\Delta \left(\frac{\gamma -s}{\gamma -1} - \frac{\rho u^{2}}{2p}\right) \Delta (\rho)+ \Delta \left(\frac{\rho u}{p}\right) \Delta (\rho u)+ \Delta \left(-\frac{\rho}{p}\right) \Delta (\rho E)
\label{eq:4_25}
\end{eqnarray} 
Thus, relative entropy in our formulation is the scalar product of change in entropy variable and change in conserved variable at the interface. This expression, derived from kinetic theoretic considerations, happens to coincide with the macroscopic entropy distance defined by Zaide and Roe \cite{zaide2009entropy}. We analyze how this relative entropy varies for Sod's shock tube problem (Toro Test Case 1). For this unsteady problem, solution at t= 0.2 is considered; it comprises of an expansion, a contact discontinuity and a shock wave. The solution (numerical and exact) is shown in Figure \ref{fig:2}. We can see that $d^{2}$ gives a positive signal ($>$0) at expansions as well as at contact discontinuities and shocks, its magnitude being much smaller at expansions. Since at expansions there is no change in entropy, we can identify expansions using the following criteria.  
\begin{equation}
\text{At expansions: }d^{2}>0, \Delta s= 0
\label{eq:4_26}
\end{equation}
However, for numerical solutions, $|\Delta s|$ at expansions is not exactly zero. Taking that into account, we use a less stringent condition to identify smooth regions, as follows:
\begin{equation}
\textrm{In smooth regions: } d^{2} > 0 \ \textrm{and} \ |\Delta s| \leq k(s_{max}-s_{min})
\label{eq:4_27}
\end{equation} 
where $s_{max}$ and $s_{min}$ are the maximum and minimum entropy in the domain at a given time level. $k$ is a fraction, taken as small as possible; $k$= 0.1 has proven to be sufficient for all our 1D and 2D Euler test cases.  Now, the modified condition accepts very small entropy changes. Therefore, we introduce a corresponding condition also for $d^{2}$, based on the term containing the entropy in its definition \eqref{eq:4_25} and conclude that non zero $\Delta s$ introduces an error $-\frac{\Delta s \Delta \rho}{\gamma -1} $ in the relative entropy $d^{2}$. Taking this into account, our modified criteria for introducing LLF type numerical diffusion is given by 
\begin{equation}
\textrm{In smooth regions: }d^{2}>-\frac{\Delta s \Delta \rho}{\gamma -1} , |\Delta s| \leq 0.1(s_{max}-s_{min})
\label{eq:4_28}
\end{equation}
Finally, we can describe the numerical coefficient of diffusion $\widetilde{\lambda}$ as follows
\begin{equation}
   \widetilde{\lambda}= \left\{\begin{array}{l}
   \sqrt{\lambda^{2}+ \frac{(\delta\lambda)^{2}}{3}}\text{, if criteria \eqref{eq:4_28} is satisfied,}\\
   \lambda\text{, otherwise}
 \end{array}\right\}  
\label{eq:4_28a}
\end{equation}

\section{Higher order accuracy}
For second order accuracy, we assume piecewise linear approximation for $\widetilde{\textbf{f}}_{i}^{eq}$ in each cell, for evaluating $(\widetilde{\textbf{f}}_{i}^{eq})_{L/R}$. However, a second order accurate scheme cannot be monotone as it would give rise to oscillations. Thus, a limiter function is used to limit the slope, as follows:
\begin{subequations}
\label{eq:5_1}
\begin{equation}
(\widetilde{\textbf{f}}_{i}^{eq})_{j+\frac{1}{2},L}= (\widetilde{\textbf{f}}_{i}^{eq})_{j}+ \frac{1}{2}\phi(\Delta^{+}_{j}\widetilde{\textbf{f}}_{i}^{eq}, \Delta^{-}_{j}\widetilde{\textbf{f}}_{i}^{eq} )
\label{eq:left-recon}
\end{equation}
\begin{equation}
(\widetilde{\textbf{f}}_{i}^{eq})_{j-\frac{1}{2},R}= (\widetilde{\textbf{f}}_{i}^{eq})_{j}- \frac{1}{2}\phi(\Delta^{-}_{j}\widetilde{\textbf{f}}_{i}^{eq}, \Delta^{+}_{j}\widetilde{\textbf{f}}_{i}^{eq} )
\label{eq:right-recon}
\end{equation}
\begin{equation} 
\Delta^{+}_{j}\widetilde{\textbf{f}}_{i}^{eq} = (\widetilde{\textbf{f}}_{i}^{eq})_{j+1}- (\widetilde{\textbf{f}}_{i}^{eq})_{j}, \Delta^{-}_{j}\widetilde{\textbf{f}}_{i}^{eq} = (\widetilde{\textbf{f}}_{i}^{eq})_{j}- (\widetilde{\textbf{f}}_{i}^{eq})_{j-1} 
\label{eq:r_j}  
\end{equation}
\end{subequations}
\begin{equation}
\phi(x,y)= \frac{x^{2}y+ x y^{2}}{x^{2}+ y^{2}}\text{ (VanAlbada limiter)}
\label{eq:phi_r}
\end{equation}
Thus,
\begin{eqnarray}
    (U_{i})_{j+\frac{1}{2},L} &=& (\widetilde{f}_{1i}^{eq})_{j+\frac{1}{2},L}+ (\widetilde{f}_{2i}^{eq})_{j+\frac{1}{2},L}\nonumber\\
    &=& (U_{i})_{j}+ \frac{1}{2}\phi(\Delta^{+}_{j}\widetilde{f}_{1i}^{eq}, \Delta^{-}_{j}\widetilde{f}_{1i}^{eq})+ \frac{1}{2}\phi(\Delta^{+}_{j}\widetilde{f}_{2i}^{eq}, \Delta^{-}_{j}\widetilde{f}_{2i}^{eq})
\label{eq:phi_U}
\end{eqnarray}
At this point, we make an approximation. It is inspired by the work of Kumar \& Dass (\cite{kumar}), who, in the continuous-velocity space, approximate the integral of limiter function of two variables by the limiter function of integral of the two variables. In our framework, integrals are replaced by summations. Thus, we approximate \eqref{eq:phi_U} by the following expression. 
\begin{eqnarray}
    (U_{i})_{j+\frac{1}{2},L} &=& (U_{i})_{j}+ \frac{1}{2}\phi(\Delta^{+}_{j}\widetilde{f}_{1i}^{eq}+ \Delta^{+}_{j}\widetilde{f}_{2i}^{eq}, \Delta^{-}_{j}\widetilde{f}_{1i}^{eq}+ \Delta^{-}_{j}\widetilde{f}_{2i}^{eq}) \nonumber\\
    &=& (U_{i})_{j}+ \frac{1}{2}\phi(\Delta^{+}_{j} U_{i}, \Delta^{-}_{j} U_{i})
    \label{eq:5_1a}
\end{eqnarray}
Similarly,
\begin{equation}
    (U_{i})_{j-\frac{1}{2},R}= (U_{i})_{j}- \frac{1}{2}\phi(\Delta^{-}_{j} U_{i}, \Delta^{+}_{j} U_{i})
    \label{eq:5_1b}
\end{equation}
%
We rewrite \eqref{eq:5_1a} and \eqref{eq:5_1b} as
\begin{subequations}
\label{eq:5_4}
\begin{equation}
(U_{i})_{j+\frac{1}{2},L}= (U_{i})_{j}+ \frac{1}{2}\phi(r_{j})\left\{(U_{i})_{j}- (U_{i})_{j-1}\right\}
\label{eq:5_4a}
\end{equation}
\begin{equation}
(U_{i})_{j-\frac{1}{2},R}= (U_{i})_{j}- \frac{1}{2}\phi(\frac{1}{r_{j}})\left\{(U_{i})_{j+1}- (U_{i})_{j}\right\}
\label{eq:5_4b}
\end{equation}
\begin{equation}
\phi(r)= \frac{r^{2}+r}{r^{2}+1}, r_{j}= \frac{\Delta^{+}_{j}U_{i}}{\Delta^{-}_{j}U_{i}}= \frac{(U_{i})_{j+1}- (U_{i})_{j}}{(U_{i})_{j}- (U_{i})_{j-1}}
\label{eq:5_4c}
\end{equation}
\end{subequations}
If the denominator of the argument of the limiter function $\phi$ in Equation \eqref{eq:5_4a} is close to zero, we 
modify $r_{j}$, to avoid numerical overflow, as follows:
\begin{equation}
r_{j}= \frac{\Delta^{+}_{j}U_{i}}{sign(\Delta^{-}_{j}U_{i})*\epsilon_{3}}\text{, if } |\Delta^{-}_{j}U_{i}| < \epsilon_{3}
\label{eq:5_2}
\end{equation}
 where $sign()$ is the standard signum function.  Here, we take $\epsilon_{3}= 10^{-12}$. Next, temporal derivative is discretized using Strong Stability Preserving Runge Kutta (SSPRK) Method \cite{gottlieb2001strong}. The update formula is,
\begin{subequations}
\label{eq:5_3}
\begin{equation}
(\textbf{f}_{i})^{1}_{j}= (\widetilde{\textbf{f}}_{i}^{eq})^{n}_{j}- \Delta t R((\widetilde{\textbf{f}}_{i}^{eq})^{n}_{j})
\end{equation}
\begin{equation}
(\textbf{f}_{i})^{2}_{j}= \frac{1}{4}(\textbf{f}_{i})^{1}_{j}+ \frac{3}{4}(\widetilde{\textbf{f}}_{i}^{eq})^{n}_{j} -\frac{1}{4}\Delta t R((\textbf{f}_{i})^{1}_{j})
\end{equation}
\begin{equation}
(\textbf{f}_{i})^{n+1}_{j}= \frac{2}{3}(\textbf{f}_{i})^{2}_{j}+ \frac{1}{3}(\widetilde{\textbf{f}}_{i}^{eq})^{n}_{j} -\frac{2}{3}\Delta t R((\textbf{f}_{i})^{2}_{j})
\end{equation}
\end{subequations}
Here, R is the residual (see \eqref{eq:4_6}). The macroscopic update formula is then obtained by taking moment of \eqref{eq:5_3} and is given by
\begin{subequations}
\label{eq:5_5}
\begin{equation}
(U_{i})^{1}_{j}= (U_{i})^{n}_{j}- \Delta t R((U_{i})^{n}_{j})
\end{equation}
\begin{equation}
(U_{i})^{2}_{j}= \frac{1}{4}(U_{i})^{1}_{j}+ \frac{3}{4}(U_{i})^{n}_{j} -\frac{1}{4}\Delta t R((U_{i})^{1}_{j})
\end{equation}
\begin{equation}
(U_{i})^{n+1}_{j}= \frac{2}{3}(U_{i})^{2}_{j}+ \frac{1}{3}(U_{i})^{n}_{j} -\frac{2}{3}\Delta t R((U_{i})^{2}_{j})
\end{equation}
\end{subequations} 
\section{Kinetic scheme for Viscous flows}
 We consider a first order approximation to f, {\em i.e.}, $f= f^{CE}$, where $f^{CE}$ is the Chapman-Enskog distribution function. 
 For a first order approximation to $f$, the moments of variable velocity Boltzmann equation give us the macroscopic Navier-Stokes equations, given by 
\begin{equation}
\frac{\partial \textbf{U}}{\partial t}+ \frac{\partial \textbf{G}_{i}}{\partial x_{i}}= \frac{\partial \textbf{G}_{vis, i}}{\partial x_{i}}
\label{eq:6_1}
\end{equation}
with
\begin{equation}
\textbf{U}= \begin{bmatrix} \rho \\ \rho u_{j} \\ \rho E \end{bmatrix}, \textbf{G}_{i}= \begin{bmatrix} \rho u_{i} \\ \rho u_{i}u_{j}+ p \delta_{ij} \\ (\rho E+ p)u_{i} \end{bmatrix}, \textbf{G}_{vis, i}= \begin{bmatrix} 0 \\ \tau_{ij} \\ \tau_{ij}u_{j} -q_{i} \end{bmatrix}
\label{eq:6_2}
\end{equation}
and
\begin{equation}
\tau_{ij}= \mu\left(\frac{\partial u_{i}}{\partial x_{j}}+ \frac{\partial u_{j}}{\partial x_{i}}\right) -\frac{2 \mu}{3}\frac{\partial u_{k}}{\partial x_{k}}\delta_{ij}, \ q_{i} =-K \frac{\partial T}{\partial x_{i}}
\label{eq:6_3}
\end{equation}
 Here, the moment relations are
\begin{equation}
\int_{\mathbb{R}^{N}}d\textbf{v} \int_{\mathbb{R}^{+}}dI \ \bm{\Psi} f^{CE} = \textbf{U}, \int_{\mathbb{R}^{N}}v_{i}d\textbf{v} \int_{\mathbb{R}^{+}}dI \ \bm{\Psi} f^{CE}= \textbf{G}_{i}- \textbf{G}_{vis,i}
\label{eq:6_4}
\end{equation}

In 1D, we define $\hat{f}^{CE}_{i}$ in the same fashion as we did $\hat{f}^{eq}_{i}$, but satisfying the moment relations \eqref{eq:6_4}, {\em i.e.}, 
\begin{subequations}
\label{eq:6_5}
\begin{equation}
\left\langle \hat{f}^{CE}_{i} \right\rangle= f^{CE}_{1i}+ f^{CE}_{2i}= U_{i}
\end{equation}
\begin{equation}
\left\langle v\hat{f}^{CE}_{i} \right\rangle= \lambda f^{CE}_{1i}- \lambda f^{CE}_{2i} = G_{i} - G_{vis, i}
\end{equation}
\end{subequations}
The moment relations \eqref{eq:6_5} give us
\begin{subequations}
\label{eq:6_6}
\begin{equation}
f^{CE}_{1i}= \frac{U_{i}}{2}+ \frac{G_{i}- G_{vis, i}}{2 \lambda}
\end{equation}
\begin{equation}
f^{CE}_{2i}= \frac{U_{i}}{2}- \frac{G_{i}- G_{vis, i}}{2 \lambda}
\end{equation}
\end{subequations}
We use the operator splitting strategy to solve a Flexible Velocity Boltzmann Equation, with the only difference being that we now relax the distribution function to the Chapman-Enskog distribution function in the collision step. The kinetic numerical flux is evaluated using Courant splitting as follows.  
\begin{eqnarray}
(\textbf{h}_{i})_{j+1/2} = (\textbf{h}_{i}^{+})_{L}+(\textbf{h}_{i}^{-})_{R}= (\Lambda^{+}\textbf{f}_{i}^{CE})_{L}+ (\Lambda^{-}\textbf{f}_{i}^{CE})_{R}
\label{eq:6_7}
\end{eqnarray}

The macroscopic flux at the interface is then evaluated to be 
\begin{eqnarray}
(G_{i})_{j+1/2} =&& \textbf{P}_{i} (\textbf{h}_{i})_{j+1/2}\nonumber\\
=&&\frac{1}{2}\left\{(G_{i})_{L} - (G_{vis,i})_{L}\right\}+ \frac{1}{2}\left\{(G_{i})_{R} - (G_{vis,i})_{R}\right\} \nonumber\\
&&- \frac{(\lambda)_{j+1/2}}{2}\left\{(U_{i})_{R}- (U_{i})_{L}\right\}
\label{eq:6_8}
\end{eqnarray}

\section{Equilibrium distribution in 2D}
\begin{figure}[h!] 
	\centering
	\includegraphics[width=0.5\textwidth]{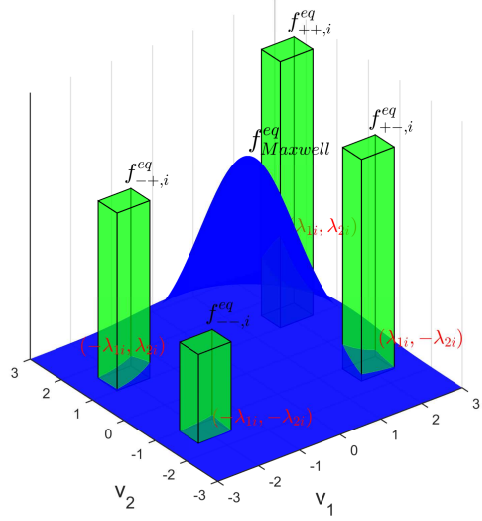}%
	\caption{Equilibrium distribution in 2D; zoomed portion of the infinite domain ($-\infty, \infty$) in both $v_{1}, v_{2}$ is shown.}%
	\label{fig:3}%
\end{figure}

	In 2D, we define the equilibrium distribution $\hat{f}^{eq}_{i}$ to take non-zero values in all four quadrants in $v_{1}-v_{2}$ plane as shown in Figure \ref{fig:3}.  As in 1-D, the average velocities ($\lambda_{1i}$,$\lambda_{2i}$) are utilized to provide the primary numerical diffusion based on R-H conditions while the $\delta \lambda_{1i}$ and $\delta \lambda_{2i}$ are used to provide the augmented numerical diffusion to avoid entropy violations.  
The 2-D distribution is given by 
	\begin{equation}
	\hat{f}^{eq}_{i} = \left\{\begin{array}{cc}
	f^{eq}_{++,i},& \lambda_{1i}-\delta \lambda_{1i} \leq v_{1} \leq  \lambda_{1i}+\delta \lambda_{1i},\\
            & \lambda_{2i}-\delta \lambda_{2i} \leq v_{2} \leq  \lambda_{2i}+\delta \lambda_{2i},\\
f^{eq}_{-+,i},& -\lambda_{1i}-\delta \lambda_{1i} \leq v_{1} \leq  -\lambda_{1i}+\delta \lambda_{1i},\\
            & \lambda_{2i}-\delta \lambda_{2i} \leq v_{2} \leq  \lambda_{2i}+\delta \lambda_{2i},\\
f^{eq}_{--,i},& -\lambda_{1i}-\delta \lambda_{1i} \leq v_{1} \leq  -\lambda_{1i}+\delta \lambda_{1i},\\
            & -\lambda_{2i}-\delta \lambda_{2i} \leq v_{2} \leq  -\lambda_{2i}+\delta \lambda_{2i},\\
f^{eq}_{+-,i},& \lambda_{1i}-\delta \lambda_{1i} \leq v_{1} \leq  \lambda_{1i}+\delta \lambda_{1i},\\
						& -\lambda_{2i}-\delta \lambda_{2i} \leq v_{2} \leq  -\lambda_{2i}+\delta \lambda_{2i},\\
0,& \text{otherwise} \end{array}\right\}
\label{eq:71}
\end{equation}
Substituting the above expressions in the moment relations given by 
\begin{subequations} 
\label{eq:2D-Moments} 
    \begin{equation} 
    U_{i} = \int_{-\infty}^{\infty} d v_{1} \int_{-\infty}^{\infty} d v_{2} \ \hat{f}^{eq}_{i} 
    = \left\langle \hat{f}^{eq}_{i}\right\rangle
    \end{equation} 
    \begin{equation}
    G_{1i} = \int_{-\infty}^{\infty} d v_{1} \int_{-\infty}^{\infty} d v_{2} \ v_{1} \hat{f}^{eq}_{i} 
    = \left\langle v_{1}\hat{f}^{eq}_{i}\right\rangle 
    \end{equation}
    \begin{equation} 
    G_{2i} = \int_{-\infty}^{\infty} d v_{1} \int_{-\infty}^{\infty} d v_{2} \ v_{2} \hat{f}^{eq}_{i} 
    = \left\langle v_{2}\hat{f}^{eq}_{i}\right\rangle 
    \end{equation} 
\end{subequations} 
we obtain 
\begin{subequations}
\label{eq:72}
\begin{equation}
4\delta \lambda_{1i}\delta \lambda_{2i} (f^{eq}_{++,i}+ f^{eq}_{-+,i}+ f^{eq}_{--,i}+ f^{eq}_{+-,i}) = U_{i}
\end{equation}
\begin{equation}
4\delta \lambda_{1i}\delta \lambda_{2i} (\lambda_{1i} f^{eq}_{++,i}-\lambda_{1i}  f^{eq}_{-+,i}-\lambda_{1i}  f^{eq}_{--,i}+\lambda_{1i}  f^{eq}_{+-,i}) = G_{1i}
\end{equation}
\begin{equation}
4\delta \lambda_{1i}\delta \lambda_{2i} (\lambda_{2i} f^{eq}_{++,i}+\lambda_{2i}  f^{eq}_{-+,i}-\lambda_{2i}  f^{eq}_{--,i}-\lambda_{2i}  f^{eq}_{+-,i}) = G_{2i}
\end{equation}
\end{subequations}
Substituting $4\delta \lambda_{1i}\delta \lambda_{2i} f^{eq}_{++,i}= f^{eq}_{1i}$, $4\delta \lambda_{1i}\delta \lambda_{2i} f^{eq}_{-+,i}= f^{eq}_{2i}$, $4\delta \lambda_{1i}\delta \lambda_{2i} f^{eq}_{--,i}= f^{eq}_{3i}$ and $4\delta \lambda_{1i}\delta \lambda_{2i} f^{eq}_{+-,i}= f^{eq}_{4i}$ in \eqref{eq:72}, the moment relations can be rewritten as
\begin{subequations}
\label{eq:72a}
\begin{equation}
f^{eq}_{1i}+ f^{eq}_{2i}+ f^{eq}_{3i}+ f^{eq}_{4i} = U_{i}
\end{equation}
\begin{equation}
\lambda_{1i} (f^{eq}_{1i}- f^{eq}_{2i}- f^{eq}_{3i}+ f^{eq}_{4i}) = G_{1i}
\end{equation}
\begin{equation}
\lambda_{2i} (f^{eq}_{1i}+ f^{eq}_{2i}- f^{eq}_{3i}- f^{eq}_{4i}) = G_{2i}
\end{equation}
\end{subequations}
Assuming $\lambda$'s and $\delta \lambda$'s as known quantities for now, we have four unknowns ($f^{eq}_{1i},...,f^{eq}_{4i}$) but only three equations. Hence, we make an assumption here that the distributions $f^{eq}_{ji}$ are a linear combination of conserved variable vector and flux vectors, {\em i.e.}, $U_{i}$, $G_{1i}$ and $G_{2i}$. One solution satisfying the moment relations then is given by 
\begin{subequations}
\label{eq:73}
\begin{equation}
f^{eq}_{1i} = \frac{1}{4}\left[U_{i}+ \frac{G_{1i}}{\lambda_{1i}}+ \frac{G_{2i}}{\lambda_{2i}}\right]
\end{equation}
\begin{equation}
f^{eq}_{2i} = \frac{1}{4}\left[U_{i}- \frac{G_{1i}}{\lambda_{1i}}+ \frac{G_{2i}}{\lambda_{2i}}\right]
\end{equation}
\begin{equation}
f^{eq}_{3i} = \frac{1}{4}\left[U_{i}- \frac{G_{1i}}{\lambda_{1i}}- \frac{G_{2i}}{\lambda_{2i}}\right]
\end{equation}
\begin{equation}
f^{eq}_{4i} = \frac{1}{4}\left[U_{i}+ \frac{G_{1i}}{\lambda_{1i}}- \frac{G_{2i}}{\lambda_{2i}}\right]
\end{equation}
\end{subequations}

For the above defined equilibrium distribution function, the computed second moments are
\begin{subequations}
\label{eq:74}
\begin{equation}
\left\langle v^{2}_{1}\hat{f}^{eq}_{i}\right\rangle= (\lambda_{1i}^{2}+ \frac{\delta\lambda^{2}_{1i}}{3})(f^{eq}_{1i}+ f^{eq}_{2i}+ f^{eq}_{3i}+ f^{eq}_{4i}) = (\widetilde{\lambda}_{1i})^{2}U_{i}
\end{equation}
\begin{equation}
\left\langle v^{2}_{2}\hat{f}^{eq}_{i}\right\rangle= (\lambda_{2i}^{2}+ \frac{\delta\lambda^{2}_{2i}}{3})(f^{eq}_{1i}+ f^{eq}_{2i}+ f^{eq}_{3i}+ f^{eq}_{4i}) = (\widetilde{\lambda}_{2i})^{2}U_{i}
\end{equation}
\begin{equation}
\left\langle v_{1}v_{2}\hat{f}^{eq}_{i}\right\rangle= \lambda_{1i}\lambda_{2i} (f^{eq}_{1i}- f^{eq}_{2i}+ f^{eq}_{3i}- f^{eq}_{4i}) = 0
\end{equation}
\end{subequations}
\section{Kinetic scheme for 2D Euler equations}
We now formulate a {\em 2D Flexible Velocity Boltzmann Equation} for $i^{th}$ distribution such that the equilibrium distribution function $\widetilde{\textbf{f}}^{eq}_{i}$ satisfies the zeroth, first and second moments given by \eqref{eq:2D-Moments} and \eqref{eq:74}. It is given by
\begin{equation}
\frac{\partial \textbf{f}_{i}}{\partial t}+ \frac{\partial (\widetilde{\Lambda}_{1i}\textbf{f}_{i})}{\partial x_{1}}+ \frac{\partial (\widetilde{\Lambda}_{2i}\textbf{f}_{i})}{\partial x_{2}} 
= - \frac{1}{\epsilon} \left[\textbf{f}_{i}-\widetilde{\textbf{f}}^{eq}_{i}\right]
\label{eq:75}
\end{equation}
Here,
\begin{equation}
\widetilde{\textbf{f}}^{eq}_{i} = \begin{bmatrix} \widetilde{f}^{eq}_{1i} \\ \widetilde{f}^{eq}_{2i} \\ \widetilde{f}^{eq}_{3i} \\ \widetilde{f}^{eq}_{4i} \end{bmatrix}, \widetilde{\Lambda}_{1i}= \begin{bmatrix} \widetilde{\lambda}_{1i} &0 &0 &0 \\ 0& -\widetilde{\lambda}_{1i} &0 &0 \\ 0& 0& -\widetilde{\lambda}_{1i} &0 \\ 0& 0& 0& \widetilde{\lambda}_{1i} \end{bmatrix}, \widetilde{\Lambda}_{2i}= \begin{bmatrix} \widetilde{\lambda}_{2i} &0 &0 &0 \\ 0& \widetilde{\lambda}_{2i} &0 &0 \\ 0& 0& -\widetilde{\lambda}_{2i} &0 \\ 0& 0& 0& -\widetilde{\lambda}_{2i} \end{bmatrix}
\label{eq:76}
\end{equation}
and
\begin{subequations}
\label{eq:77}
\begin{equation}
\widetilde{f}^{eq}_{1i} = \frac{1}{4}\left[U_{i}+ \frac{G_{1i}}{\widetilde{\lambda}_{1i}}+ \frac{G_{2i}}{\widetilde{\lambda}_{2i}}\right]
\end{equation}
\begin{equation}
\widetilde{f}^{eq}_{2i} = \frac{1}{4}\left[U_{i}- \frac{G_{1i}}{\widetilde{\lambda}_{1i}}+ \frac{G_{2i}}{\widetilde{\lambda}_{2i}}\right]
\end{equation}
\begin{equation}
\widetilde{f}^{eq}_{3i} = \frac{1}{4}\left[U_{i}- \frac{G_{1i}}{\widetilde{\lambda}_{1i}}- \frac{G_{2i}}{\widetilde{\lambda}_{2i}}\right]
\end{equation}
\begin{equation}
\widetilde{f}^{eq}_{4i} = \frac{1}{4}\left[U_{i}+ \frac{G_{1i}}{\widetilde{\lambda}_{1i}}- \frac{G_{2i}}{\widetilde{\lambda}_{2i}}\right]
\end{equation}
\end{subequations}
Defining $\textbf{P}_{i}= \left[ 1, 1, 1, 1\right]$, we have
\begin{subequations}
\label{eq:78}
\begin{equation}
\textbf{P}_{i}\widetilde{\textbf{f}}_{i}^{eq}= U_{i},
\end{equation}
\begin{equation}
\textbf{P}_{i}\widetilde{\Lambda}_{1i}\widetilde{\textbf{f}}_{i}^{eq}= G_{1i}, \textbf{P}_{i}\widetilde{\Lambda}_{2i}\widetilde{\textbf{f}}_{i}^{eq}= G_{2i}
\end{equation}
\begin{equation}
\textbf{P}_{i}\widetilde{\Lambda}^{2}_{1i}\widetilde{\textbf{f}}_{i}^{eq}= \widetilde{\lambda}^{2}_{1i}U_{i}, \textbf{P}_{i}\widetilde{\Lambda}^{2}_{2i}\widetilde{\textbf{f}}_{i}^{eq}= \widetilde{\lambda}^{2}_{2i}U_{i}, \textbf{P}_{i}\widetilde{\Lambda}_{1i}\widetilde{\Lambda}_{2i}\widetilde{\textbf{f}}_{i}^{eq}= 0
\end{equation}
\end{subequations}
Thus, all the moments are satisfied. Next, we solve the $i^{th}$ component of the Boltzmann equations \eqref{eq:75} by using operator splitting, leading to relaxation and advection steps for $(j,k)^{th}$ cell, as follows:\\\\
Relaxation step:\hspace{1cm} Instantaneous, {\em i.e.}, $\epsilon\rightarrow$ 0. Thus,
\begin{equation}
(\textbf{f}_{i})^{n}_{j,k}=  (\widetilde{\textbf{f}}_{i}^{eq})^{n}_{j,k}
\label{eq:79}
\end{equation}
Advection step:\hspace{1cm} The advection part of Boltzmann equation is
\begin{equation}
\frac{\partial \textbf{f}_{i}}{\partial t}+ \frac{\partial \textbf{h}_{1i}}{\partial x_{1}}+ \frac{\partial \textbf{h}_{2i}}{\partial x_{2}}= 0; \textbf{h}_{1/2i}= \widetilde{\Lambda}_{1/2i}\widetilde{\textbf{f}}_{i}^{eq}
\label{eq:80}
\end{equation}
Rewriting the advection equation in integral form for $(j,k)^{th}$ cell,
\begin{subequations}
\label{eq:81}
\begin{equation}
A_{j,k}\frac{d \textbf{f}_{i}}{dt}+  \oint \textbf{h}_{n}dl =0; \textbf{h}_{ni}= \widetilde{\Lambda}_{ni}\widetilde{\textbf{f}}_{i}^{eq}, \widetilde{\Lambda}_{ni}= \widetilde{\Lambda}_{1i}n_{1}+ \widetilde{\Lambda}_{2i}n_{2}
\end{equation}
\begin{equation}
\Rightarrow A_{j,k}\frac{d \textbf{f}_{i}}{dt}+  \sum_{s=1}^{4} (\textbf{h}_{ni})_{s}l_{s} =0\text{ (mid-point quadrature,)}\nonumber\\
\end{equation}
\begin{equation}
\textrm{where,} \ (\textbf{h}_{ni})_{s} = (\textbf{h}_{ni}^{+})_{L}+ (\textbf{h}_{ni}^{-})_{R}= (\widetilde{\Lambda}_{ni}^{+}\widetilde{\textbf{f}}_{i}^{eq})_{L}+(\widetilde{\Lambda}_{ni}^{-}\widetilde{\textbf{f}}_{i}^{eq})_{R}
\end{equation}
\end{subequations}
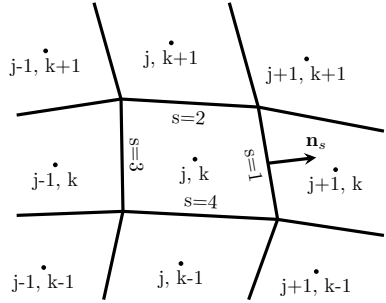
\begin{figure}[h!] 
\centering
\begin{tikzpicture}[scale=.75]
\small\begin{axis}
 [every axis plot post/.append style={
  mark=none,domain=0:4,samples=50,smooth},
  xmin=0,xmax=4,ymin=0,ymax= 4,
	axis x line*=bottom, 
  axis y line*=left,
	axis line style={draw=none},
	xticklabels={\empty},
	yticklabel={\empty},
	tick style={draw=none}
	]
\draw [ultra thick](axis cs:1,0.2) --(axis cs:1.2,1.3);
\draw [ultra thick](axis cs:1.2, 1.3) --(axis cs:1.18,2.7) node [midway, above, sloped] {s=3};
\draw [ultra thick](axis cs:1.18, 2.7) --(axis cs:0.9,3.9);

\draw [ultra thick](axis cs:2.5,0.2) --(axis cs:2.8, 1.2);
\draw [ultra thick](axis cs:2.8,1.2) --(axis cs:2.6, 2.6)node [midway, below, sloped] {s=1};
\draw [ultra thick](axis cs:2.6,2.6) --(axis cs:2.3,3.9);

\draw [ultra thick](axis cs:0.1, 2.5) --(axis cs:1.18,2.7);
\draw [ultra thick](axis cs:1.18,2.7) --(axis cs:2.6, 2.6)node [midway, below, sloped] {s=2};
\draw [ultra thick](axis cs:2.6, 2.6) --(axis cs:3.9, 2.3);

\draw [ultra thick](axis cs:0.1,1.25) --(axis cs:1.2,1.3);
\draw [ultra thick](axis cs:1.2,1.3) --(axis cs:2.8, 1.2)node [midway, above, sloped] {s=4};
\draw [ultra thick](axis cs:2.8, 1.2) --(axis cs:3.9, 1);

\draw [ultra thick,-stealth](axis cs:2.7,1.9) --(axis cs:3.2,1.971)node [right, above, sloped] {$\textbf{n}_{s}$};
\fill(axis cs:1.945,1.95) circle (1.2pt) node[anchor=north]{j, k};
\fill(axis cs:3.4,1.82) circle (1.2pt) node[anchor=north]{j+1, k};
\fill(axis cs:0.5,1.88) circle (1.2pt) node[anchor=north]{j-1, k};
\fill(axis cs:1.7,3.4) circle (1.2pt) node[anchor=north]{j, k+1};
\fill(axis cs:0.4,3.3) circle (1.2pt) node[anchor=north]{j-1, k+1};
\fill(axis cs:3.1,3.2) circle (1.2pt) node[anchor=north]{j+1, k+1};
\fill(axis cs:0.38,0.6) circle (1.2pt) node[anchor=north]{j-1, k-1};
\fill(axis cs:1.8,0.65) circle (1.2pt) node[anchor=north]{j, k-1};
\fill(axis cs:3.2,0.55) circle (1.2pt) node[anchor=north]{j+1, k-1};
\end{axis}
\end{tikzpicture}
\caption{Interior cell and associated unit normal vector for a structured grid in 2D.}
\label{fig:3a}%
\end{figure}
In the above expressions, we have assumed a structured grid with quadrilateral finite volumes (see Fig\ref{fig:3a}).   Multiplying  $(\textbf{h}_{ni})_{s}$ by $\textbf{P}_{i}$ gives us the macroscopic normal flux $(\textbf{G}_{ni})_{s}$.  
\begin{eqnarray} 
\begin{array}{lcl} 
(\textbf{G}_{ni})_{s} &=& \textbf{P}_{i}(\textbf{h}_{ni})_{s}= \textbf{P}_{i}(\widetilde{\Lambda}_{ni}^{+}\widetilde{\textbf{f}}_{i}^{eq})_{L}+\textbf{P}_{i}(\widetilde{\Lambda}_{ni}^{-}\widetilde{\textbf{f}}_{i}^{eq})_{R} \\[2mm]  
&=& \frac{1}{2}\left[ (G_{ni})_{L}+ (G_{ni})_{R}\right]- \frac{1}{2}\frac{|\widetilde{\lambda}_{1i}n_{1}+ \widetilde{\lambda}_{2i}n_{2}|+ |-\widetilde{\lambda}_{1i}n_{1}+ \widetilde{\lambda}_{2i}n_{2}|}{2}\Delta U_{i} 
\end{array} 
\label{eq:82}
\end{eqnarray}
or 
\begin{equation}
(\textbf{G}_{ni})_{s}= \frac{1}{2}\left[ (G_{ni})_{L}+ (G_{ni})_{R}\right]- \frac{\widetilde{\lambda}_{ni}}{2}\Delta U_{i}
\label{eq:84}
\end{equation}
where 
\begin{equation}
 \widetilde{\lambda}_{ni}= \frac{|\widetilde{\lambda}_{1i}n_{1}+ \widetilde{\lambda}_{2i}n_{2}|+ |-\widetilde{\lambda}_{1i}n_{1}+ \widetilde{\lambda}_{2i}n_{2}|}{2}
\label{eq:84a}    
\end{equation}

\subsection{\texorpdfstring{Fixing $\lambda_{n}$}{Fixing ?n}}
We take $\lambda_{1i}=\lambda_{1}, \lambda_{2i}=\lambda_{2}, \delta \lambda_{1i}= \delta\lambda_{1}, \delta \lambda_{2i}= \delta\lambda_{2}$. Thus, we obtain the scalar diffusion as $\widetilde{\lambda}_{ni}=\widetilde{\lambda}_{n}= \sqrt{\lambda_{n}^{2}+ (\delta\lambda)^{2}/3}$. Before defining $\lambda_{n}$, we define the following wave speeds.  
\begin{subequations}
\label{eq:85}
\begin{equation}
\lambda_{s}= min_{i}(\frac{\left|\Delta G_{n,i}\right|}{\left|\Delta U_{i}\right|+ \epsilon_{0}}),\text{ }\Delta = ()_{R}- ()_{L} 
\label{eq:85a}
\end{equation}
\begin{equation}
\lambda_{min}= max\left\{ min(|u_{n}-a|,|u_{n}|,|u_{n}+a|)_{L}, min(|u_{n}-a|,|u_{n}|,|u_{n}+a|)_{R}\right\}
\label{eq:85b}
\end{equation}
\end{subequations} 
Note that the expression \eqref{eq:85a} is based on enforcing Rankine-Hugoniot jump conditions across a cell-interface. 
Then, we define $\lambda_{n}$ as follows
\begin{equation}
\lambda_{n}=\left\{\begin{array}{l}
\lambda_{min},\text{ if }\left\|\Delta \textbf{U}\right\| \leq \epsilon_{1}\text{ (uniform flow)}\\
\lambda_{s},\text{ if }\left\|\Delta \textbf{U}\right\| > \epsilon_{1}, \left\|\Delta \textbf{G}_{n}\right\| \leq \epsilon_{2}\text{ (steady discontinuity)}\\
max(\lambda_{min}, \lambda_{s}),\text{ otherwise} 
\end{array}\right\}
\label{eq:86}
\end{equation}
Here too, we take $\epsilon_{1}= 10^{-5}$, $\epsilon_{2}= 10^{-8}$ and $\epsilon_{0}= 10^{-14}$ for all the test cases.

\subsection{Relative entropy in 2D}
Following the same procedure as in 1-D, the relative entropy $d^{2}$ in 2D can be derived (see \ref{appendix:a2}) and is given by
\begin{eqnarray}
d^{2}=&& \left(\Delta \frac{\partial \eta}{\partial \textbf{U}}\right)^{T} \cdot \Delta \textbf{U}\nonumber\\
=&&\Delta \left(\frac{\gamma -s}{\gamma -1} - \frac{\rho u_{1}^{2}}{2p} - \frac{\rho u_{2}^{2}}{2p}\right) \Delta (\rho)+ \Delta \left(\frac{\rho u_{1}}{p}\right) \Delta (\rho u_{1}) + \nonumber \\ && \Delta \left(\frac{\rho u_{2}}{p}\right) \Delta (\rho u_{2})+ \Delta \left(-\frac{\rho}{p}\right) \Delta (\rho E)
\label{eq:87}
\end{eqnarray}
The relative entropy based criterion given by \eqref{eq:4_28} is then used to provide LLF type numerical diffusion in smooth flow regions.

\section{Boundary Conditions based on a Discrete Kinetic System}  
Boundary conditions for 2D Euler equations are non-trivial and substantial research is done in developing them at the macroscopic level.  For boundary conditions based on molecular velocities and the classical Maxwellians, we refer to \cite{mandal1994kinetic} for {\em flow tangency boundary conditions} (based on specular reflection model) and \cite{ramesh2001} for far-field kinetic boundary conditions.  We utilize similar strategies in a novel way, based on discrete kinetic systems, to arrive at flow tangency and farfield boundary conditions in this section.

\begin{figure}[h!] 
\centering
\begin{tikzpicture}[scale=.75]
\small\begin{axis}
 [every axis plot post/.append style={
  mark=none,domain=0:4,samples=50,smooth},
  xmin=0,xmax=4,ymin=0,ymax= 4,
	axis x line*=bottom, 
  axis y line*=left,
	axis line style={draw=none},
	xticklabels={\empty},
	yticklabel={\empty},
	tick style={draw=none}
	]

\draw [ultra thick](axis cs:1.2, 1.2) --(axis cs:1.18,2.6);
\draw [ultra thick](axis cs:1.18, 2.6) --(axis cs:0.9,3.9);

\draw [ultra thick](axis cs:2.8,1.25) --(axis cs:2.6, 2.7);
\draw [ultra thick](axis cs:2.6,2.7) --(axis cs:2.3,3.9);

\draw [ultra thick](axis cs:0.1, 2.4) --(axis cs:1.18,2.6);
\draw [ultra thick](axis cs:1.18,2.6) --(axis cs:2.6, 2.7);
\draw [ultra thick](axis cs:2.6, 2.7) --(axis cs:3.9, 2.4);

\draw [red, ultra thick](axis cs:0.1,1) --(axis cs:1.2,1.2);
\draw [red, ultra thick](axis cs:1.2,1.2) --(axis cs:2.8, 1.25);
\draw [red, ultra thick](axis cs:2.8, 1.25) --(axis cs:3.9, 1);

\draw [ultra thick,-stealth](axis cs:2,1.225) --(axis cs:2.0156,0.725)node [anchor=east] {$\textbf{n}_{w}$};
\fill(axis cs:1.945,1.95) circle (1.2pt);
\fill(axis cs:3.3,1.82) circle (1.2pt);
\fill(axis cs:0.5,1.8) circle (1.2pt);
\fill(axis cs:1.7,3.4) circle (1.2pt);
\fill(axis cs:0.4,3.2) circle (1.2pt);
\fill(axis cs:3.1,3.2) circle (1.2pt);

\end{axis}
\end{tikzpicture}
\caption{Boundary cell and associated unit normal vector for a structured grid in 2D.}
\label{fig:3b}%
\end{figure}
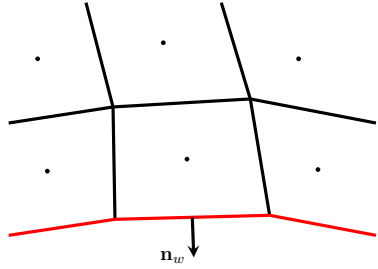
At boundaries, we assume that $\delta\lambda\rightarrow 0$ for simplicity, {\em i.e.}, our equilibrium distribution function simplifies to Dirac-delta distribution functions.  We then proceed to formulate {\em Discrete Kinetic Flow Tangency Boundary Condition} (DK-FTBC) and {\em Discrete Kinetic Farfield Boundary Condition} (DK-FBC) based on our kinetic model.  To start, we assume that the unit normal $\textbf{n}(=n_{1}\textbf{e}_{1}+n_{2}\textbf{e}_{2})$ at boundary surfaces points outward(see Fig\ref{fig:3b}). Now, rewriting our discrete velocity equilibrium distribution function $\hat{f}^{eq}_{i}$ in terms of normal and tangential components of velocities, we have
\begin{eqnarray}
\hat{f}^{eq}_{i}=&& f^{eq}_{1i}\delta\left(v_{1}-\lambda_{1i}\right)\delta\left(v_{2}-\lambda_{2i}\right)+ f^{eq}_{2i}\delta\left(v_{1}+\lambda_{1i}\right)\delta\left(v_{2}-\lambda_{2i}\right)+ \nonumber\\ 
&&f^{eq}_{3i}\delta\left(v_{1}+\lambda_{1i}\right)\delta\left(v_{2}+\lambda_{2i})+f^{eq}_{4i}\delta(v_{1}-\lambda_{1i}\right)\delta\left(v_{2}+\lambda_{2i}\right)\nonumber\\
=&&f^{eq}_{1i}\delta\left(v_{n}-\left[\lambda_{1i}n_{1}+ \lambda_{2i}n_{2}\right]\right)\delta\left(v_{t}-\left[-\lambda_{1i}n_{2}+\lambda_{2i}n_{1}\right]\right)+\nonumber\\
&&f^{eq}_{2i}\delta\left(v_{n}-\left[-\lambda_{1i}n_{1}+ \lambda_{2i}n_{2}\right]\right)\delta\left(v_{t}-\left[\lambda_{1i}n_{2}+ \lambda_{2i}n_{1}\right]\right)+\nonumber\\
&&f^{eq}_{3i}\delta\left(v_{n}-\left[-\lambda_{1i}n_{1}- \lambda_{2i}n_{2}\right]\right)\delta\left(v_{t}-\left[\lambda_{1i}n_{2}- \lambda_{2i}n_{1}\right]\right)+\nonumber\\
&&f^{eq}_{4i}\delta\left(v_{n}-\left[\lambda_{1i}n_{1}- \lambda_{2i}n_{2}\right]\right)\delta\left(v_{t}-\left[-\lambda_{1i}n_{2}- \lambda_{2i}n_{1}\right]\right) 
\label{eq:b1}
\end{eqnarray}
where, $v_{n}= v_{1}n_{1}+ v_{2}n_{2}, v_{t}= -v_{1}n_{2}+ v_{2}n_{1}$. The split normal fluxes at surface $s$ for our scheme, which are formed by Courant splitting of normal component of the (discrete) kinetic velocities, can be written as
\begin{subequations}
\label{eq:b2}
\begin{equation}
\textbf{G}^{+}_{ns,i}= \int^{\infty}_{0}v_{n}dv_{n}\int^{\infty}_{-\infty}dv_{t}\int^{\infty}_{0}dI\Psi_{i}f^{eq}= \int^{\infty}_{0}v_{n}dv_{n}\int^{\infty}_{-\infty}dv_{t}\hat{f}^{eq}_{i}
\end{equation}
\begin{equation}
\textbf{G}^{-}_{ns,i}= \int^{0}_{-\infty}v_{n}dv_{n}\int^{\infty}_{-\infty}dv_{t}\int^{\infty}_{0}dI\Psi_{i}f^{eq}= \int^{0}_{-\infty}v_{n}dv_{n}\int^{\infty}_{-\infty}dv_{t}\hat{f}^{eq}_{i}
\end{equation}
\end{subequations}
The split flux $\textbf{G}^{+}_{ns,i}$  with $v_{n}\geq 0$ corresponds to outgoing information and depends on interior data,  whereas the split flux $\textbf{G}^{-}_{ns,i}$ with $v_{n}\leq 0$ corresponds to information coming into the domain from the boundary.  \\\\
\textbf{\textit{Discrete kinetic flow tangency boundary condition:}} 
This is the inviscid wall boundary condition, and is based on the specular reflection model of kinetic theory of gases. According to specular reflection model, at inviscid wall the normal velocity of particles gets reversed while the tangential velocity remains unchanged. Thus, we take
\begin{subequations}
\label{eq:b3}
\begin{equation}
\textbf{G}^{+}_{nw,i}= \int^{\infty}_{0}v_{n}dv_{n}\int^{\infty}_{-\infty}dv_{t}\int^{\infty}_{0}dI\Psi_{i}f^{eq}(v_{n},v_{t})
\end{equation}
\begin{equation}
\textbf{G}^{-}_{nw,i}= \int^{0}_{-\infty}v_{n}dv_{n}\int^{\infty}_{-\infty}dv_{t}\int^{\infty}_{0}dI\Psi_{i}f^{eq}(-v_{n},v_{t})
\end{equation}
\end{subequations}
Now, with the above definitions, the split fluxes can be written as 
\begin{subequations}
\label{eq:b4}
\begin{eqnarray}
\textbf{G}^{+}_{nw,1}=&& \int^{\infty}_{0}v_{n}dv_{n}\int^{\infty}_{-\infty}dv_{t}\int^{\infty}_{0}dI \ f^{eq}(v_{n},v_{t})\nonumber\\
=&& \int^{\infty}_{0}v_{n}dv_{n}\int^{\infty}_{-\infty}dv_{t} \ \hat{f}^{eq}_{1}(v_{n},v_{t})\nonumber\\
=&& (\lambda_{11}n_{1}+ \lambda_{21}n_{2})^{+}f^{eq}_{11}+ (-\lambda_{11}n_{1}+ \lambda_{21}n_{2})^{+}f^{eq}_{21}+ \nonumber\\
&& (-\lambda_{11}n_{1}- \lambda_{21}n_{2})^{+}f^{eq}_{31}+ (\lambda_{11}n_{1}- \lambda_{21}n_{2})^{+}f^{eq}_{41} 
\end{eqnarray}
\begin{eqnarray}
\textbf{G}^{-}_{nw,1}=&& \int^{0}_{-\infty}v_{n}dv_{n}\int^{\infty}_{-\infty}dv_{t}\int^{\infty}_{0}dI \ f^{eq}(-v_{n},v_{t})\nonumber\\
=&& \int^{0}_{-\infty}v_{n}dv_{n}\int^{\infty}_{-\infty}dv_{t} \ \hat{f}^{eq}_{1}(-v_{n},v_{t})\nonumber\\
=&& (-\lambda_{11}n_{1}- \lambda_{21}n_{2})^{-}f^{eq}_{11}+ (\lambda_{11}n_{1}- \lambda_{21}n_{2})^{-}f^{eq}_{21}+ \nonumber\\
&& (\lambda_{11}n_{1}+ \lambda_{21}n_{2})^{-}f^{eq}_{31}+ (-\lambda_{11}n_{1}+ \lambda_{21}n_{2})^{-}f^{eq}_{41}\nonumber\\
=&& -(\lambda_{11}n_{1}+ \lambda_{21}n_{2})^{+}f^{eq}_{11}- (-\lambda_{11}n_{1}+ \lambda_{21}n_{2})^{+}f^{eq}_{21}- \nonumber\\
&& (-\lambda_{11}n_{1}- \lambda_{21}n_{2})^{+}f^{eq}_{31}- (\lambda_{11}n_{1}- \lambda_{21}n_{2})^{+}f^{eq}_{41}
\end{eqnarray}
\begin{eqnarray}
\textbf{G}_{nw,1}= \textbf{G}^{+}_{nw,1}+ \textbf{G}^{+}_{nw,1}= 0
\end{eqnarray}
\end{subequations}

\begin{subequations}
\label{eq:b5}
\begin{eqnarray}
\textbf{G}^{+}_{nw,2}=&& \int^{\infty}_{0}v_{n}dv_{n}\int^{\infty}_{-\infty}dv_{t}\int^{\infty}_{0}dI \ v_{1}f^{eq}(v_{n},v_{t})\nonumber\\
=&& \int^{\infty}_{0}v_{n}dv_{n}\int^{\infty}_{-\infty}dv_{t}\int^{\infty}_{0}dI \ (v_{n}n_{1}- v_{t}n_{2})f^{eq}(v_{n},v_{t})\nonumber\\
=&& \int^{\infty}_{0}v_{n}dv_{n}\int^{\infty}_{-\infty}dv_{t} \ \hat{f}^{eq}_{2}(v_{n},v_{t})\nonumber\\
=&& (\lambda_{12}n_{1}+ \lambda_{22}n_{2})^{+}f^{eq}_{12}+ (-\lambda_{12}n_{1}+ \lambda_{22}n_{2})^{+}f^{eq}_{22}+ \nonumber\\
&& (-\lambda_{12}n_{1}- \lambda_{22}n_{2})^{+}f^{eq}_{32}+ (\lambda_{12}n_{1}- \lambda_{22}n_{2})^{+}f^{eq}_{42}
\end{eqnarray}
\begin{eqnarray}
\textbf{G}^{-}_{nw,2}=&& \int^{0}_{-\infty}v_{n}dv_{n}\int^{\infty}_{-\infty}dv_{t}\int^{\infty}_{0}dI \ v_{1}f^{eq}(-v_{n},v_{t})\nonumber\\
=&& \int^{0}_{-\infty}v_{n}dv_{n}\int^{\infty}_{-\infty}dv_{t}\int^{\infty}_{0}dI \ (v_{n}n_{1}- v_{t}n_{2})f^{eq}(-v_{n},v_{t})\nonumber\\
=&& \int^{0}_{-\infty}v_{n}dv_{n}\int^{\infty}_{-\infty}dv_{t}\int^{\infty}_{0}dI \ (-v_{n}n_{1}- v_{t}n_{2})f^{eq}(-v_{n},v_{t})+\nonumber\\
&& \int^{0}_{-\infty}v_{n}dv_{n}\int^{\infty}_{-\infty}dv_{t}\int^{\infty}_{0}dI \ (2v_{n}n_{1})f^{eq}(-v_{n},v_{t})\nonumber\\
=&& \int^{0}_{-\infty}v_{n}dv_{n}\int^{\infty}_{-\infty}dv_{t} \ \hat{f}^{eq}_{2}(-v_{n},v_{t})+\nonumber\\
&& 2n_{1} \int^{0}_{-\infty}v^{2}_{n}dv_{n}\int^{\infty}_{-\infty}dv_{t} \ f^{eq}(-v_{n},v_{t})\nonumber\\
=&& \left\{-(\lambda_{12}n_{1}+ \lambda_{22}n_{2})\right\}^{-}f^{eq}_{12}+ \left\{- (-\lambda_{12}n_{1}+ \lambda_{22}n_{2})\right\}^{-}f^{eq}_{22}+\nonumber\\
&& \left\{-(-\lambda_{12}n_{1}- \lambda_{22}n_{2})\right\}^{-}f^{eq}_{32}+ \left\{- (\lambda_{12}n_{1}- \lambda_{22}n_{2})\right\}^{-}f^{eq}_{42}+\nonumber\\
&& 2n_{1} \int^{0}_{-\infty}v^{2}_{n}dv_{n}\int^{\infty}_{-\infty}dv_{t}f^{eq}(-v_{n},v_{t})\nonumber\\
=&& -(\lambda_{12}n_{1}+ \lambda_{22}n_{2})^{+}f^{eq}_{12}- (-\lambda_{12}n_{1}+ \lambda_{22}n_{2})^{+}f^{eq}_{22}-\nonumber\\
&& (-\lambda_{12}n_{1}- \lambda_{22}n_{2})^{+}f^{eq}_{32}- (\lambda_{12}n_{1}- \lambda_{22}n_{2})^{+}f^{eq}_{42}+\nonumber\\
&& 2n_{1} \int^{0}_{-\infty}v^{2}_{n}dv_{n}\int^{\infty}_{-\infty}dv_{t}f^{eq}(-v_{n},v_{t})
\end{eqnarray}
\begin{eqnarray}
\textbf{G}_{nw,2}=&& \textbf{G}^{+}_{nw,2}+ \textbf{G}^{-}_{nw,2}= 2n_{1} \int^{0}_{-\infty}v^{2}_{n}dv_{n}\int^{\infty}_{-\infty}dv_{t} \ f^{eq}(-v_{n},v_{t})\nonumber\\
=&& \frac{1}{2}[\{|\lambda_{12}n_{1}+ \lambda_{22}n_{2}|+|\lambda_{12}n_{1}- \lambda_{22}n_{2}|\}\rho u_{1}n^{2}_{1}+\nonumber\\
&& \{|\lambda_{13}n_{1}+ \lambda_{23}n_{2}|+|\lambda_{13}n_{1}- \lambda_{23}n_{2}|\}\rho u_{2}n_{1}n_{2}]+\nonumber\\
&& \left(\rho u^{2}_{nw}+p \right)n_{1}\nonumber\\
=&& pn_{1}, \ \textrm{for} \ \lambda_{1i}= \lambda_{1},  \lambda_{2i}= \lambda_{2}
\end{eqnarray} 
\textrm{Similarly,} \ 
\begin{eqnarray}
\textbf{G}_{nw,3}=&& 2n_{2} \int^{0}_{-\infty}v^{2}_{n}dv_{n}\int^{\infty}_{-\infty}dv_{t} \ f^{eq}(-v_{n},v_{t})\nonumber\\
=&& \frac{1}{2}[\{|\lambda_{12}n_{1}+ \lambda_{22}n_{2}|+|\lambda_{12}n_{1}- \lambda_{22}n_{2}|\}\rho u_{1}n_{1}n_{2}+\nonumber\\
&& \{|\lambda_{13}n_{1}+ \lambda_{23}n_{2}|+|\lambda_{13}n_{1}- \lambda_{23}n_{2}|\}\rho u_{2}n^{2}_{2}]+\nonumber\\
&& \left(\rho u^{2}_{nw}+p \right)n_{2}\nonumber\\
=&& pn_{2}, \ \textrm{for} \ \lambda_{1i}= \lambda_{1},  \lambda_{2i}= \lambda_{2}
\end{eqnarray}
\end{subequations}

\begin{subequations}
\label{eq:b6}
\begin{eqnarray}
\textbf{G}^{+}_{nw,4}=&& \int^{\infty}_{0}v_{n}dv_{n}\int^{\infty}_{-\infty}dv_{t}\int^{\infty}_{0}dI \ \left[I+ \frac{v^{2}_{n}+ v^{2}_{t}}{2}\right]f^{eq}(v_{n},v_{t})\nonumber\\
=&& \int^{\infty}_{0}v_{n}dv_{n}\int^{\infty}_{-\infty}dv_{t} \ \hat{f}^{eq}_{4}(v_{n},v_{t})\nonumber\\
=&& (\lambda_{14}n_{1}+ \lambda_{24}n_{2})^{+}f^{eq}_{14}+ (-\lambda_{14}n_{1}+ \lambda_{24}n_{2})^{+}f^{eq}_{24}+ \nonumber\\
&& (-\lambda_{14}n_{1}- \lambda_{24}n_{2})^{+}f^{eq}_{34}+ (\lambda_{14}n_{1}- \lambda_{24}n_{2})^{+}f^{eq}_{44} 
\end{eqnarray}
\begin{eqnarray}
\textbf{G}^{-}_{nw,4}=&& \int^{0}_{-\infty}v_{n}dv_{n}\int^{\infty}_{-\infty}dv_{t}\int^{\infty}_{0}dI \ \left[I+ \frac{v^{2}_{n}+ v^{2}_{t}}{2}\right]f^{eq}(-v_{n},v_{t})\nonumber\\
=&& \int^{0}_{-\infty}v_{n}dv_{n}\int^{\infty}_{-\infty}dv_{t} \ \hat{f}^{eq}_{4}(-v_{n},v_{t})\nonumber\\
=&& -(\lambda_{14}n_{1}+ \lambda_{24}n_{2})^{+}f^{eq}_{14}- (-\lambda_{14}n_{1}+ \lambda_{24}n_{2})^{+}f^{eq}_{24}-\nonumber\\
&& (-\lambda_{14}n_{1}- \lambda_{24}n_{2})^{+}f^{eq}_{34}- (\lambda_{14}n_{1}- \lambda_{24}n_{2})^{+}f^{eq}_{44} 
\end{eqnarray}
\begin{eqnarray}
\textbf{G}_{nw,4}= \textbf{G}^{+}_{nw,4}+ \textbf{G}^{-}_{nw,4}= 0
\end{eqnarray}
\end{subequations}

\noindent 
\textbf{\textit{Discrete kinetic farfield boundary condition:}}This boundary condition is applicable at farfield boundary, at a very large distance from the body. The positive split flux at farfield boundary depends on interior data, whereas negative split flux depends on boundary data. Thus,
\begin{subequations}
\label{eq:b7}
\begin{eqnarray}
\textbf{G}^{+}_{nf,i}=&&\int^{\infty}_{0}v_{n}dv_{n}\int^{\infty}_{-\infty}dv_{t}\int^{\infty}_{0}dI \ \Psi_{i}(f^{eq})_{interior} \nonumber\\
=&&\int^{\infty}_{0}v_{n}dv_{n}\int^{\infty}_{-\infty}dv_{t} \ (\hat{f}^{eq}_{i})_{interior}
\end{eqnarray}
\begin{eqnarray}
\textbf{G}^{-}_{nf,i}=&&\int^{0}_{-\infty}v_{n}dv_{n}\int^{\infty}_{-\infty}dv_{t}\int^{\infty}_{0}dI \ \Psi_{i}(f^{eq})_{boundary}\nonumber\\
=&&\int^{0}_{-\infty}v_{n}dv_{n}\int^{\infty}_{-\infty}dv_{t} \ (\hat{f}^{eq}_{i})_{boundary}
\end{eqnarray}
\begin{equation}
\textbf{G}_{nf,i}= \textbf{G}^{+}_{nf,i}+ \textbf{G}^{-}_{nf,i}
\end{equation}
\end{subequations}
Thus, the macroscopic flux at the farfield boundary interface is computed using the same scheme as that at the interior interfaces, using the appropriate interior and boundary values.  

\section{Results and discussion}

\subsection{Experimental Order of Convergence}
To determine the {\em Experimental Order of Convergence} of the presented scheme, a simple 1D Euler test is considered, for which the exact solution is known. The computational domain taken is x$\in$[0,2]. The initial conditions are
\begin{subequations}
\label{eq:91}
\begin{equation}
\rho(x,0)= \rho_{0}(x)= 1+ 0.2 sin(\pi x)
\end{equation}
\begin{equation}
u(x,0)= 0.1, \ p(x,0)= 0.5
\end{equation}
\end{subequations}
Thus, velocity and pressure are initially constant, while the density varies sinusoidally. Periodic boundary conditions are applied at the two ends. For this problem, $u$ and $p$ remain constant while exact solution for density is given by
\begin{equation}
\rho(x,t)= \rho_{0}(x- ut)= 1+ 0.2 sin\left(\pi \left[x- 0.1 t\right]\right)
\label{eq:92}
\end{equation}
The problem is solved numerically, and its solution is considered at t= 0.5. The numerical solution is computed for varying grid sizes, {\em i.e.}, $Nx$ (=$\frac{2}{\Delta x}$)= 40, 80, 160, .. and so on. Next, the $L_{1}$ and $L_{2}$ errors in solution are computed as follows.  
\begin{subequations}
\label{eq:93}
\begin{equation}
\left\|\varepsilon_{K}\right\|_{L_{1}}= \Delta x \sum^{K}_{i=1}|\rho^{i}- \rho^{i}_{e}|
\end{equation}
\begin{equation}
\left\|\varepsilon_{K}\right\|_{L_{2}}= \sqrt{\Delta x \sum^{K}_{i=1}(\rho^{i}- \rho^{i}_{e})^{2}}
\end{equation}
\end{subequations}
Here K is the number of cells, $\rho^{i}$ and $\rho^{i}_{e}$ are the numerical and exact solution in the $i^{th}$ cell. Now, for a $p^{th}$ order accurate scheme,
\begin{subequations}
\label{eq:94}
\begin{equation}
\left\|\varepsilon_{K}\right\|= C \Delta x^{p}+ O(\Delta x^{p+1})\text{. Similarly,}
\end{equation}
\begin{equation}
\left\|\varepsilon_{K/2}\right\|= C (2 \Delta x)^{p}+ O(\Delta x)^{p+1}\text{, }(K\propto\frac{1}{\Delta x})
\end{equation}
\end{subequations}
Thus,
\begin{equation}
\frac{\left\|\varepsilon_{K/2}\right\|}{\left\|\varepsilon_{K}\right\|}= 2^{p}+ O(\Delta x) \Rightarrow log_{2}\left(\frac{\left\|\varepsilon_{K/2}\right\|}{\left\|\varepsilon_{K}\right\|}\right)= p+ O(\Delta x)
\label{eq:95}
\end{equation}
The experimental order of convergence (EOC) of the scheme is then given by
\begin{equation}
\textrm{EOC} = log_{2}\left(\frac{\left\|\varepsilon_{K/2}\right\|}{\left\|\varepsilon_{K}\right\|}\right)
\label{eq:96}
\end{equation}

\begin{table}
\centering
\begin{tabular}{ |c|c|c|c|c|c| }
\hline
Nx& $\Delta$x& $L_{1}$ Error & EOC & $L_{2}$ Error & EOC\\
\hline
40&	0.05&	0.0191413808&	&	0.0154857921&	\\
80&	0.025&	0.0085164603&	1.168369&	0.0069153845&	1.163063\\
160&	0.0125&	0.0045451439&	0.905928&	0.0036194091&	0.934055\\
320&	0.00625&	0.0022895408&	0.989267&	0.0018024477&	1.005796\\
640&	0.003125&	0.0011552003&	0.986915&	0.0009209818&	0.968713\\
1280&	0.0015625&	0.0005665738&	1.027807&	0.0004564290&	1.012782\\
2560&	0.00078125&	0.0002874274&	0.979066&	0.0002298604&	0.989632 \\
\hline
\end{tabular}
\centering
\caption{EOC using $L_{1}$ and $L_{2}$ error norms for first order accuracy}
\label{table:1}
\end{table}

\begin{table}
\centering
\begin{tabular}{ |c|c|c|c|c|c| }
\hline
Nx& $\Delta$x& $L_{1}$ Error & EOC & $L_{2}$ Error & EOC\\
\hline
40&	0.05&	0.0012055970&	&	0.0013636619&	\\
80&	0.025&	0.0003310609&	1.864579&	0.0004857523&	1.489193\\
160&	0.0125&	0.0000765373&	2.112862&	0.0001403528&	1.791163\\
320&	0.00625&	0.0000187416&	2.029916&	0.0000433965&	1.693408\\
640&	0.003125&	0.0000045506&	2.042103&	0.0000135291&	1.681510\\
1280&	0.0015625&	0.0000010846&	2.068911&	0.0000040809&	1.729078\\
2560&	0.00078125&	0.0000002438&	2.153041&	0.000001174&	1.797770\\
\hline
\end{tabular}
\centering
\caption{EOC using $L_{1}$ and $L_{2}$ error norms for second order accuracy with Van Albada limiter}
\label{table:2}
\end{table}

\begin{table}
\centering
\begin{tabular}{ |c|c|c|c|c|c| }
\hline
Nx& $\Delta$x& $L_{1}$ Error & EOC & $L_{2}$ Error & EOC\\
\hline
40&	0.05&	0.0003656949&	&	0.0002969886&	\\  
80&	0.025&	0.0000851224&	2.103030&	0.0000671913&	2.144061 \\
160&	0.0125&	0.0000207283&	2.037941&	0.0000163095&	2.042564\\
320&	0.00625&	0.0000051489&	2.009272&	0.0000040460&	2.011155\\
640&	0.003125&	0.0000012847&	2.002791&	0.0000010091&	2.003356\\
1280&	0.0015625&	0.0000003207&	2.002197&	0.0000002519&	2.002316\\
2560&	0.00078125&	0.0000000797&	2.008218&	0.0000000626&	2.008252\\
\hline
\end{tabular}
\centering
\caption{EOC using $L_{1}$ and $L_{2}$ error norms for second order accuracy without limiter}
\label{table:2a}
\end{table}

\begin{figure}
\centering
\begin{tabular}{cc}
\includegraphics[width=0.45\textwidth]{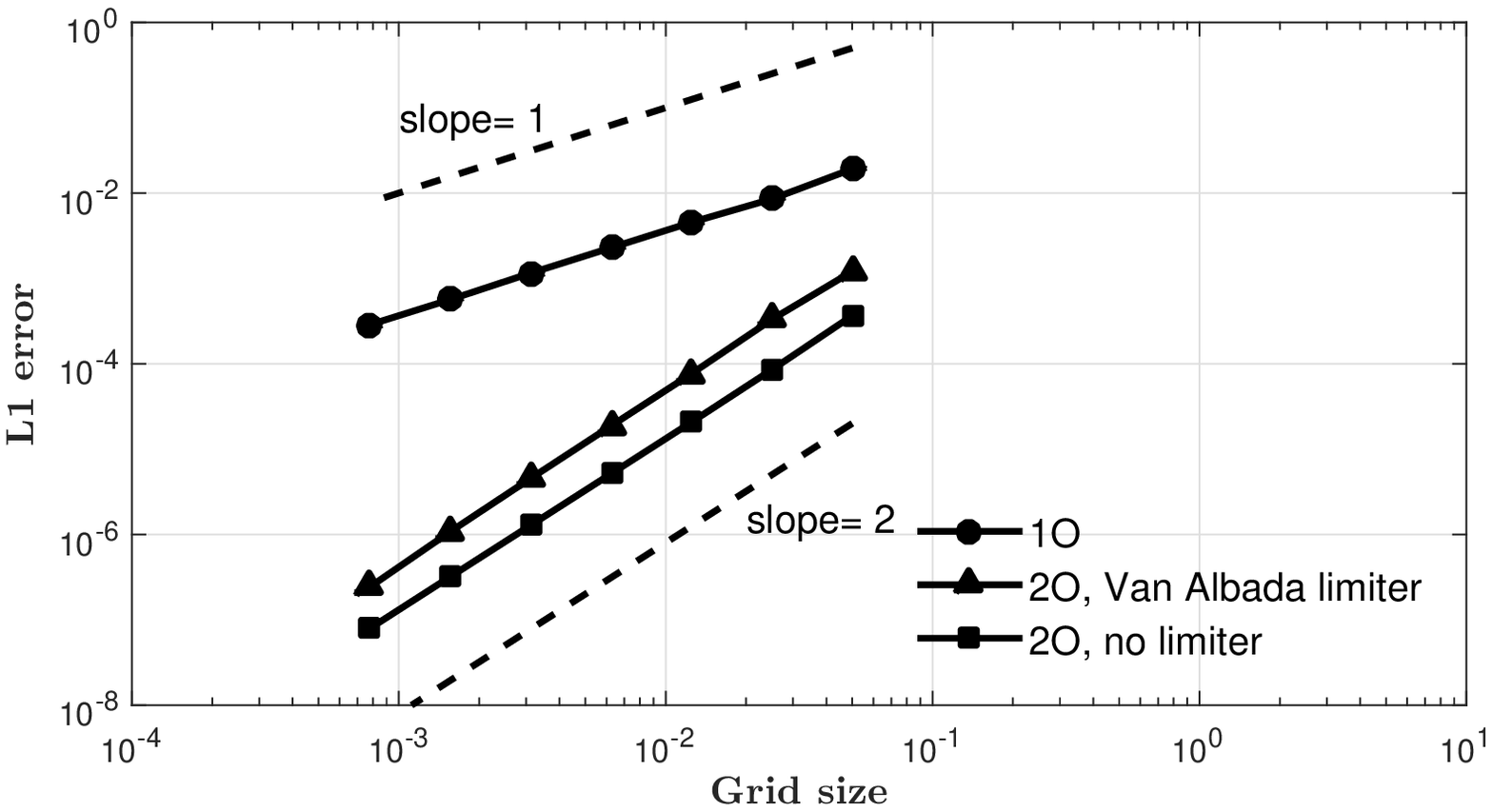} & \includegraphics[width=0.45\textwidth]{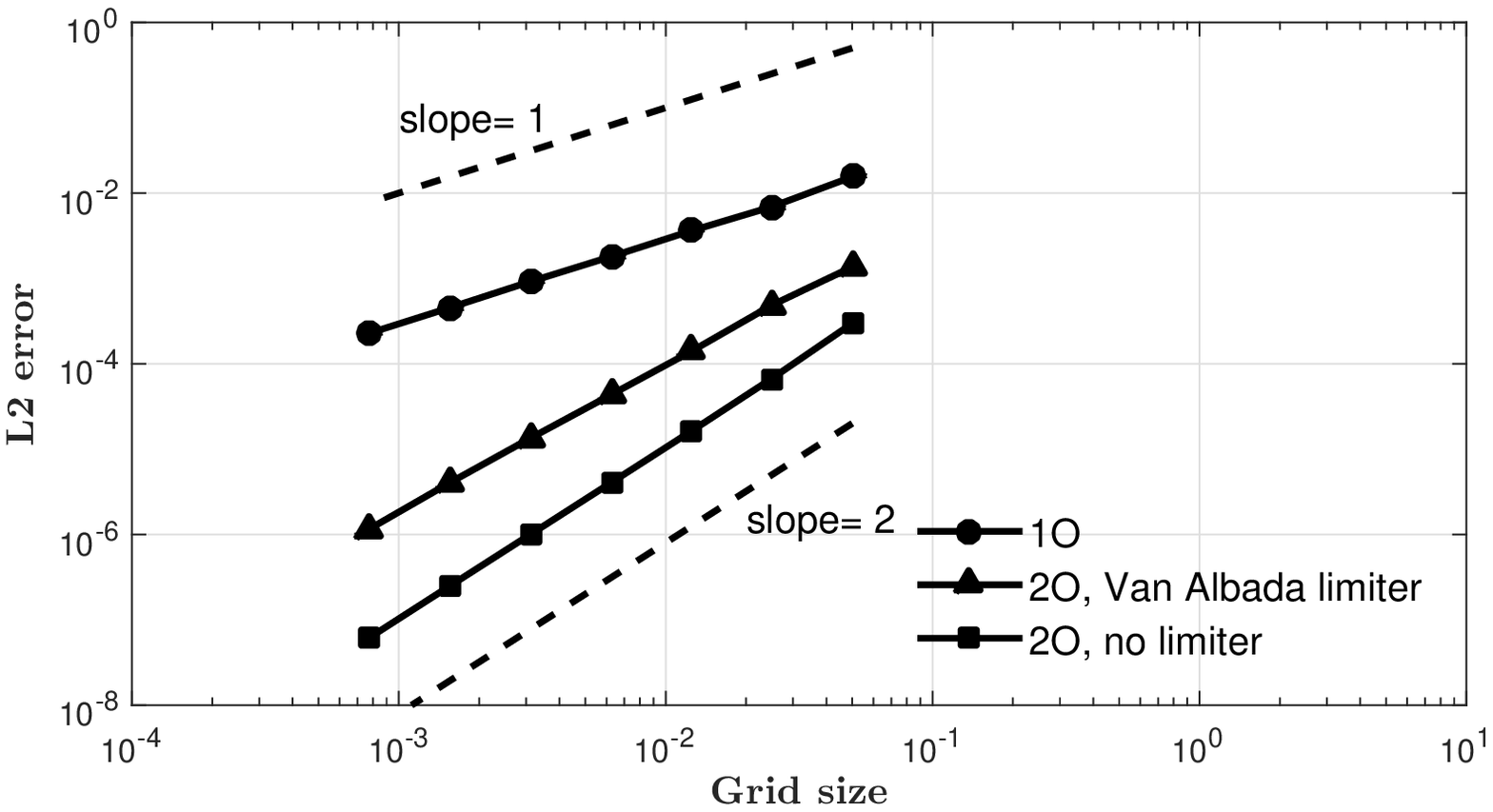}
\end{tabular}
\caption{(a) $L_{1}$ error norm vs grid size, (b) $L_{2}$ error norm vs grid size}
\label{fig:4}
\end{figure}
The $L_{1}$ and $L_{2}$ errors of the present scheme for I order accuracy are tabulated in Table \ref{table:1}. The II order results are tabulated in Table \ref{table:2} (with limiter) and Table \ref{table:2a} (without limiter) respectively. The  log-log plots comparing the EOC with slopes 1 and 2 are shown in Figure \ref{fig:4}.  

\subsection{1D Euler tests}
\begin{table}
\centering
\begin{tabular}{ |c|c|c|c|c|c|c|c| }
\hline
$x_{0}$& $\rho_{L}$& $u_{L}$& $p_{L}$& $\rho_{R}$& $u_{R}$& $p_{R}$& $t_{final}$\\
\hline
0.5& 1.4& 0& 1.0 &1.0 &0.0 &1.0 &2.0 \\
0.5& 1.0& 1.0& $\frac{1}{\gamma M(=2)^{2}}$& $\frac{\frac{\gamma +1}{\gamma -1}\frac{p_{R}}{p_{l}}+1}{\frac{\gamma +1}{\gamma -1}+ \frac{p_{R}}{p_{L}}}$& $\sqrt{\frac{\gamma (2+ (\gamma -1)M^{2})p_{R}}{(2\gamma M^{2}+1-\gamma)\rho_{R}}}$& $p_{L}\frac{2\gamma M^{2}- (\gamma -1)}{\gamma +1}$ &1.5\\
0.5& 1.4& 0.1& 1.0& 1.0& 0.1& 1.0& 1.0\\
0.5& 3.86& -0.81& 10.33& 1.0& -3.44& 1.0& 1.0\\
0.3& 1.0& 0.75& 1.0& 0.125& 0.0& 0.1& 0.2 \\
0.5& 1.0& -2.0& 0.4& 1.0& 2.0& 0.4& 0.15 \\
0.5& 1.0& 0.0& 1000.0& 1.0& 0.0& 0.01& 0.012\\
0.4& 5.99924& 19.5975& 460.894& 5.99242& -6.19633& 46.0950& 0.035\\
0.8& 1.0& -19.59745& 1000.0& 1.0& -19.59745& 0.01& 0.012\\
\hline
\end{tabular}
\caption{Initial condition for 1D test cases}
\label{table:3}
\end{table}

\begin{figure}
\centering
\includegraphics[width=14cm]{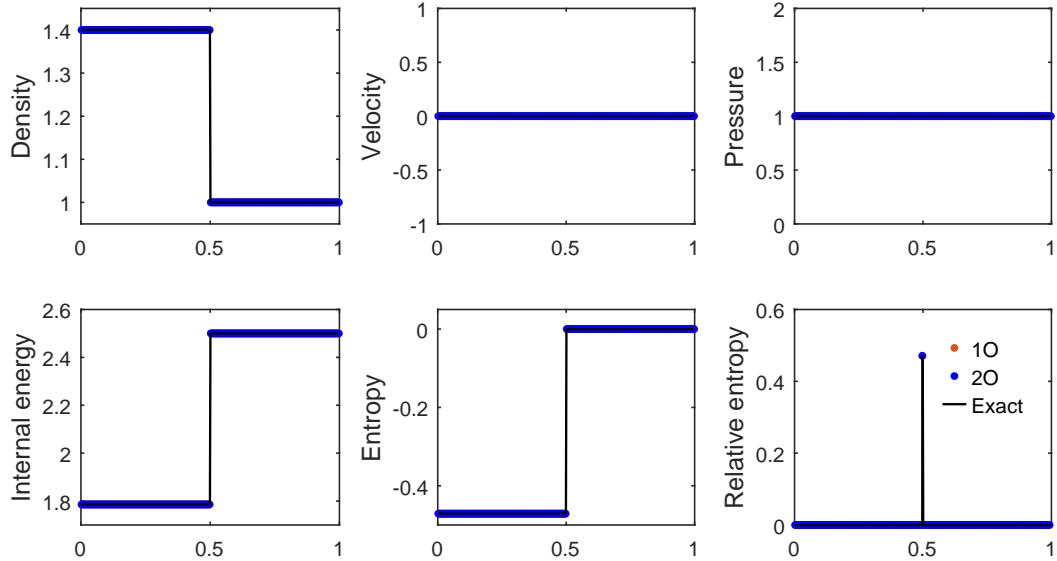}
\caption{\label{fig:5} Test case 1: Steady contact discontinuity}
\end{figure}

\begin{figure}
\centering
\includegraphics[width=14cm]{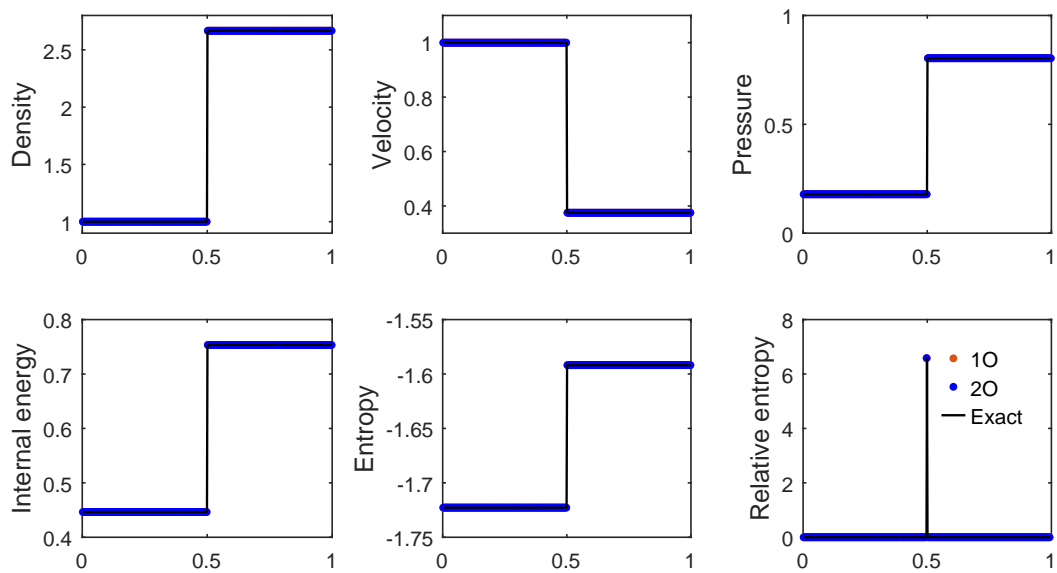}
\caption{\label{fig:6} Test case 2: Steady shock}
\end{figure}

\begin{figure}
\centering
\includegraphics[width=14cm]{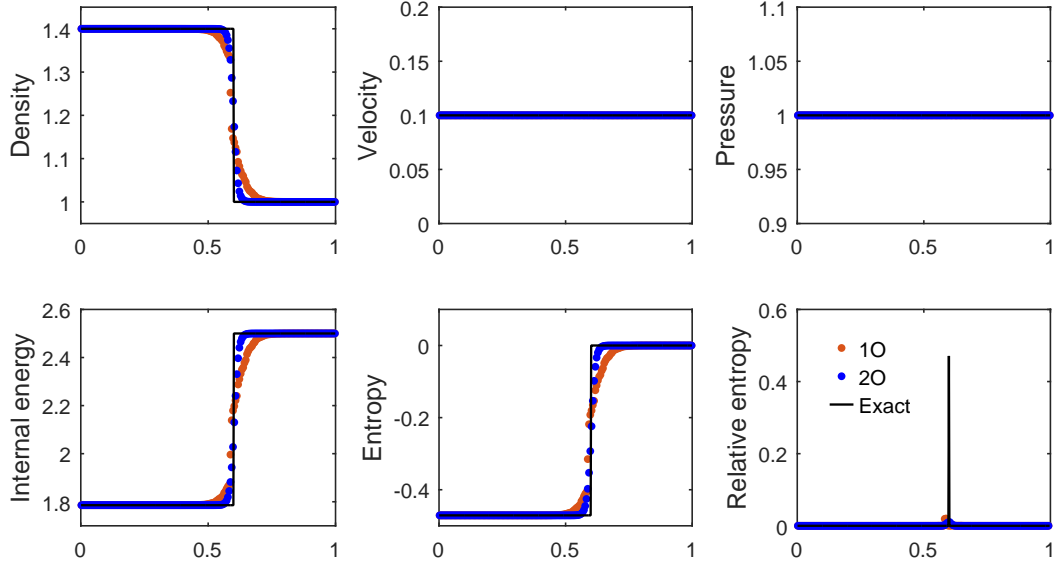}
\caption{\label{fig:7} Test case 3: Slowly moving contact discontinuity}
\end{figure}

\begin{figure}
\centering
\includegraphics[width=14cm]{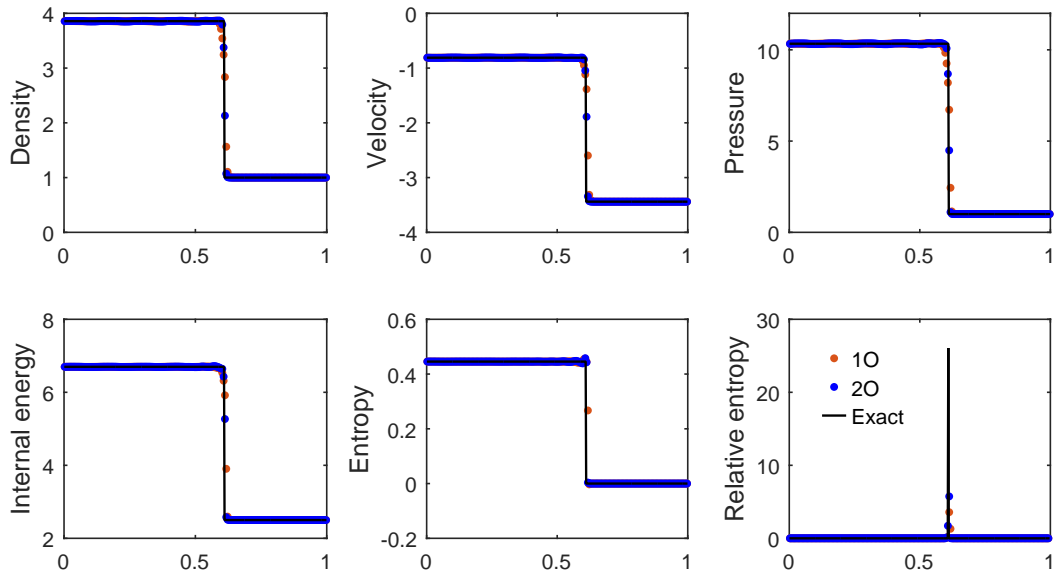}
\caption{\label{fig:8} Test case 4: Slowly moving shock}
\end{figure}

\begin{figure}
\centering
\includegraphics[width=14cm]{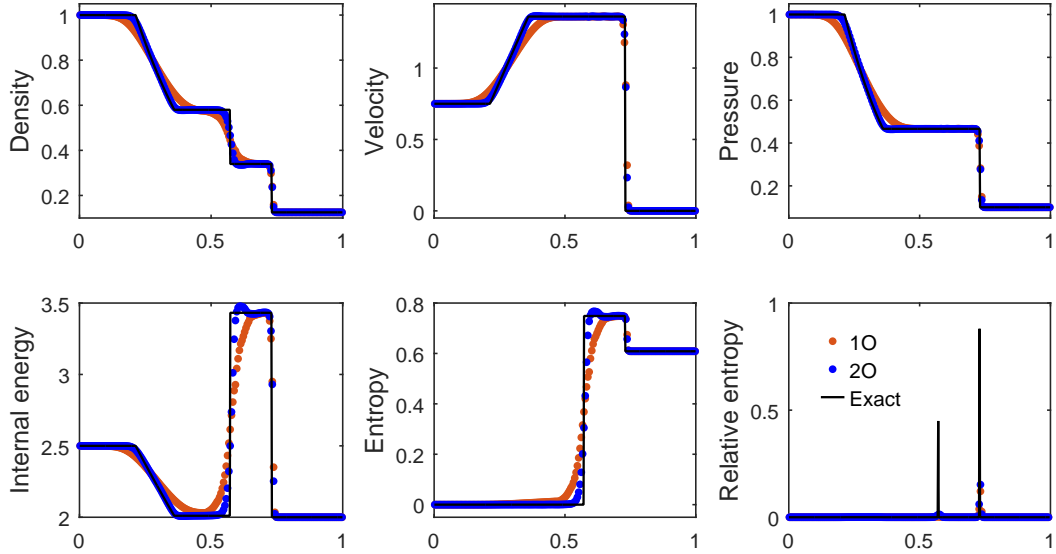}
\caption{\label{fig:9} Test case 5: Sod's shock tube problem, t=0.2}
\end{figure}

\begin{figure}
\centering
\includegraphics[width=14cm]{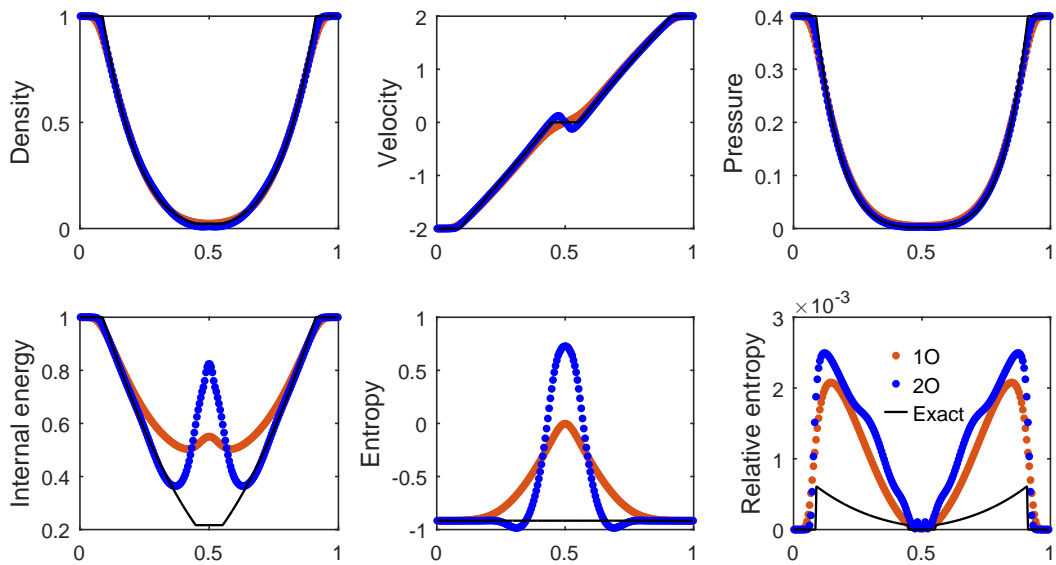}
\caption{\label{fig:10} Test case 6: Overheating problem}
\end{figure}

\begin{figure}
\centering
\includegraphics[width=14cm]{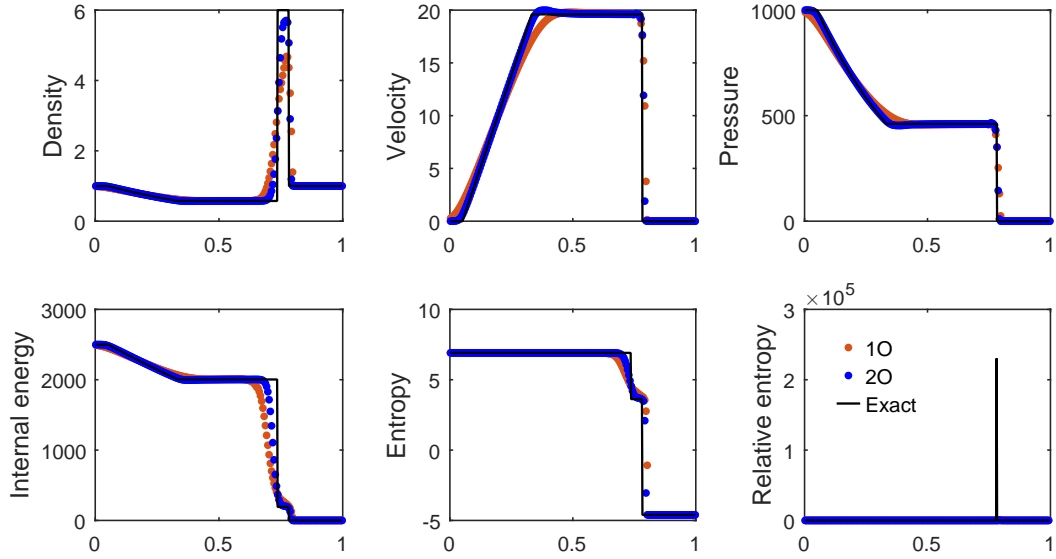}
\caption{\label{fig:11} Test case 7: Left half portion of Woodward and Colella problem}
\end{figure}

\begin{figure}
\centering
\includegraphics[width=14cm]{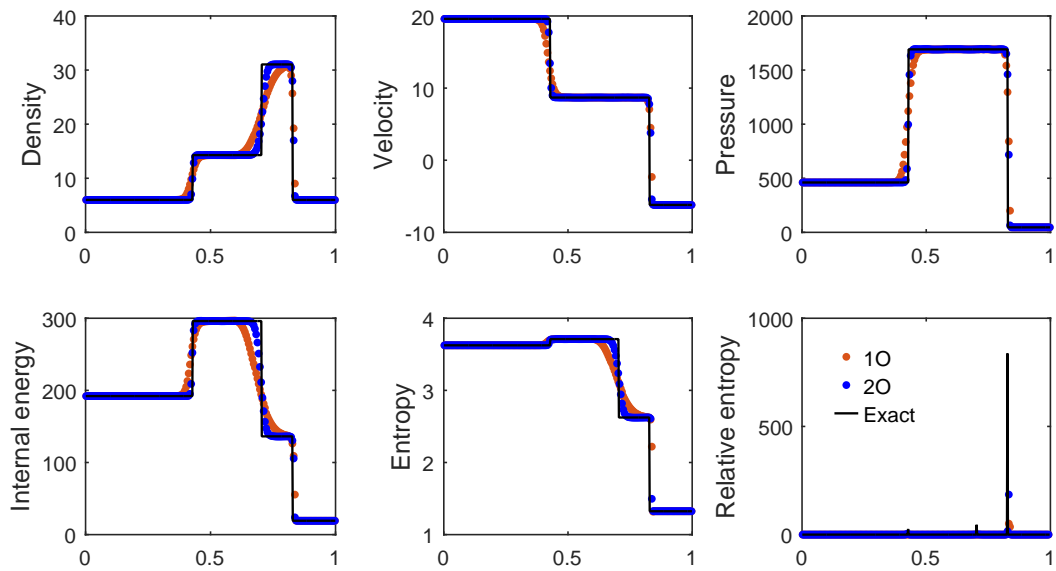}
\caption{\label{fig:12} Test case 8: Colliding strong shocks}
\end{figure}

\begin{figure}
\centering
\includegraphics[width=14cm]{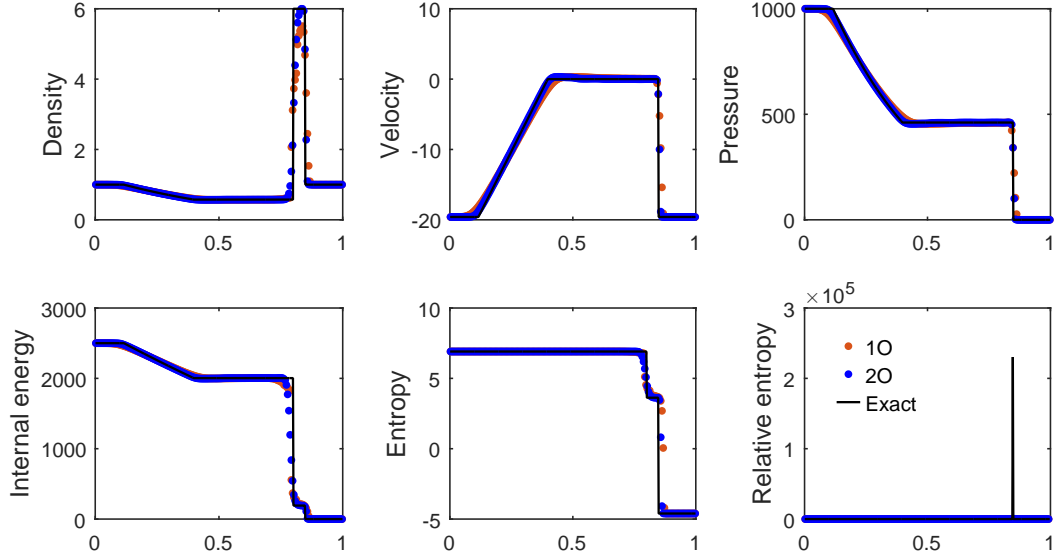}
\caption{\label{fig:13} Test case 9}
\end{figure}

An extensive list of 1D Euler test cases are solved to test the robustness and accuracy of our scheme. The initial conditions are given in Table \ref{table:3}. For all the problems, domain is x$\in[0,1]$, Nx= 200, CFL= 0.8 and Neumann boundary conditions are applied at the two ends. The global time step is computed as: $\Delta t$= CFL $\frac{\Delta x}{max_{i} (|u|+ a)_{i}}$. Test case 1 has a stationary contact discontinuity as initial discontinuity. Test case 2 has a stationary shock as initial condition, with freestream Mach no= 2. Post shock conditions are obtained using gas dynamics relations for the stationary shock test case. As the results in Figure \ref{fig:5} and \ref{fig:6} show, the stationary discontinuities for both these test cases are, by design, captured exactly. Test case 3 (Figure \ref{fig:7}) comprises of a slowly moving contact discontinuity. The moving contact discontinuity is diffused as expected; the result is reasonably accurate. Test case 4 comprises of a slowly moving shock wave. The results in Figure \ref{fig:8} show the shock captured over few cells, with very minor post-shock oscillations. Test cases 5 to 9 are taken from Toro (Chapter 6, test cases 1 to 5 \cite{toro2013riemann}). Test case 5 is Sod's shock tube problem; the solution consists of a right shock wave, a right traveling contact discontinuity and a left sonic expansion wave. The results in Figure \ref{fig:9} show that no entropy violating expansion shocks are formed. Test case 6 (overheating problem) comprises of two strong symmetric expansions at left and right, with a contact discontinuity of vanishing strength in the middle. The pressure at the center reaches near vacuum. Thus, this problem is suitable for assessing performance of a scheme for low-density flows. The results for this test case in Figure \ref{fig:10} show that our scheme does not fail even at low densities. However, an increase in internal energy at the center is observed, which is common to many numerical methods due to numerical overheating. Test cases 7 to 9 test the robustness of a scheme in handling large gradients. Test case 7 is the left half of the blast wave problem of Woodward and Colella. Its solution comprises of a strong shock to the left, a contact discontinuity in middle and expansion fan to the right. Test case 8 involves collision of two strong shocks; its solution consists of a left facing shock (traveling very slowly to the right), a right traveling contact discontinuity and a right traveling shock. Test case 9 consists of a left rarefaction wave, a right traveling shock wave and a stationary contact discontinuity. The results for these test cases are shown in Figures \ref{fig:11}- \ref{fig:13}. The results are reasonably accurate.

\subsection{2D Euler tests}
Some standard 2D inviscid test cases are solved to showcase the accuracy and robustness of the scheme.  Time step is computed as $\Delta t= \text{CFL} \ min_{j,k}\frac{Area_{j,k}}{(|u_{\xi}|+a)l_{\xi}+ (|u_{\eta}|+a)l_{\eta}}$, where ($\eta, \xi$) are the grid-coordinate directions.  For all test cases, we take CFL= 0.8 unless specified otherwise.  For steady test cases, the solution is evolved in time until a minimum residual (of density) of $10^{-10}$ or maximum time steps of 50000 is reached, whichever happens earlier.  

\subsubsection{Oblique shock reflection}
\begin{figure}[h!] 
\centering
\begin{tabular}{cc}
\includegraphics[width=0.45\textwidth]{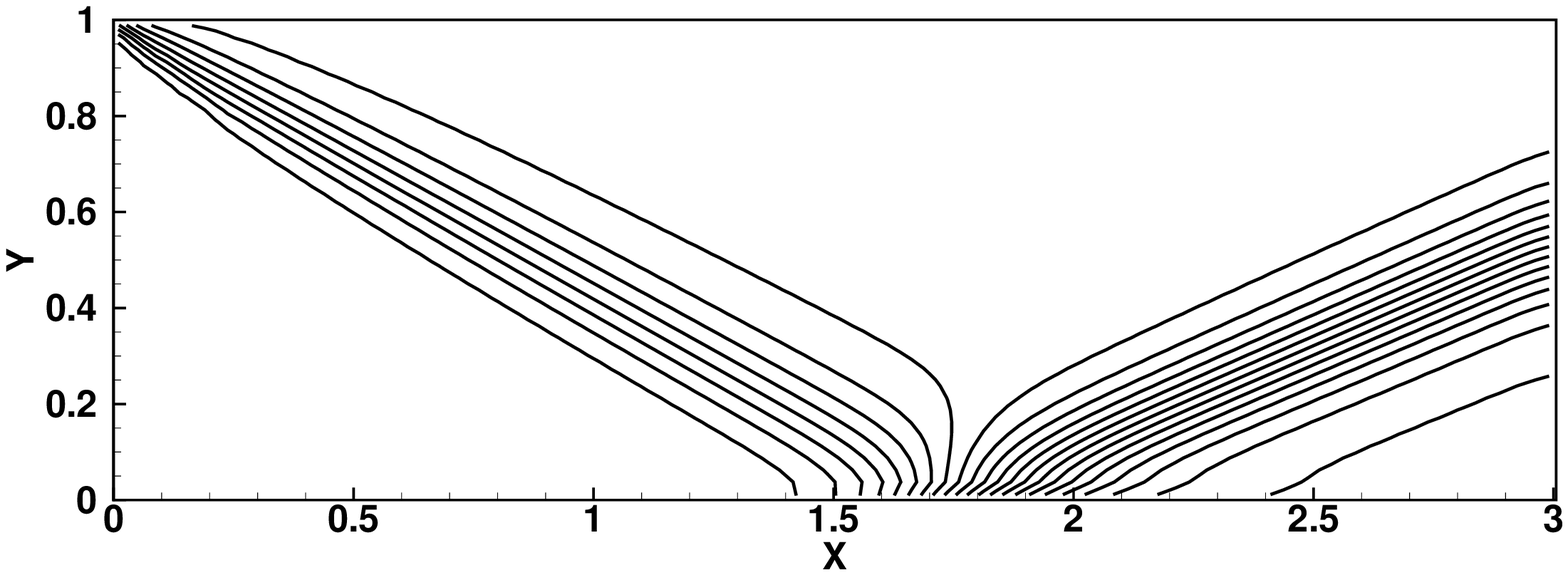} & \includegraphics[width=0.45\textwidth]{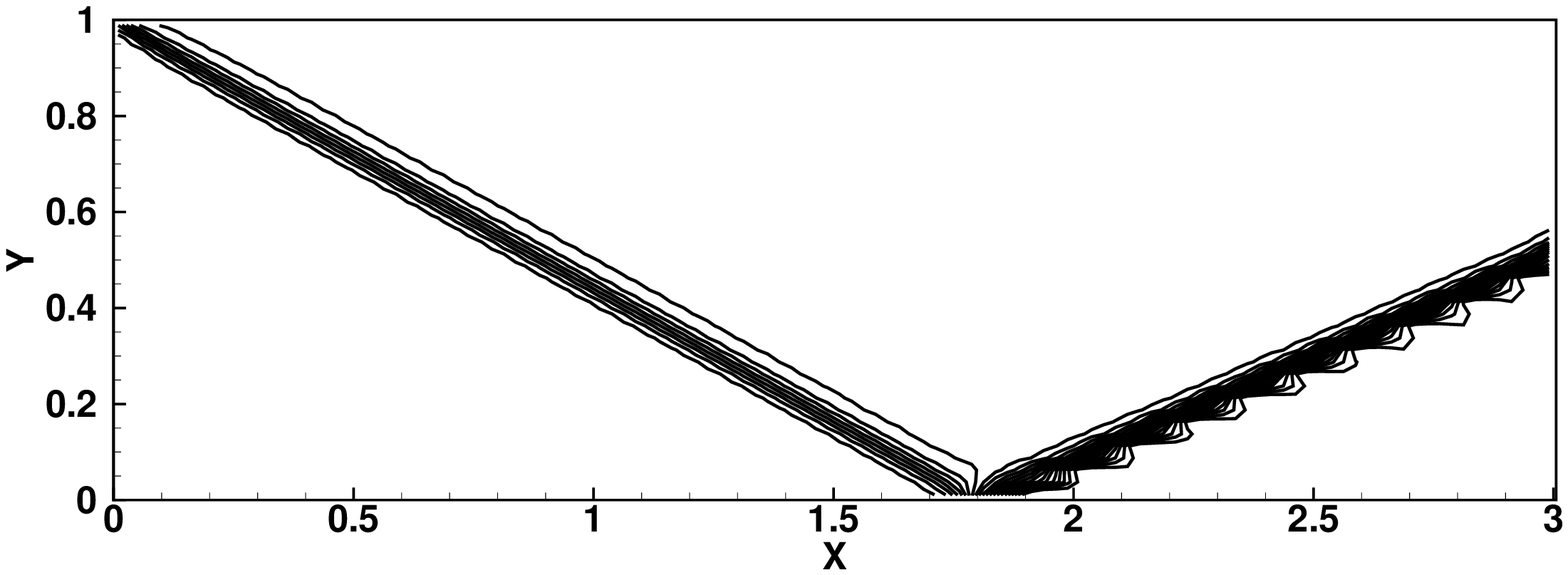}\\
\includegraphics[width=0.45\textwidth]{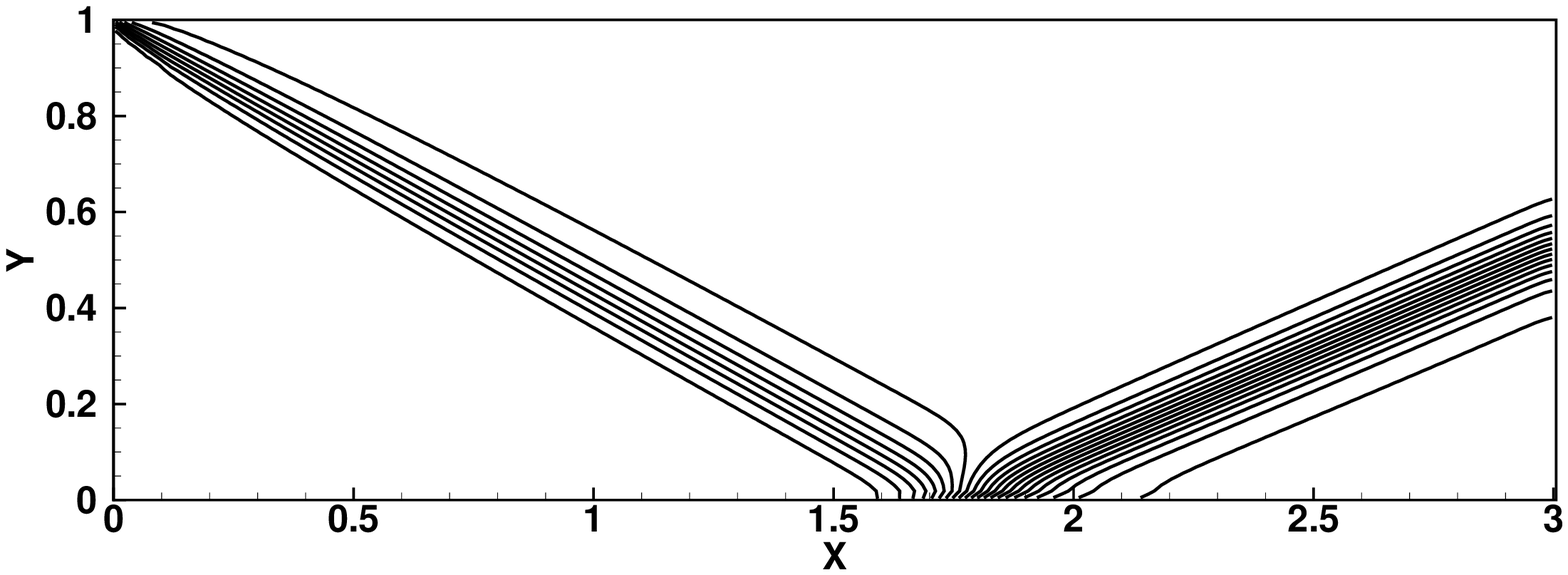} & \includegraphics[width=0.45\textwidth]{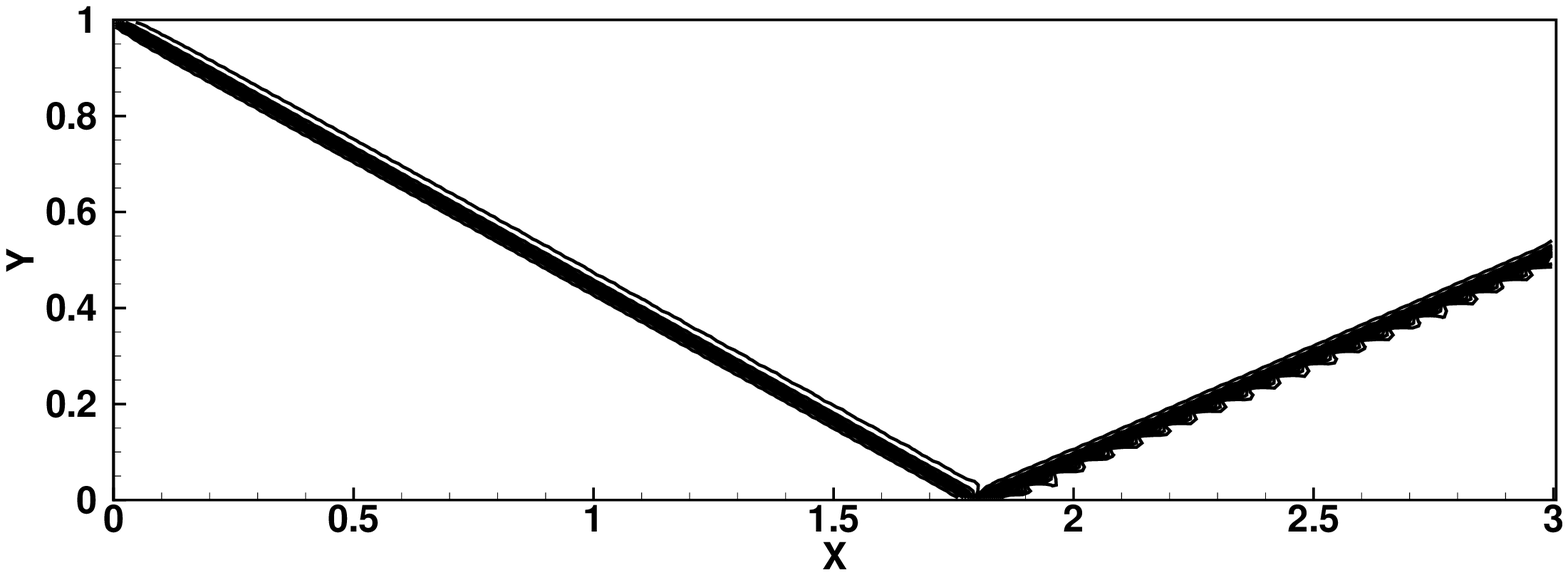} 
\end{tabular}
\caption{Oblique shock reflection with shock angle $29^\circ$, inflow Mach no. 2.9 - Pressure contours (0.7:0.1:2.9), Top) I order and II order accurate results for $120 \times 40$ grid, Bottom) I order and II order accurate results on $240 \times 80$ grid} 
\label{fig:16}
\end{figure}
\begin{figure}[h!] 
\centering
\begin{tabular}{cc}
\includegraphics[width=0.45\textwidth]{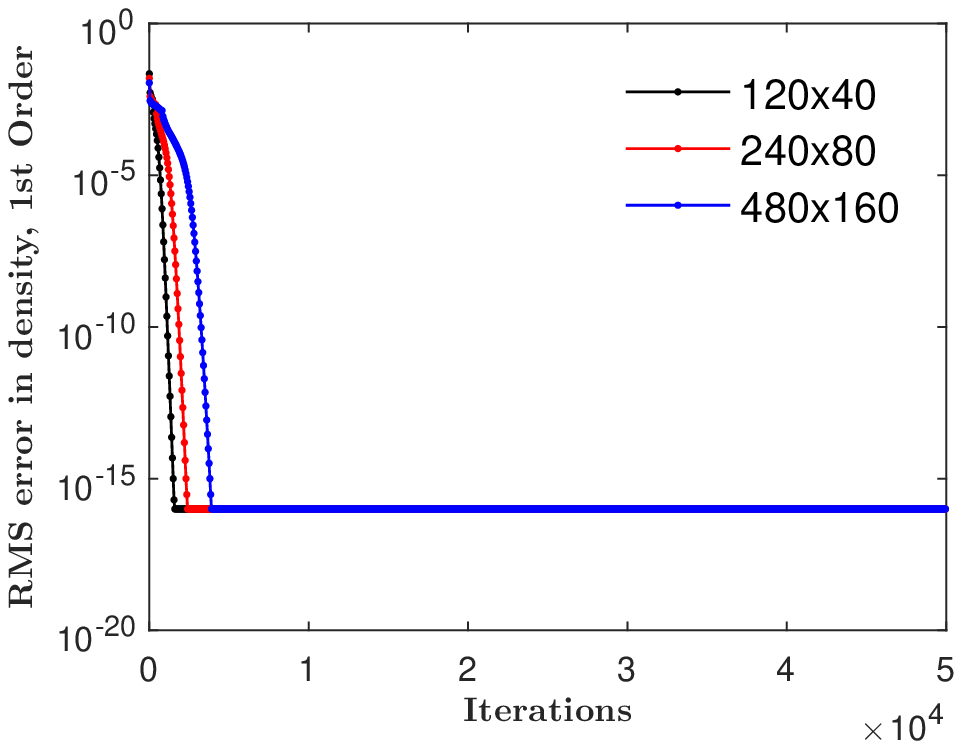} & \includegraphics[width=0.45\textwidth]{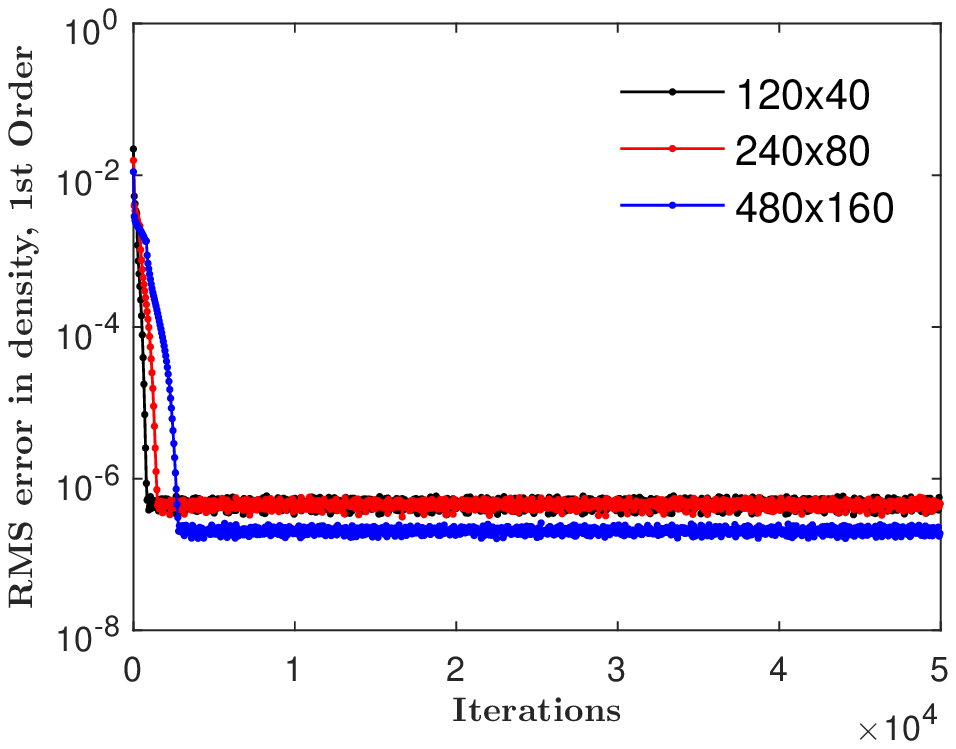} 
\end{tabular}
\caption{Oblique shock reflection test case: RMS error in density vs number of iterations for I order result, a) Criterion \eqref{eq:4_27} used for smooth flow regions, b) Criterion \eqref{eq:4_28} used for smooth flow regions}
\label{fig:17}
\end{figure}

\begin{figure}[h!] 
\centering
\includegraphics[width=0.45\textwidth]{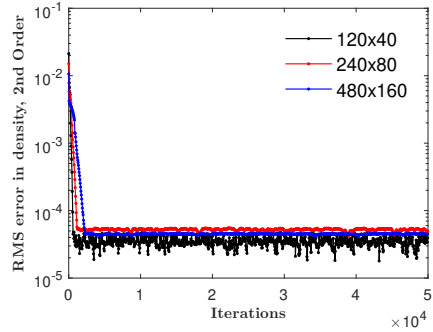}
\caption{Oblique shock reflection test case: RMS error in density vs number of iterations for II order result}
\label{fig:17a}
\end{figure}
This test case \cite{yee1982high} comprises of an oblique shock, striking and reflecting from a solid wall. The incident shock angle is $29^\circ$ and freestream Mach no. is 2.9.  The computational domain is $[0,3] \times [0,1]$, with Cartesian cells. At the left boundary, freestream conditions are applied; flow tangency (wall) conditions are applied at the bottom. Post shock conditions obtained using compressible flow relations are applied at the top, and supersonic outflow conditions are applied at the right boundary. Freestream initial conditions are used. Figure \ref{fig:16} shows the pressure contours of first and second order accurate steady state solutions. The shock profile for second order accurate result is much sharper compared to the first order accurate result.

Figure \ref{fig:17} shows the variation in RMS error in density with number of iterations for the first order result with varying grid sizes. Two plots are shown; one where criteria  \eqref{eq:4_27} is used to identify smooth flow regions, and the other where the much stricter criteria \eqref{eq:4_28} is used to identify smooth flow regions. The criteria \eqref{eq:4_27} leads to better convergence as it is a less strict criteria to introduce LLF type numerical diffusion. However, it makes the scheme more diffusive, hence we use \eqref{eq:4_28} for our test cases. Figure \ref{fig:17a} shows the variation in RMS error in density with iterations for second order result with varying grid sizes. 

\subsubsection{Supersonic flow over a compression ramp}
\begin{figure}[h!] 
\centering
\begin{tabular}{cc}
\includegraphics[width=0.45\textwidth]{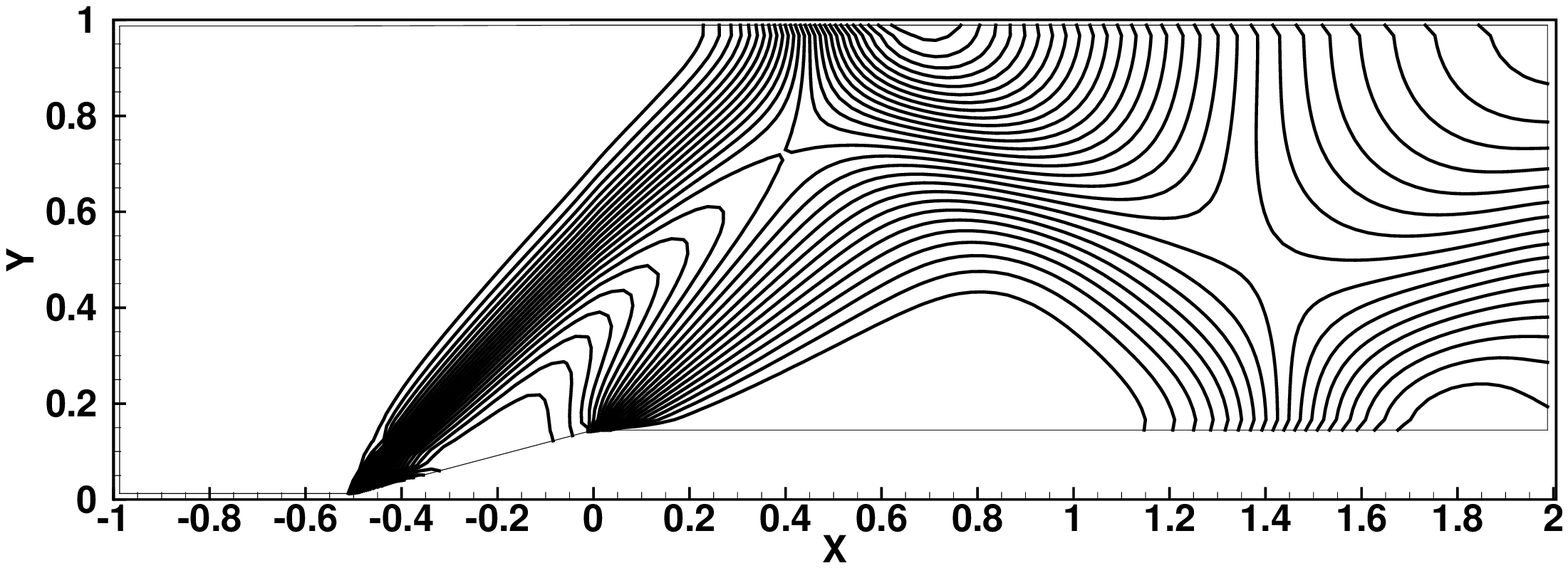} & \includegraphics[width=0.45\textwidth]{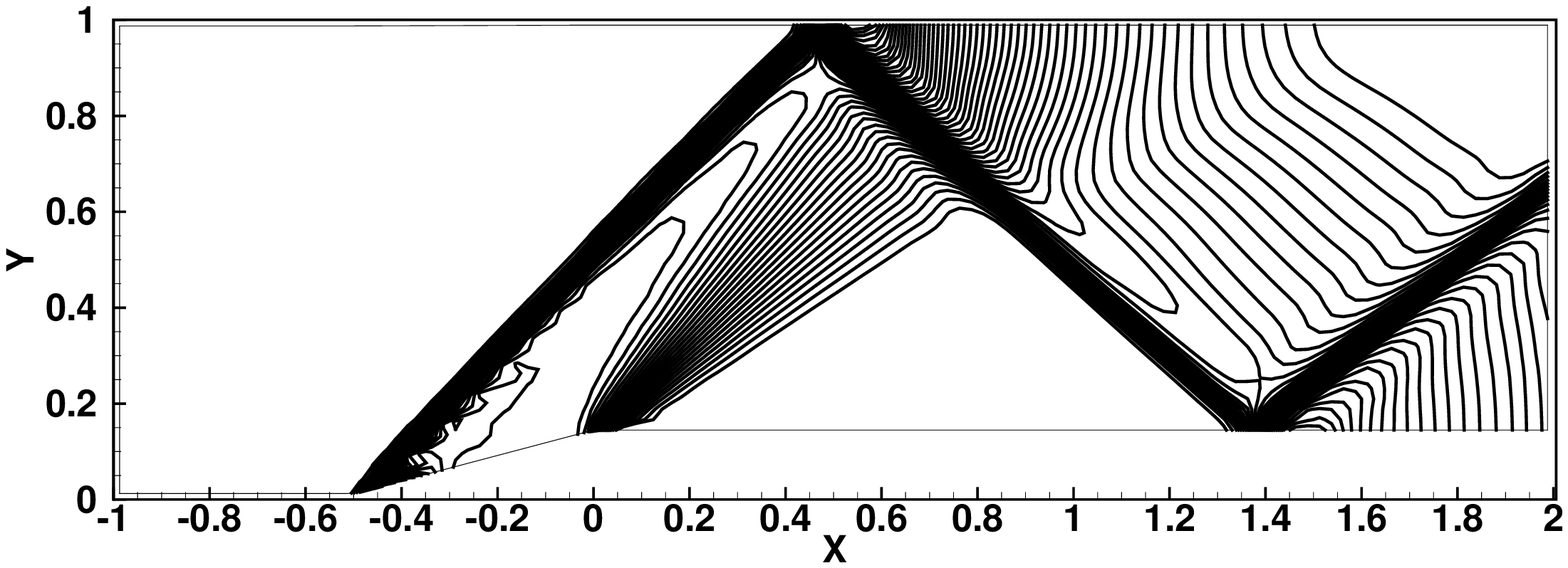}\\
\includegraphics[width=0.45\textwidth]{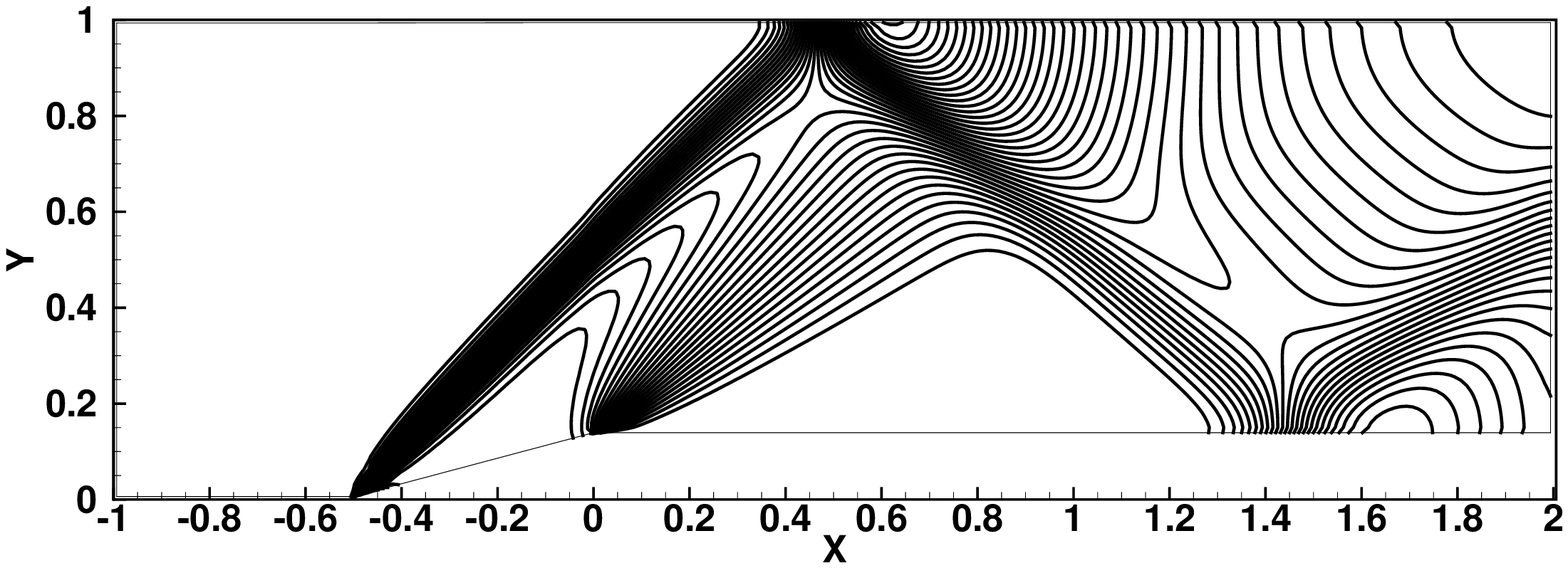} & \includegraphics[width=0.45\textwidth]{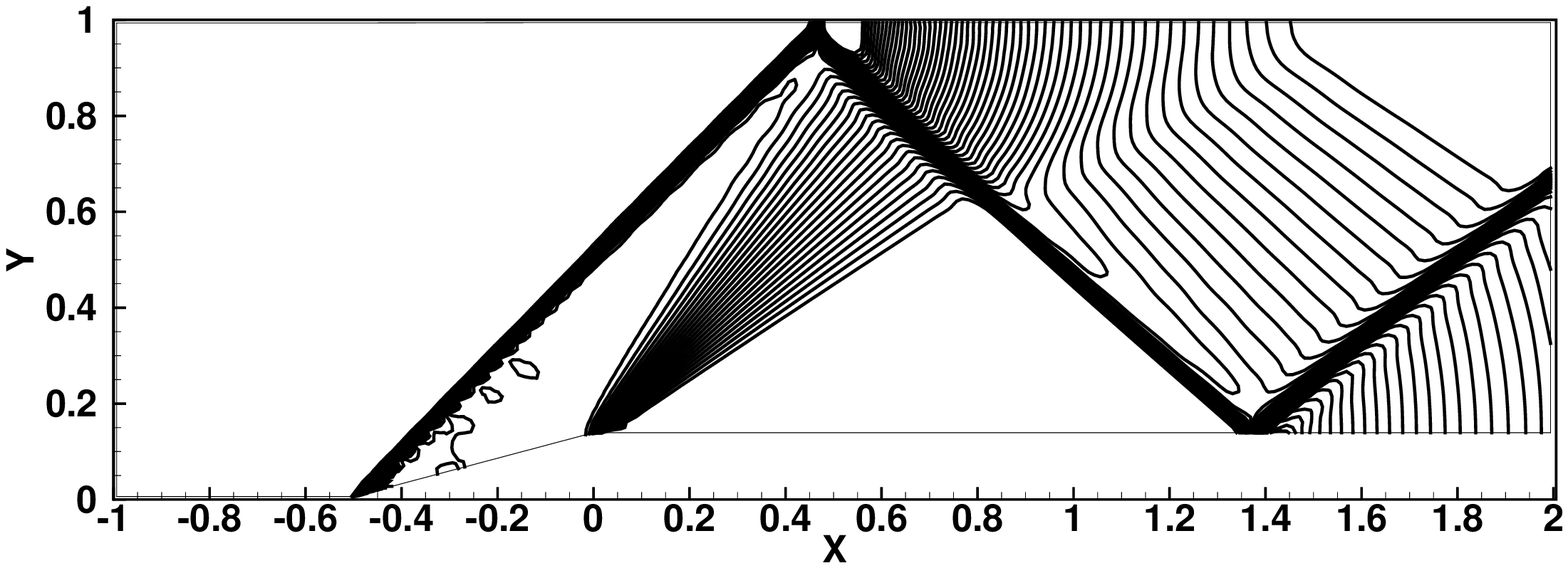} 
\end{tabular}
\caption{Mach 2 flow over a $15^\circ$ ramp, - Pressure contours (1.1:0.05:3.8), Top) I order and II order accurate results on $120 \times 40$ grid, Bottom) I order and II order accurate results on $240 \times 80$ grid}
\label{fig:18}
\end{figure}
This test consists of a Mach 2 flow over a $15^\circ$ compression ramp in a wind tunnel \cite{levy1993use}. The dimensions of the computational domain are $[-1,2] \times [0,1]$, with a $15^\circ$ ramp at the bottom from $x= -0.5$ to $x=0$.  Freestream conditions are applied at the left boundary, flow tangency boundary conditions are applied at top and bottom walls and supersonic outflow conditions are applied at the right boundary. Freestream initial conditions are used throughout the interior domain.  The steady state solution comprises of an oblique shock originating at the concave corner (start of the ramp) and expansion fans  starting from the convex corner (end of ramp). The oblique shock strikes and reflects from the top and bottom walls and also interacts with the emerging and reflected expansion fans. As the contours in Figure \ref{fig:18} show, no entropy violating expansion shocks are formed.  

\subsubsection{Horizontal Slip flow}
\begin{figure}[h!] 
\centering
\begin{tabular}{cc}
\includegraphics[width=0.4\textwidth]{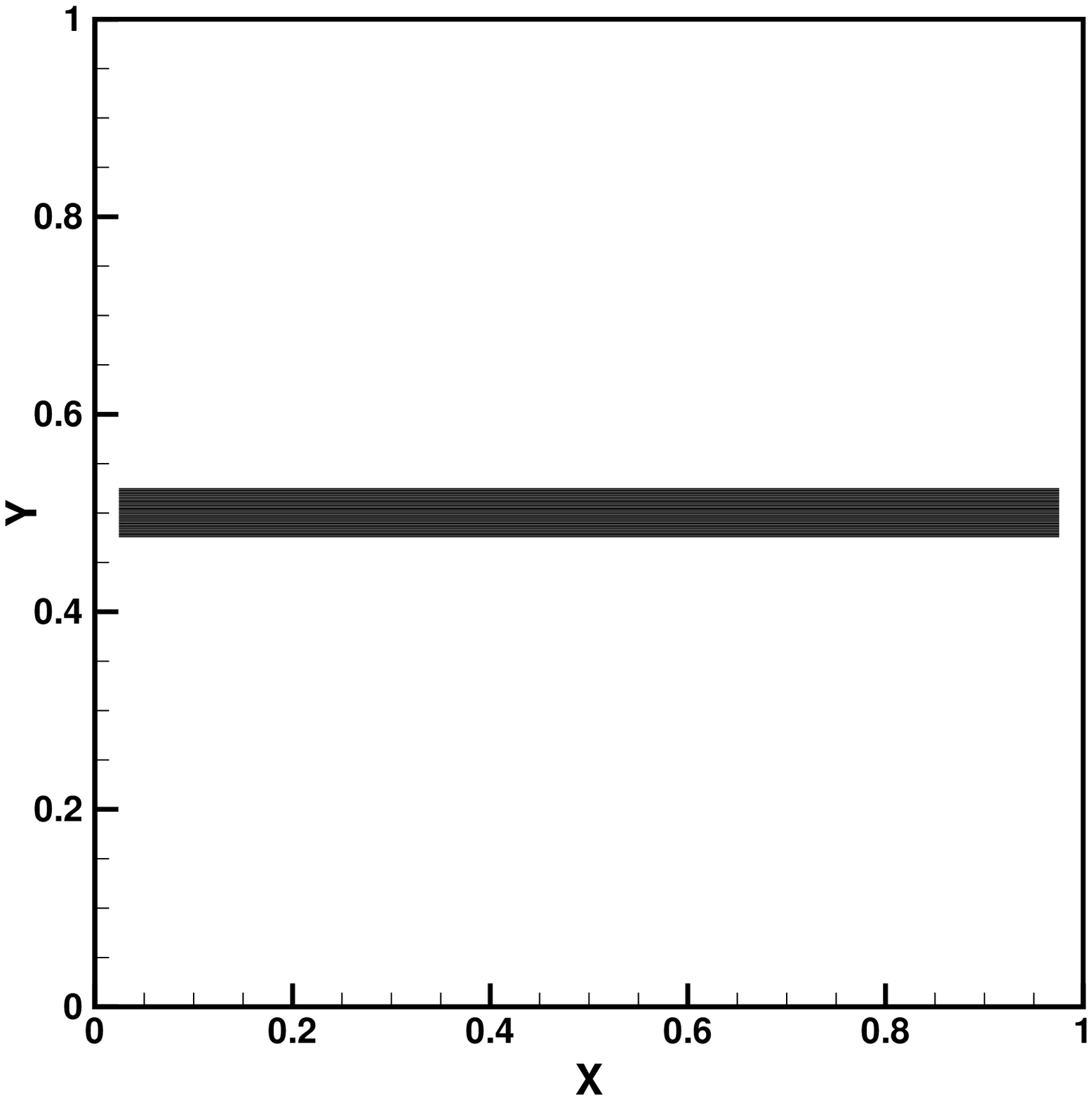} & \includegraphics[width=0.4\textwidth]{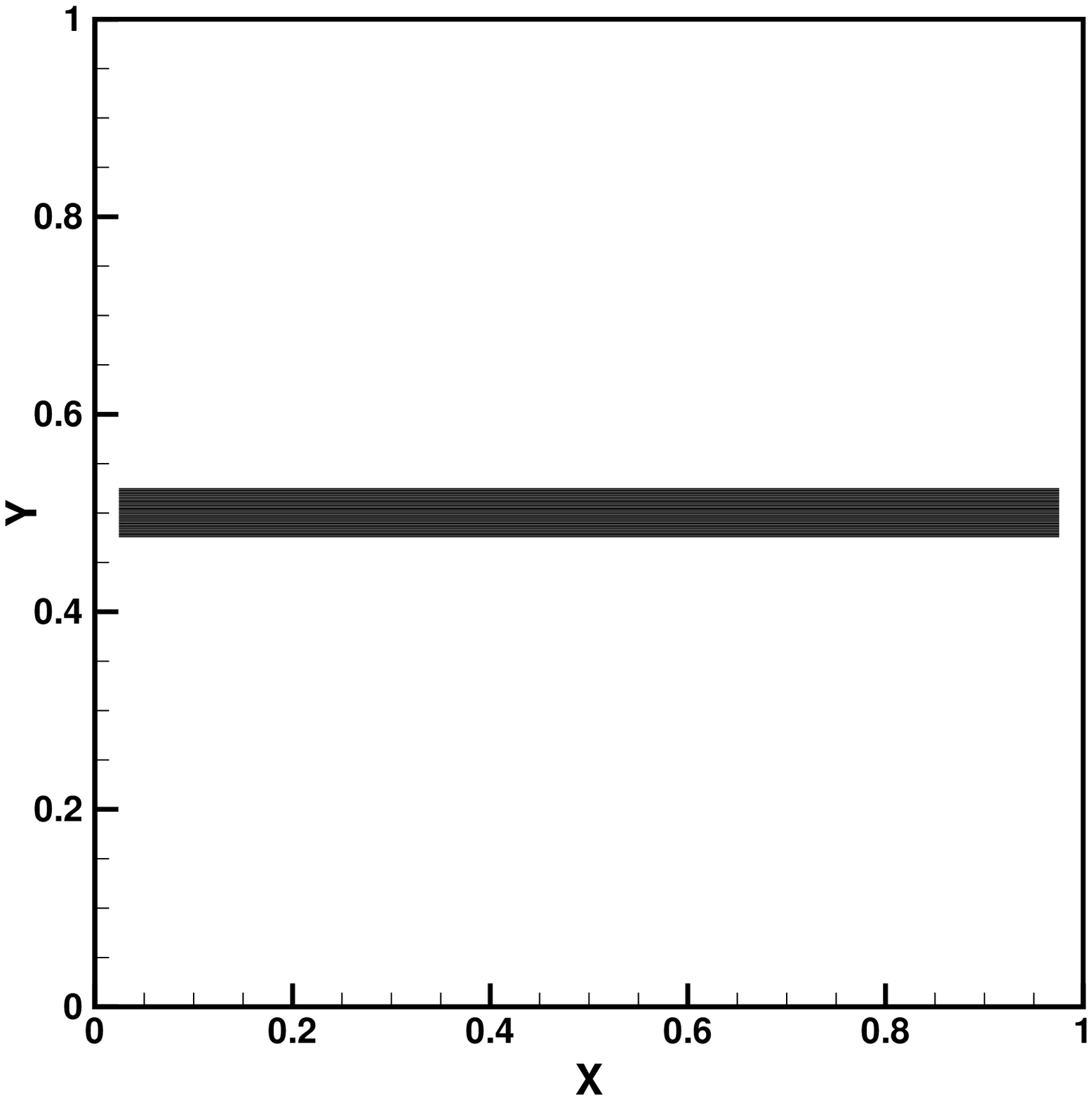}
\end{tabular}
\caption{Mach 3 flow slipping on a Mach 2 flow, $u_{1}$ contours (2:0.033:3)(a) I order, $20 \times 20$ grid (b) II order, $20 \times 20$ grid}
\label{fig:19}
\end{figure}
In this test case, a Mach 3 flow slips on top of a Mach 2 flow \cite{manna1992three}. The computational domain is $[0,1] \times [01]$ with Cartesian cells. The flow is initialized with a horizontal Mach 3 flow for  $y \geq 0.5$ and Mach 2 flow for $y < 0.5 $, keeping density and pressure same for both flows.  Supersonic inflow conditions are used at the left boundary and supersonic outflow conditions at the right boundary.  Neumann boundary conditions are used at top and bottom boundaries. The steady state solution shown in Figure \ref{fig:19} shows that our scheme captures the grid aligned slip-stream exactly without any numerical diffusion.
\subsubsection{Hypersonic flow over a half-cylinder}
\begin{figure}
\centering
\begin{tabular}{cccc}
\includegraphics[width=0.2\textwidth]{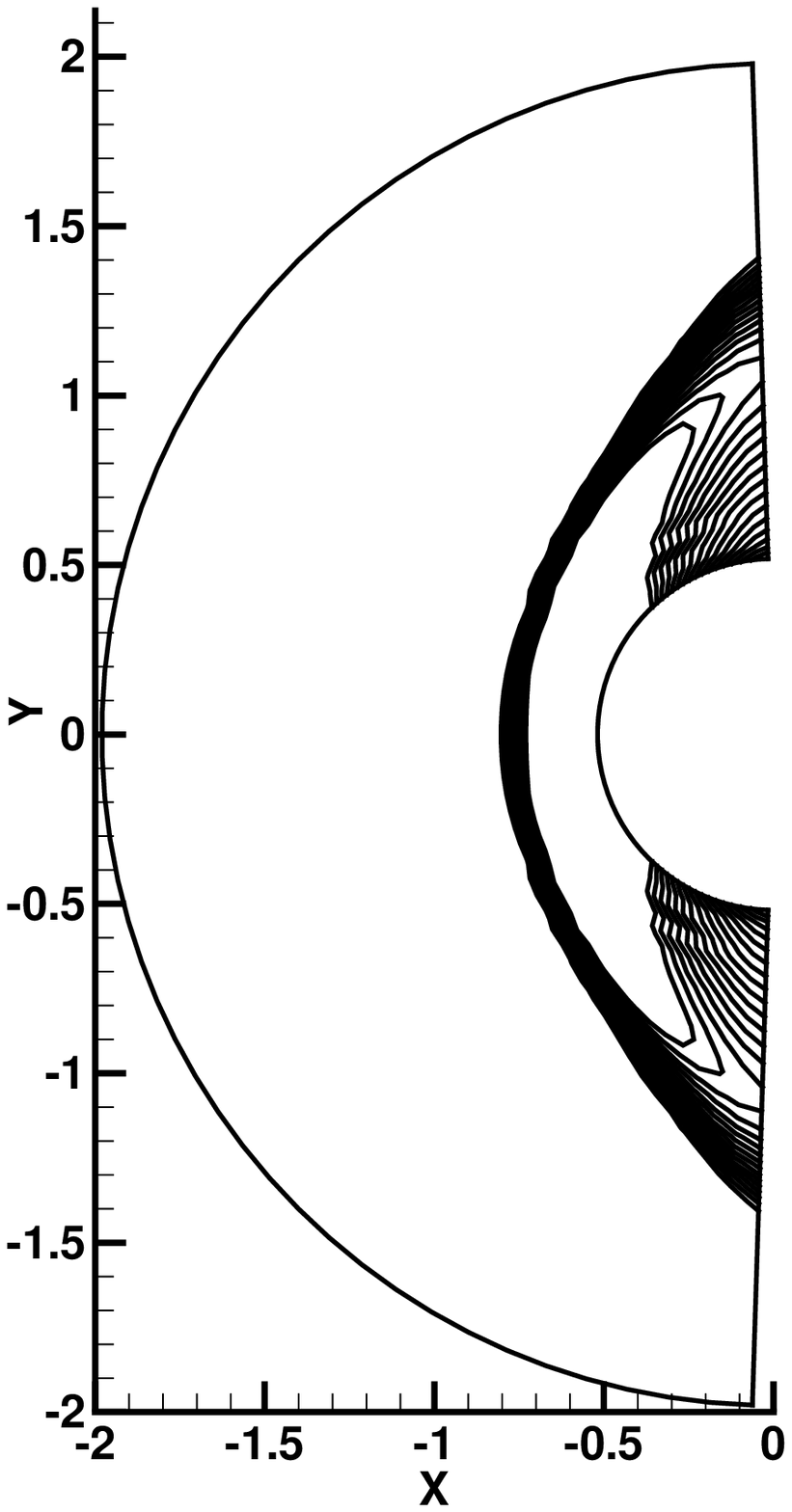} & \includegraphics[width=0.2\textwidth]{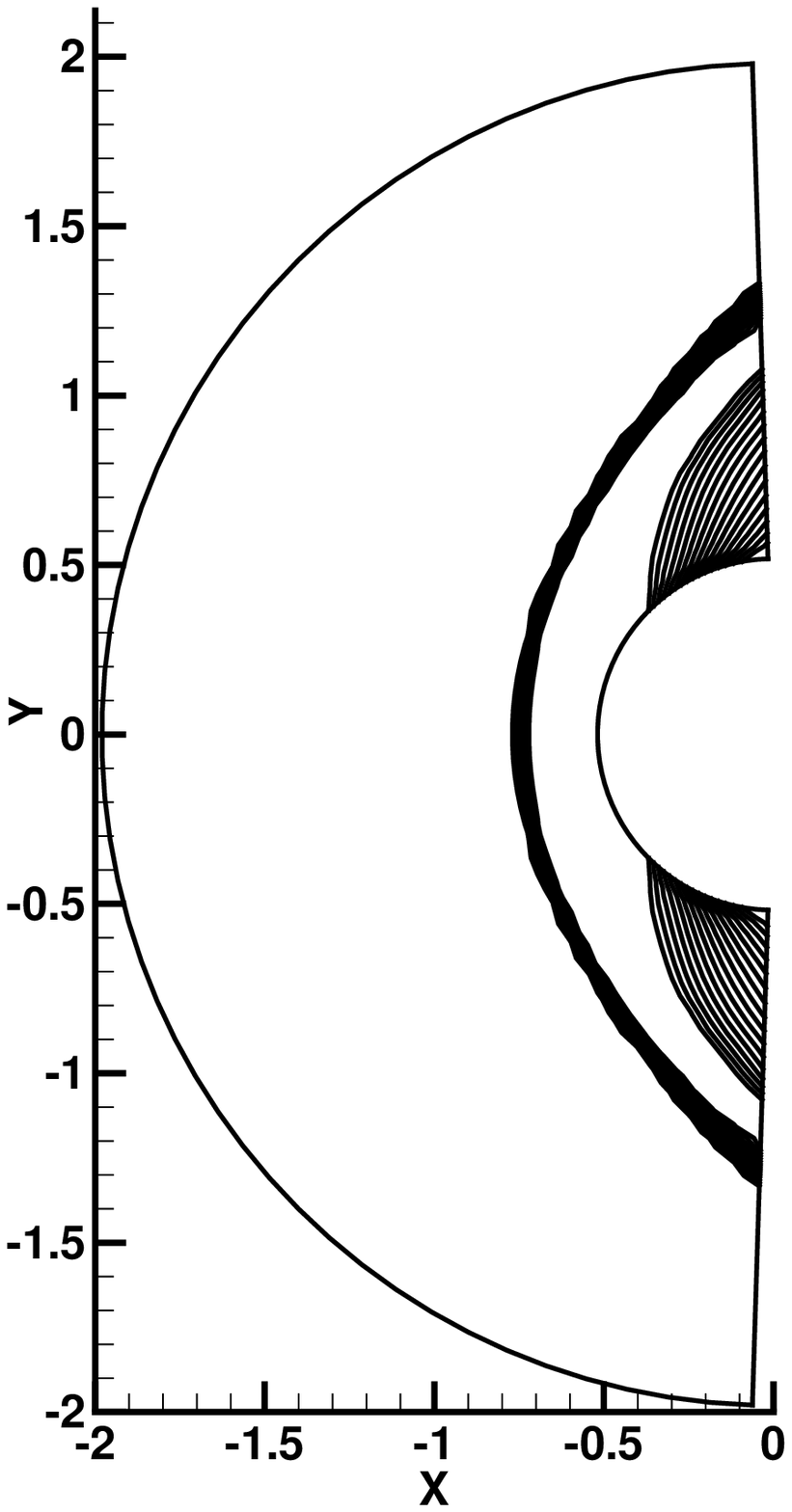} & \includegraphics[width=0.2\textwidth]{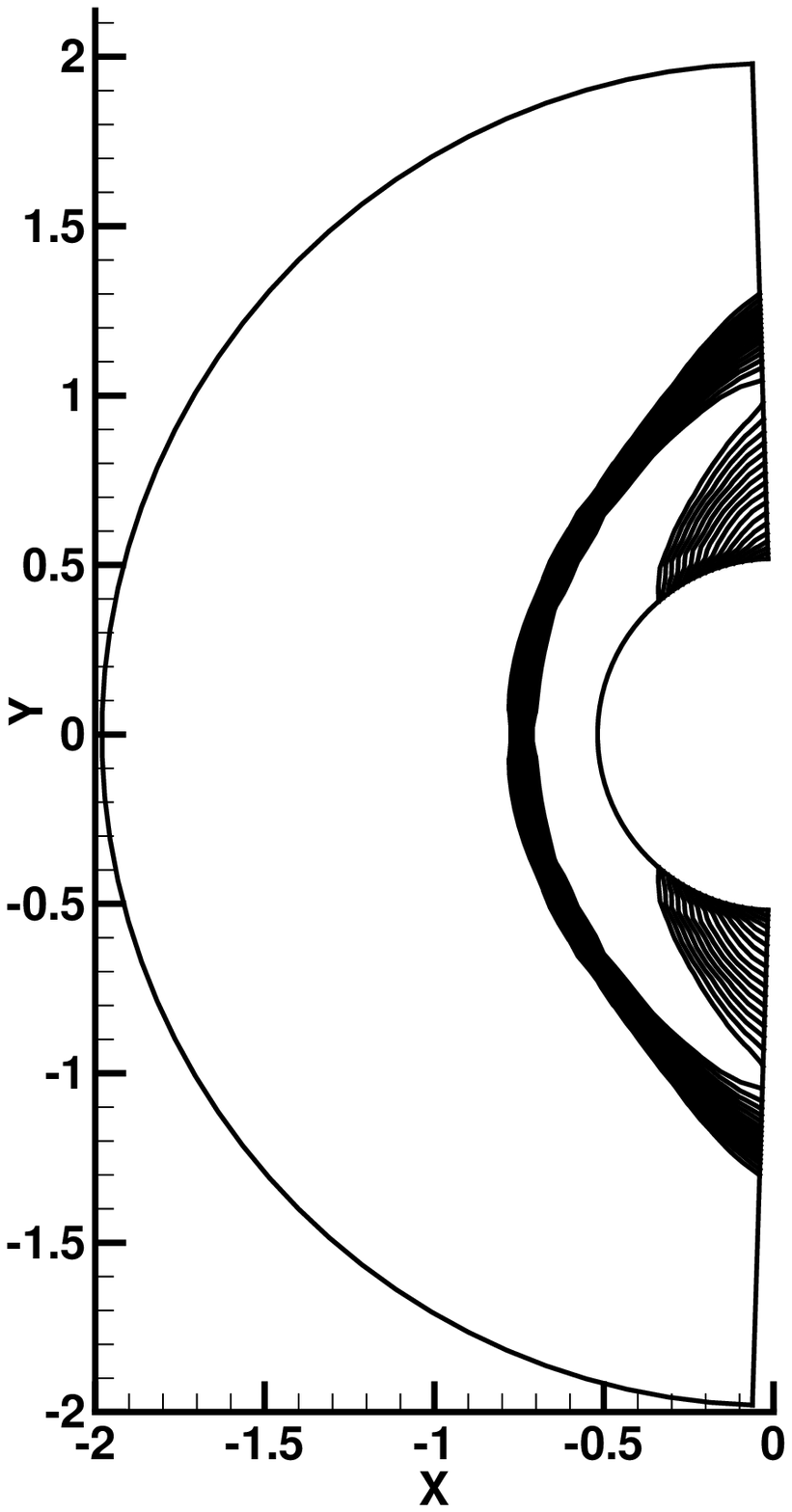} & \includegraphics[width=0.2\textwidth]{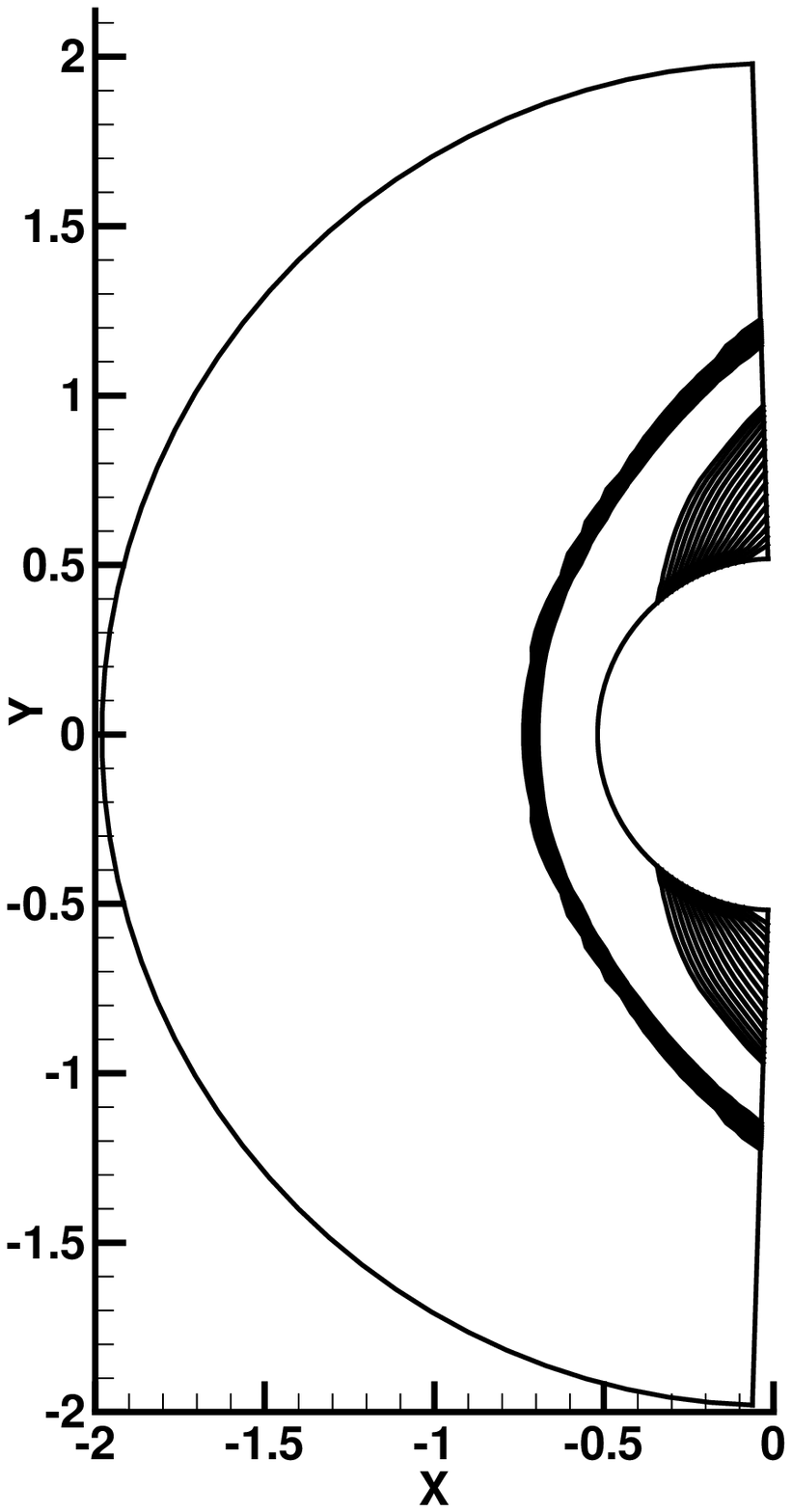}
\end{tabular}
\caption{Hypersonic flow past a half-cylinder, density contours (2:0.2:5) on $50 \times 40$ grid (a) Mach 6 flow, I order (b) Mach 6 flow, II order, (c) Mach 20 flow, I order (d) Mach 20 flow, II order}
\label{fig:20}
\end{figure}
This test consists of Mach 6 and Mach 20 flows over a half cylinder. The computational domain is $r, \theta \in  [0.5,2] \times [\frac{\pi}{2},\frac{3\pi}{2}]$, with constant grid spacing along $r$ and $\theta$ directions. Boundary conditions used are: supersonic inflow and flow tangency conditions at $r=2$ and $r= 0.5$ and supersonic outflow conditions at $\theta=\frac{\pi}{2}$ and $\frac{3\pi}{2}$ boundaries.  The flow is initialized with freestream conditions. Further, for the Mach 20 flow test case, a lower CFL of 0.5 is used to prevent pressure from taking negative values. The steady state solution consists of a bow shock formed in front of, and detached from, the half cylinder. For this test case, many low diffusive schemes like Riemann solvers give rise to carbuncle phenomenon leading to solution failures due to (numerical) shock instability \cite{quirk1997contribution}. The results for this test case are shown in Figure \ref{fig:20}. No carbuncle shock is observed in our solutions.

\subsubsection{Supersonic flow over a forward-facing step}
\begin{figure}[h!] 
\centering
\begin{tabular}{cc}
\includegraphics[width=0.45\textwidth]{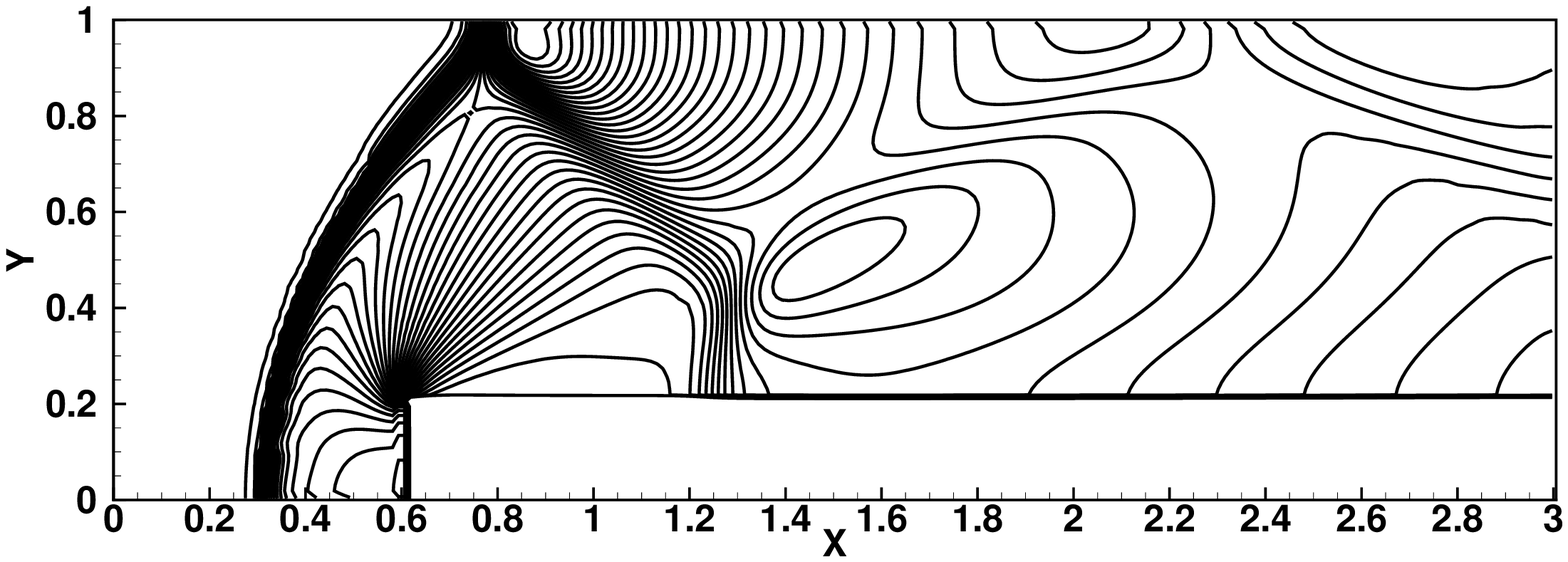} & \includegraphics[width=0.45\textwidth]{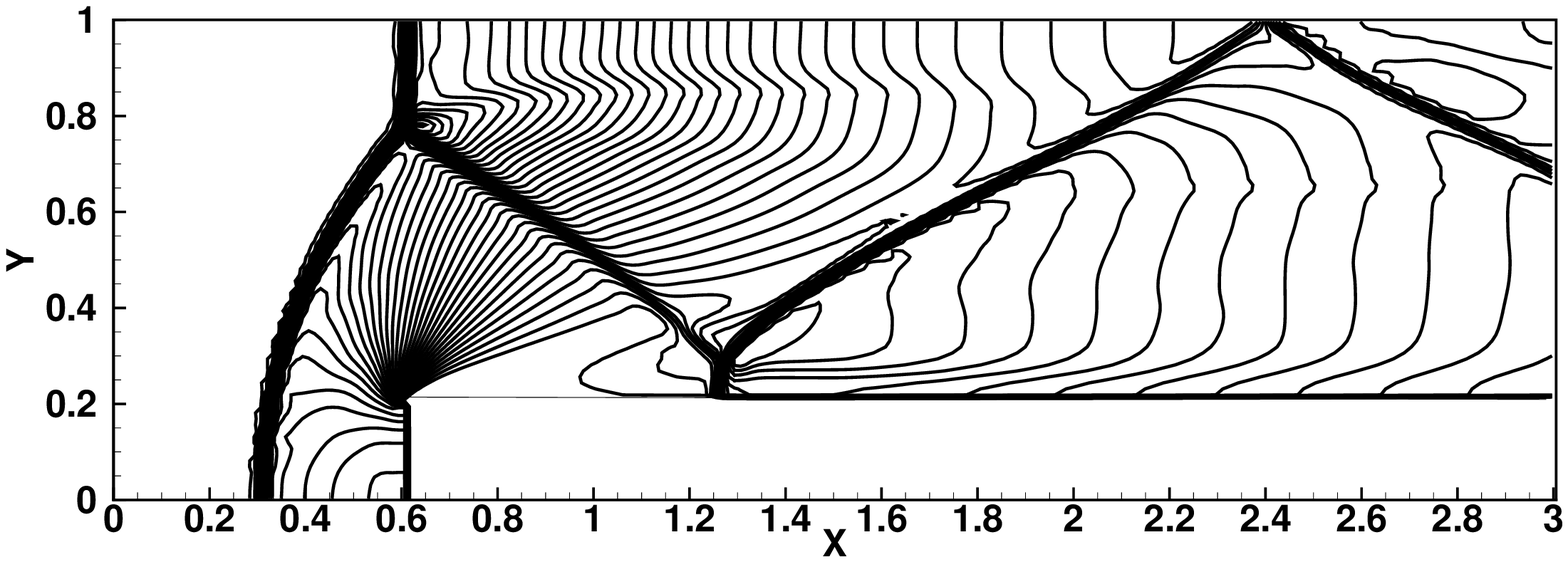}\\
\includegraphics[width=0.45\textwidth]{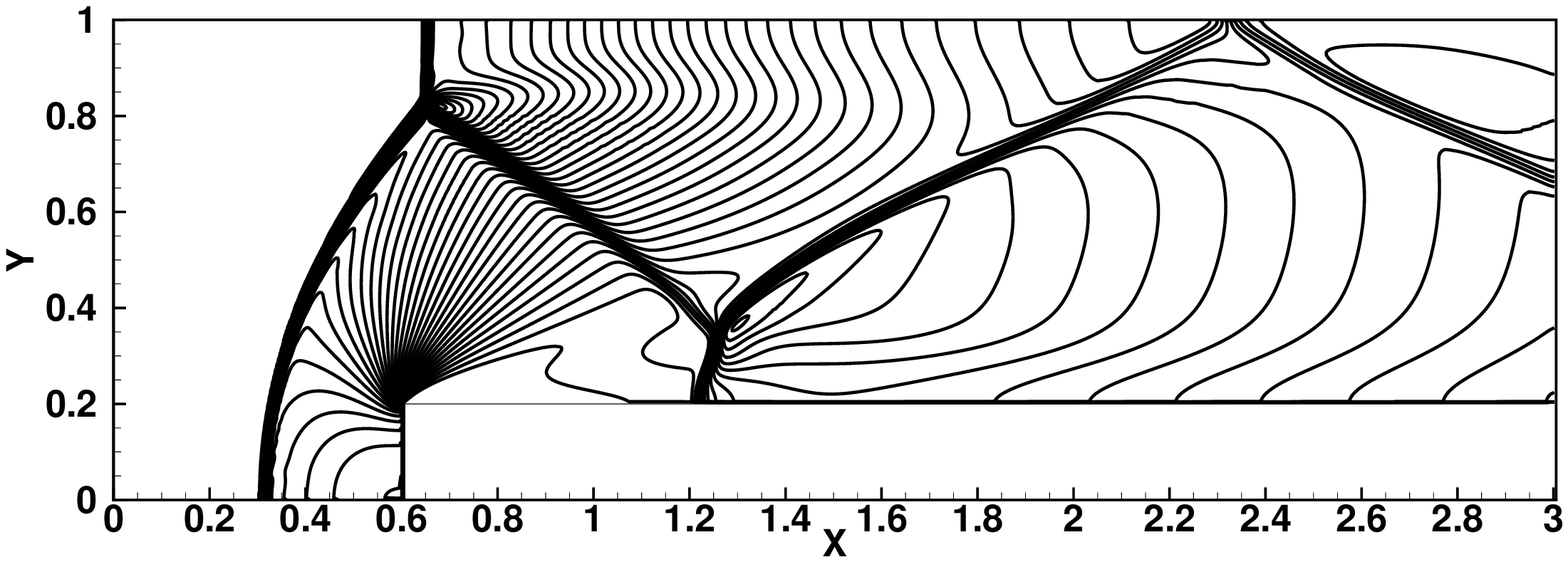} & \includegraphics[width=0.45\textwidth]{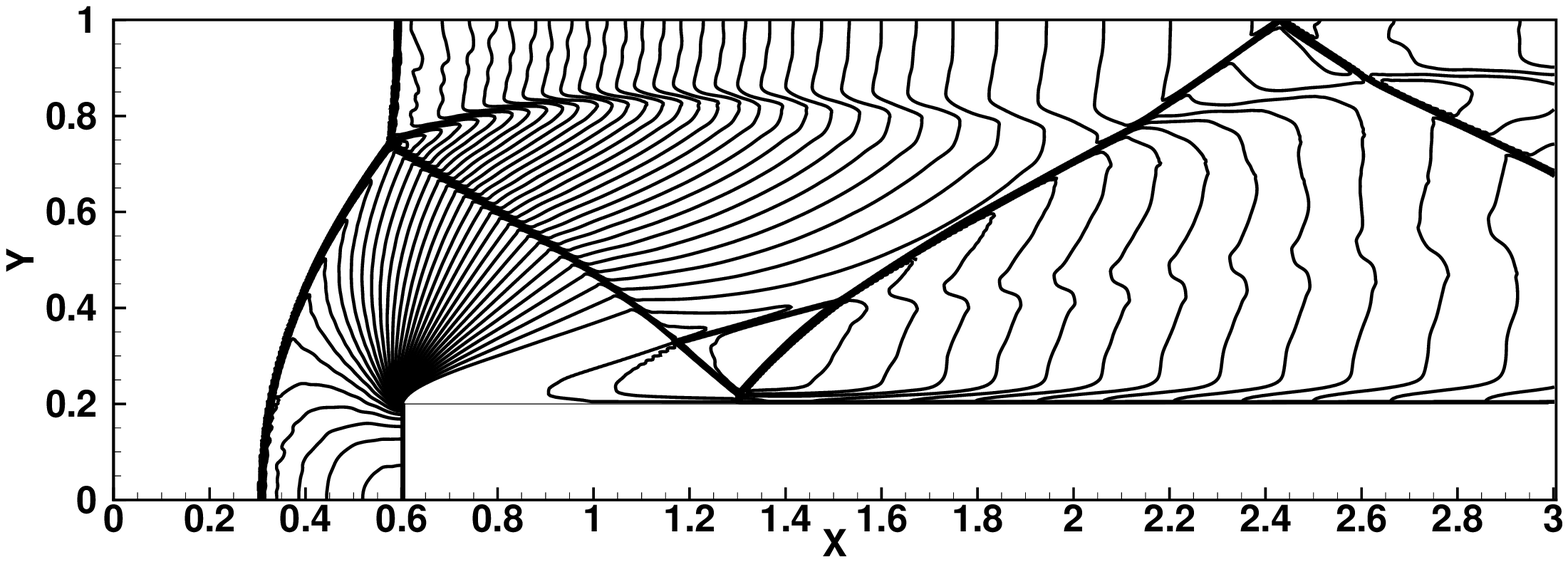} 
\end{tabular}
\caption{Mach 3 flow over a forward-facing step in wind tunnel, $t=4$, density contours (1:0.15:6.5); Top) I order and II order accurate results on $240 \times 80$ grid, Bottom) I order and II order accurate results on $960 \times 320$ grid}
\label{fig:21}
\end{figure}
In this test case, a Mach 3 flow enters the wind tunnel containing a step, from the left \cite{woodward1984numerical}. The dimensions of the wind tunnel are $[0,3] \times [0,1]$. The step is 0.2 units high and located at distance of 0.6 units from the left end. The boundary conditions are: supersonic inflow at the left boundary, flow tangency at the top and bottom (including the step) boundaries and supersonic outflow at the right boundary.  Freestream initial conditions are used. At $t= 4$, a lambda shock is developed.  A clear slip stream can be seen beyond the triple point, which can be captured well only by low diffusion schemes.  The first order and second order accurate results for this test case are shown in Figure \ref{fig:21}.   

\subsubsection{Odd-even decoupling}
\begin{figure}[h!] 
\centering
\begin{tabular}{c}
\includegraphics[width=0.8\textwidth]{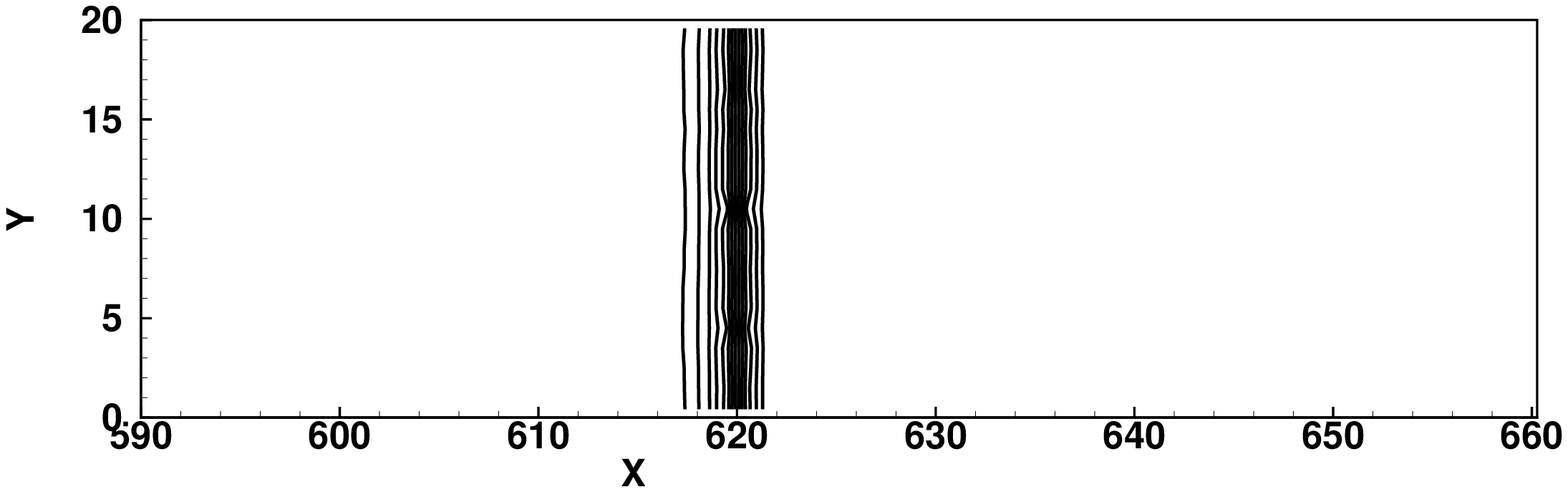} \\ \includegraphics[width=0.8\textwidth]{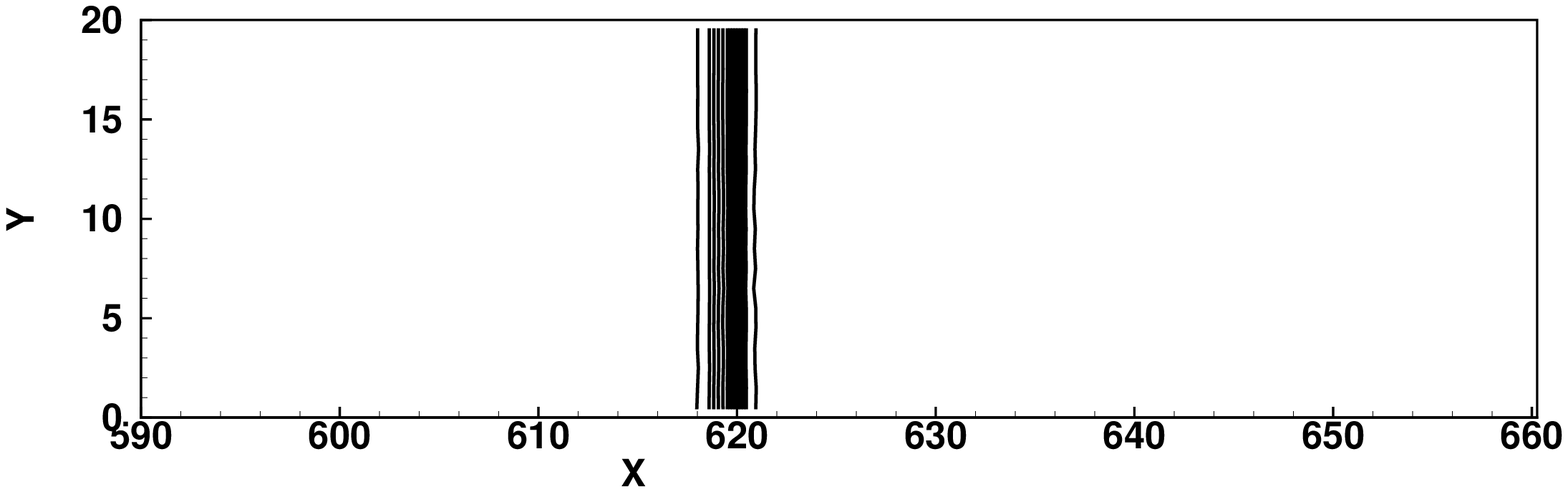} 
\end{tabular}
\caption{Mach 6 shock wave through a rectangular duct at $t=100$, $800 \times 20$ grid, density contours; (a) I order and (b) II order}
\label{fig:22}
\end{figure}
This test case assesses a scheme for a form of numerical instability called odd-even decoupling. The test case comprises of a planar shock wave of $M_{s} = 6$ propagating through a rectangular duct \cite{quirk1997contribution}. A Cartesian mesh of $800 \times 200$ square cells is used. The center-line of the grid is perturbed as
\begin{equation}
(y_{i,j})_{mid}= \left\{\begin{array}{c}
(y_{i,j})_{mid}+ 10^{-3}, \ i \ \text{is even}\\
(y_{i,j})_{mid}- 10^{-3}, \ i \ \text{is odd}
\end{array}\right\}
\label{eq:97}
\end{equation} 
Low diffusion schemes and approximate Riemann solvers develop oscillations, which then destroy the solution. However, as our results in Figure \ref{fig:22} show, our scheme is not prone to this form of instability.

\subsubsection{Double Mach reflection}
\begin{figure}[h!] 
\centering
\begin{tabular}{cc}
\includegraphics[width=0.3\textwidth]{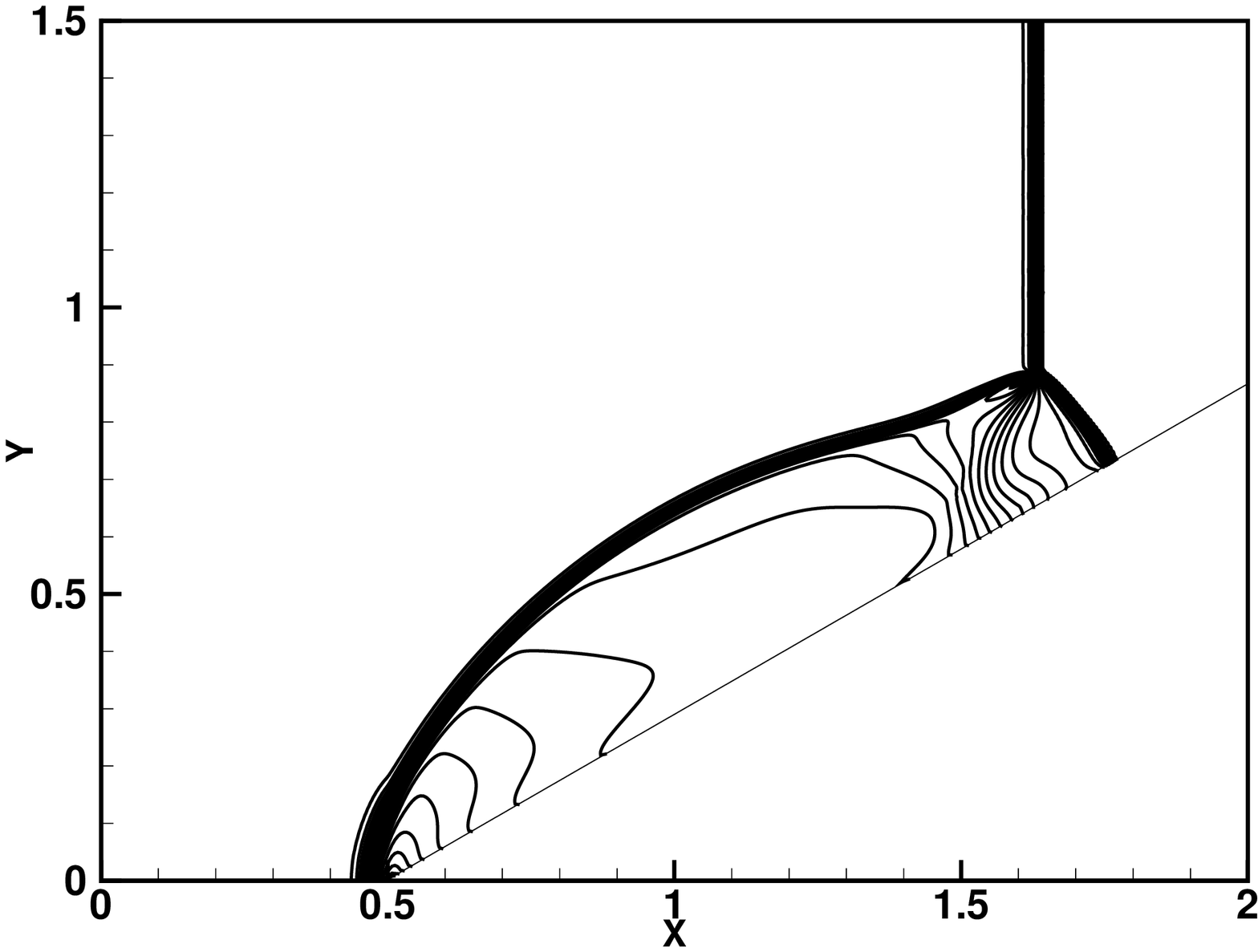} & \includegraphics[width=0.3\textwidth]{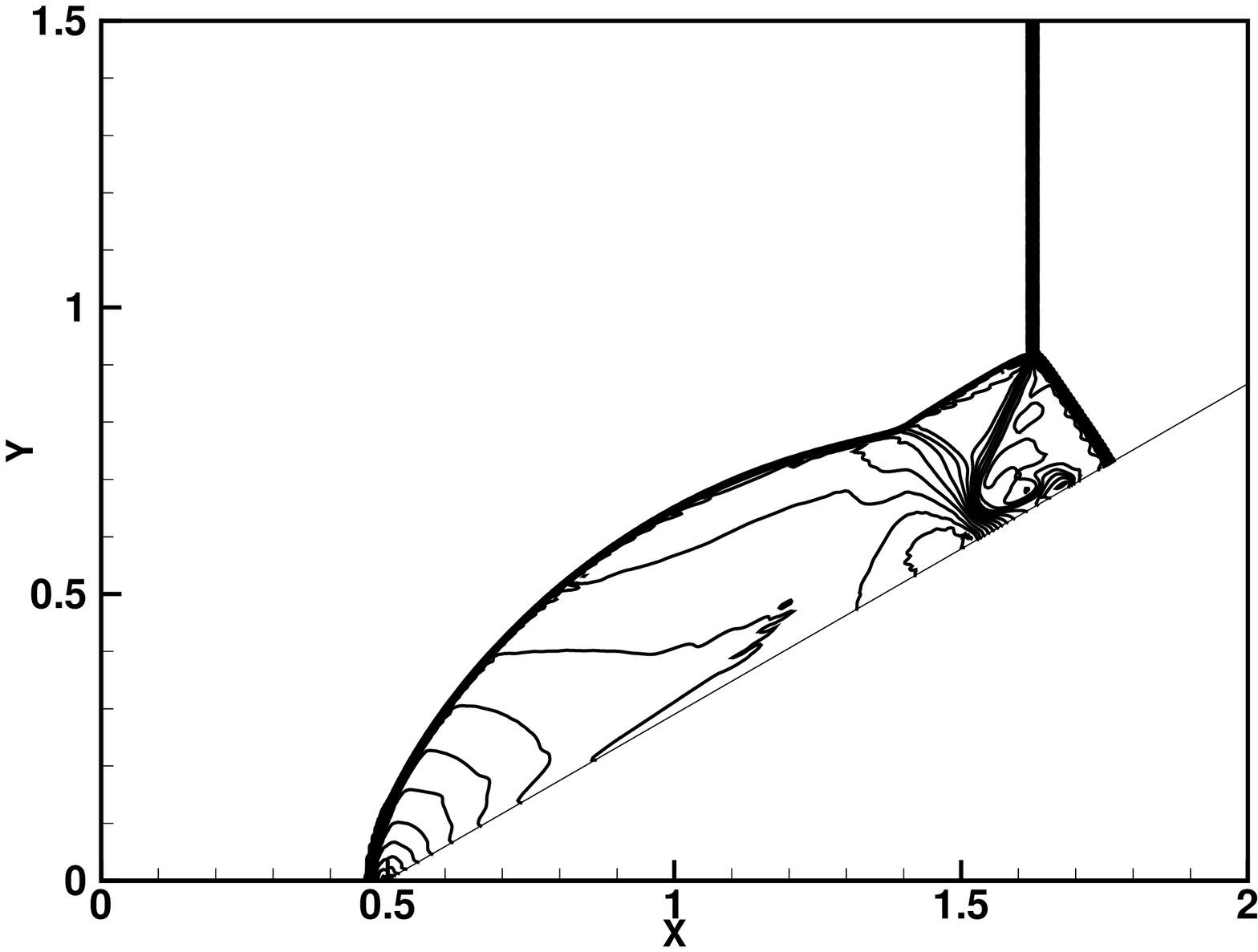}\\
\includegraphics[width=0.3\textwidth]{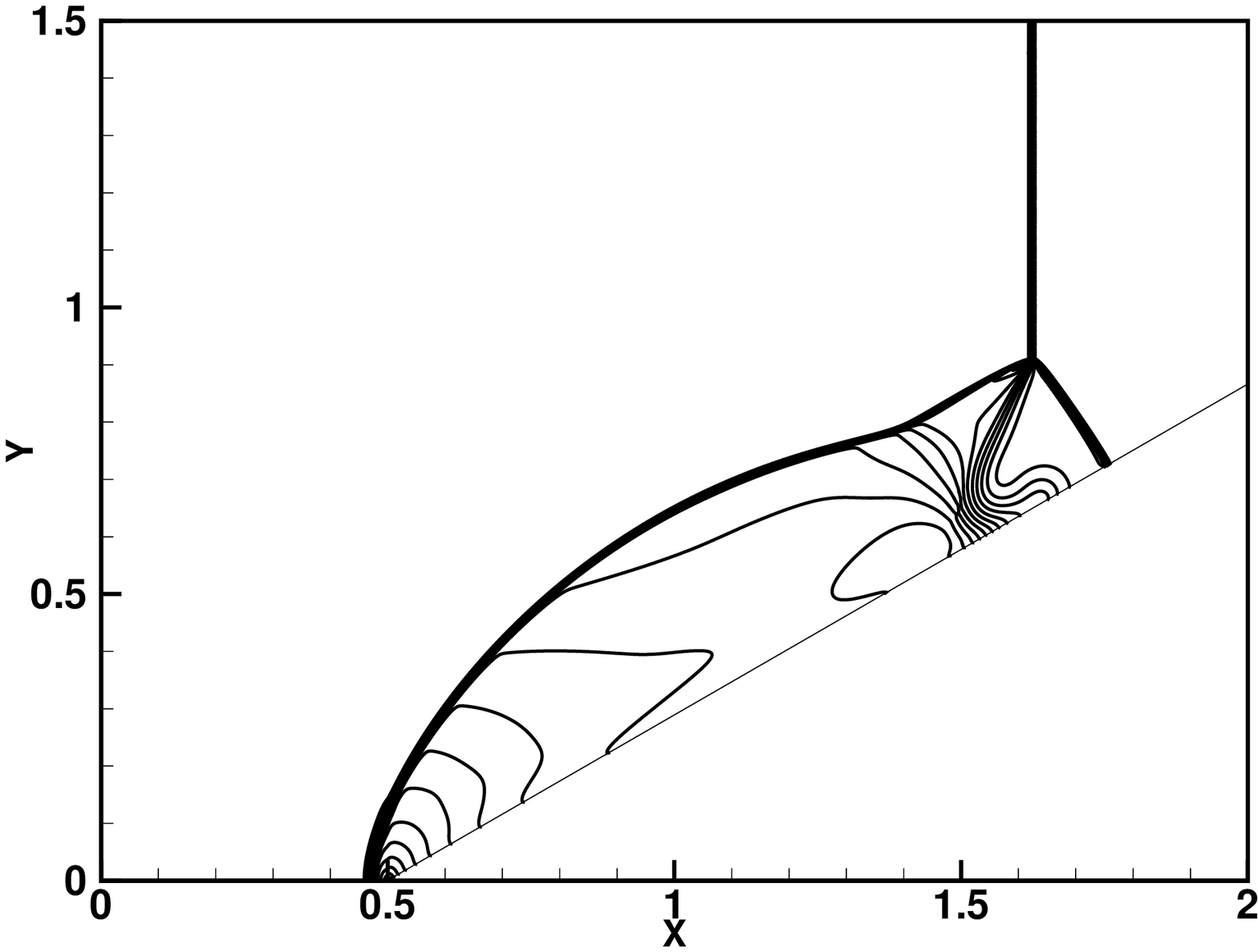} & \includegraphics[width=0.3\textwidth]{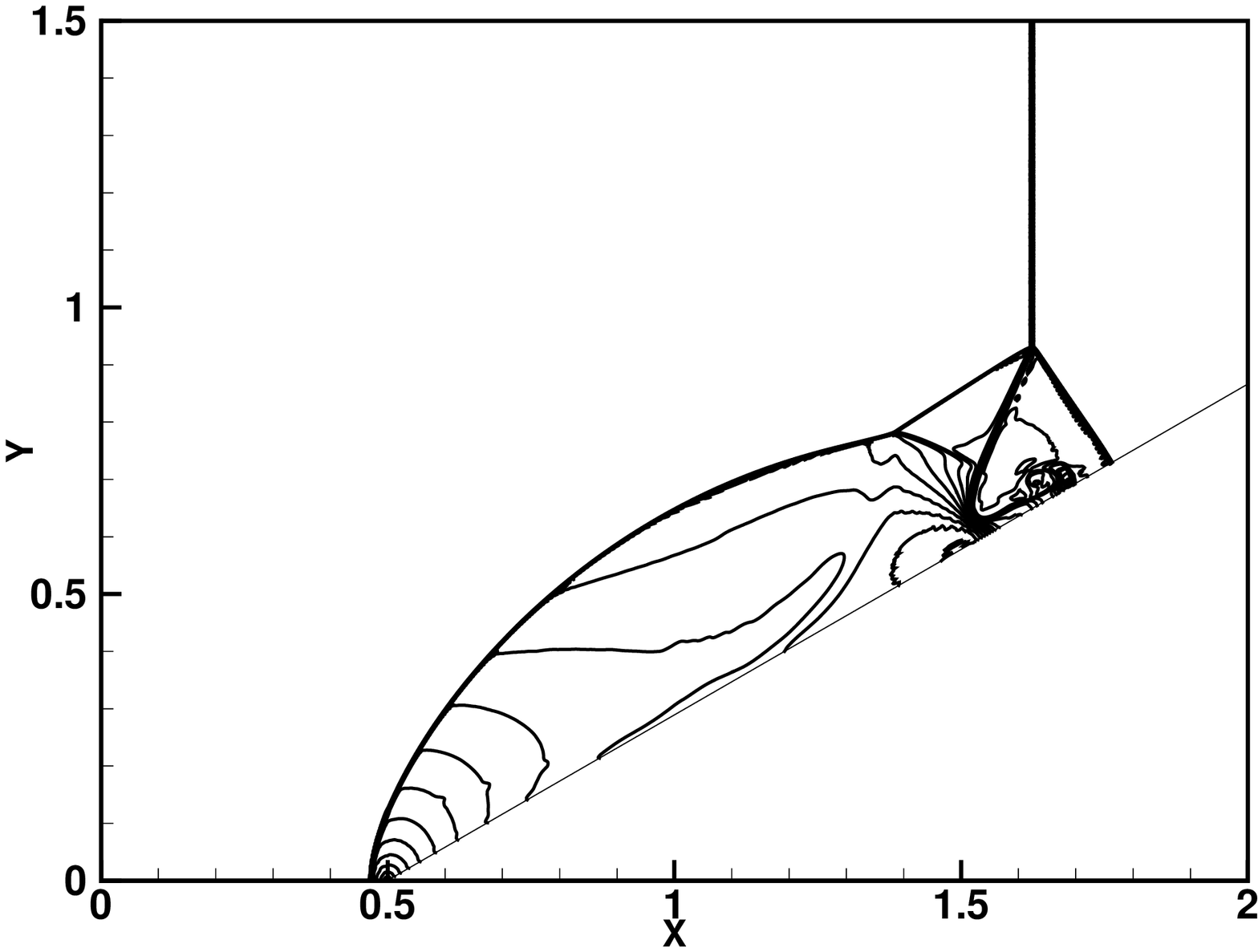} 
\end{tabular}
\caption{Double Mach reflection for Mach 5.5 shock across a $30^\circ$ wedge, $t=0.25$, density contours (1.5:0.5:19); Top) I order and II order accurate results on $400 \times 400$ grid, Bottom) I order and II order accurate results on $1200 \times 1200$ grid}
\label{fig:23}
\end{figure}
This test case comprises of a planar Mach 5.5 shock across a $30^\circ$ wedge \cite{quirk1997contribution}. The domain is $[0,2] \times [0,1.5]$ with a $30^\circ$ wedge at the bottom starting at $x= 0.5$. The initial conditions consist of a planar Mach 5.5 shock at $x= 0.25$, with stationary medium to its right. Inviscid wall boundary conditions are applied at the top and bottom boundaries, supersonic inflow conditions are used at the left boundary and constant extrapolation is done at the right boundary. When the planar Mach 5.5 shock collides with the $30^\circ$ ramp, it reflects over the surface as Mach reflection. The wave configuration consists of four discontinuities: the initial shock, the reflected shock, one Mach stem and one slip stream, all of which meet at a single triple point. Figure \ref{fig:23} shows the numerical solution of this unsteady problem at $t=0.25$. Some low diffusion schemes produce an unphysical kinked Mach stem.  No such kinked Mach stem is seen in the above figure.    

\subsubsection{Shock diffraction}
\begin{figure}[h!] 
\centering
\begin{tabular}{cc}
\includegraphics[width=0.3\textwidth]{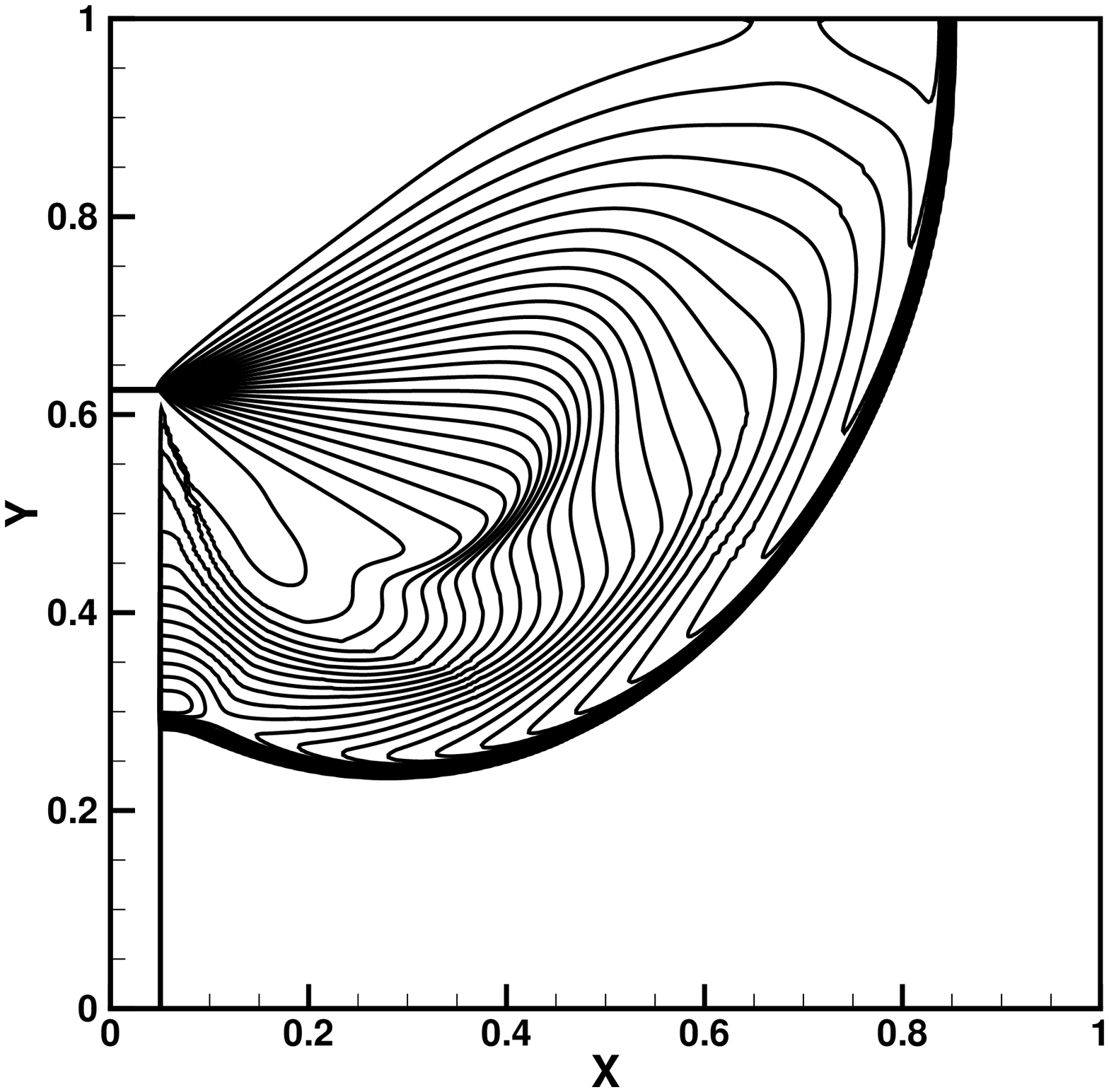} & \includegraphics[width=0.3\textwidth]{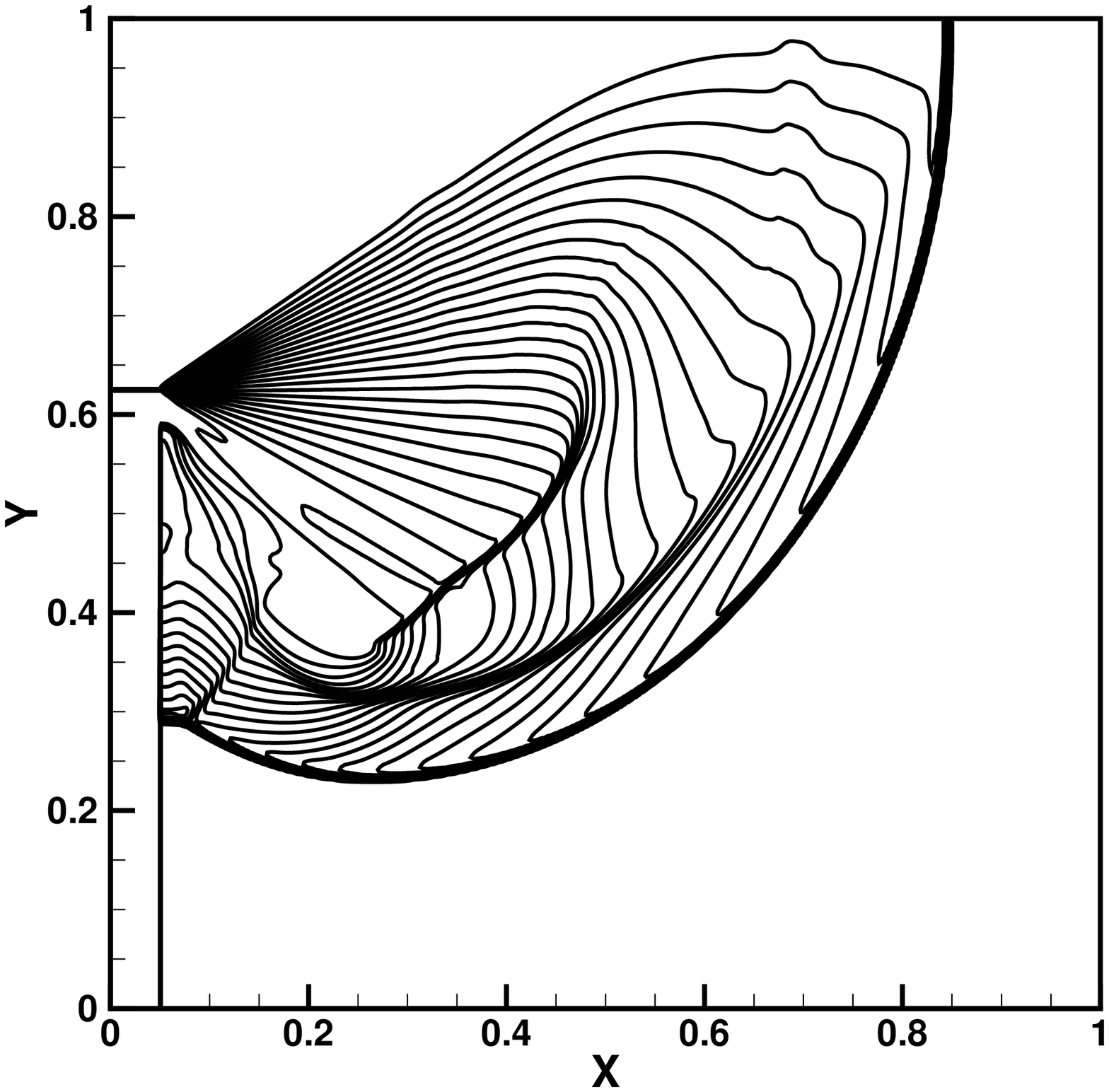}\\
\includegraphics[width=0.3\textwidth]{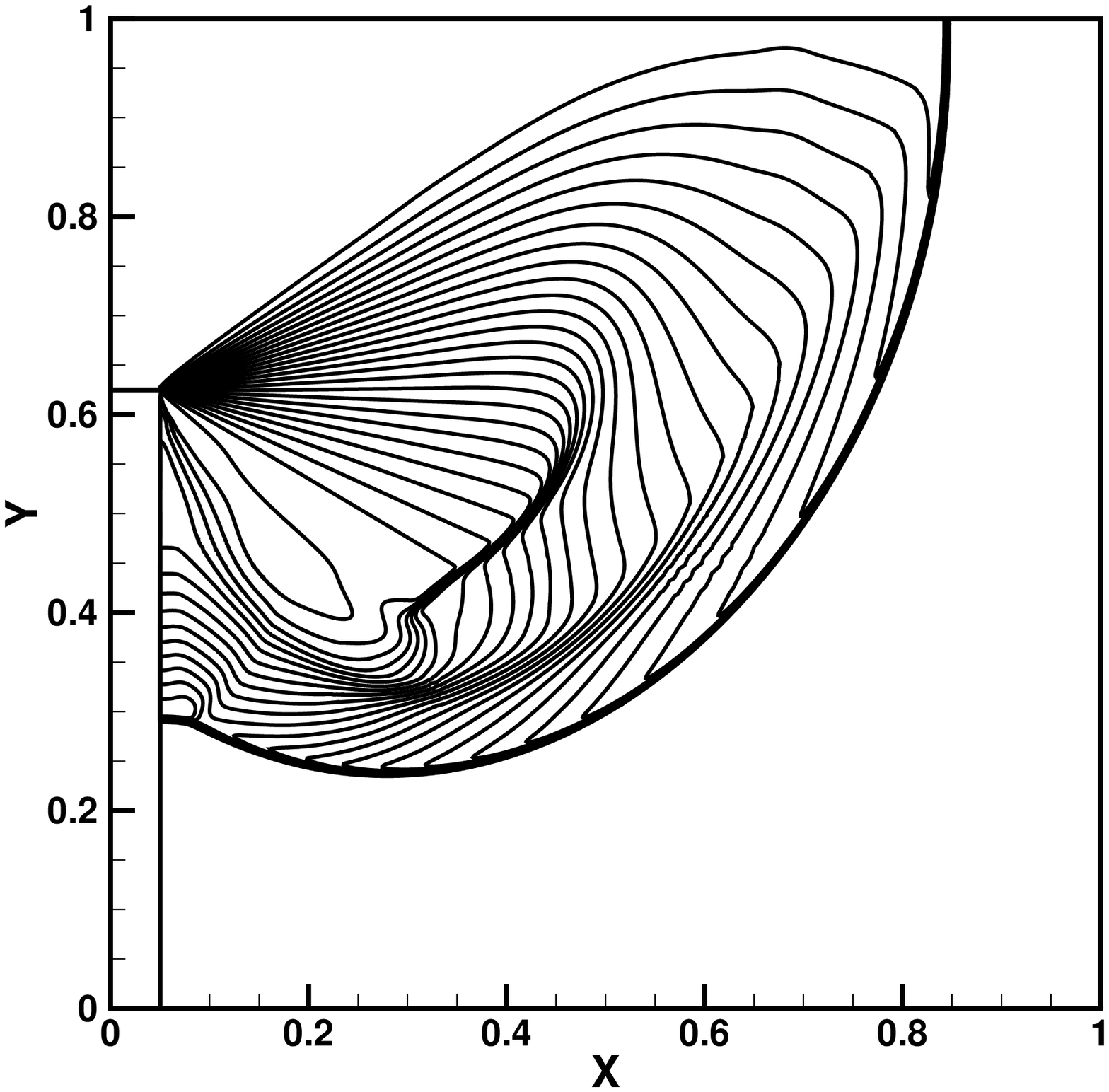} & \includegraphics[width=0.3\textwidth]{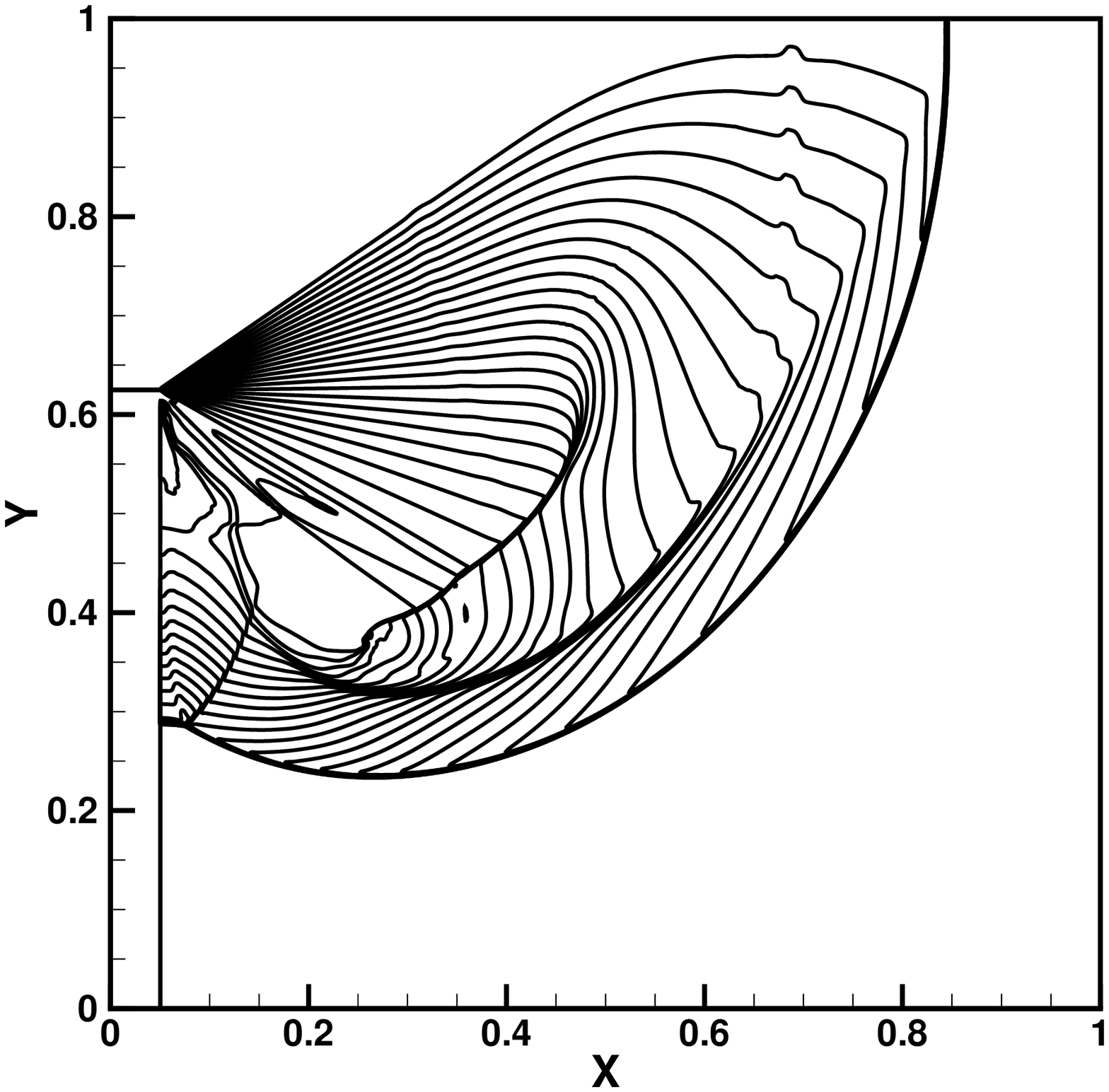} 
\end{tabular}
\caption{Shock diffracting around a $90^\circ$ corner, $t=0.1561$, density contours (0.5:0.25:6.75); Top) I order and II order accurate results on $400 \times 400$ grid, Bottom) I order and II order accurate results on $1200 \times 1200$ grid}
\label{fig:24}
\end{figure}
In this test case, a planar Mach 5.09 shock diffracts around a $90^\circ$ corner \cite{quirk1997contribution}. The computational domain is $[0,1] \times [0,1]$, with a corner at the bottom left end of width 0.05 and height 0.625 units. The initial conditions consist of a planar Mach 5.09 shock at $x= 0.05$ moving towards a stationary medium to the right. The boundary conditions used are: flow tangency conditions at the top and for the corner, constant extrapolation at the right and bottom boundaries, and supersonic inflow conditions at the left boundary. Figure \ref{fig:24} shows the numerical solution to this unsteady problem at $t=0.1561$. The solution has a complex wave structure comprising of the incident planar shock, the diffracted shock, a strong expansion fan and a slip stream. Without an entropy fix, several low-diffusive schemes give rise to unphysical expansion shocks. Our results are free of expansion shocks and all flow features are captured well.  

\subsubsection{NACA0012 airfoil test cases} 
\begin{figure}[h!] 
\centering
\begin{tabular}{cc}
\includegraphics[width=0.25\textwidth]{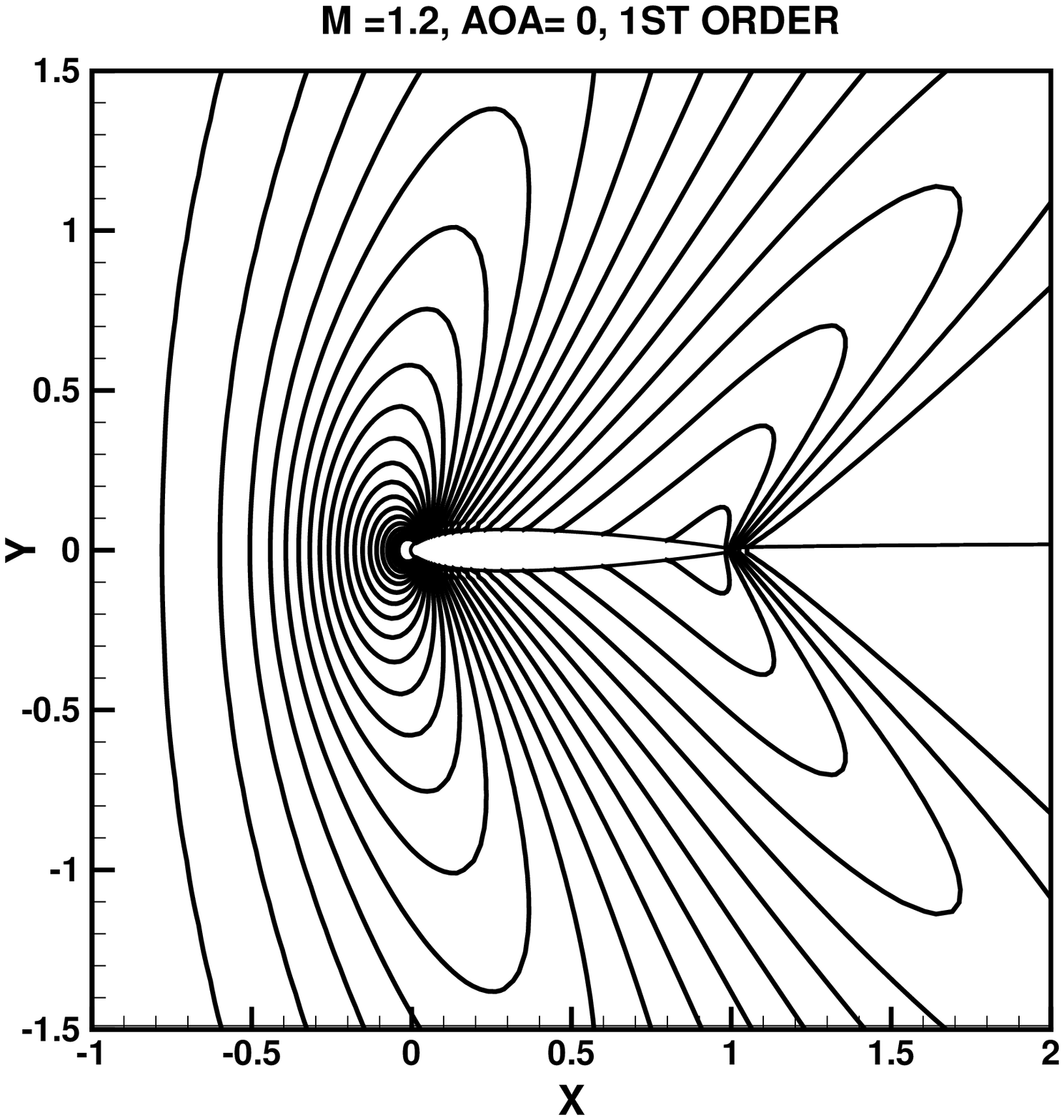} & \includegraphics[width=0.25\textwidth]{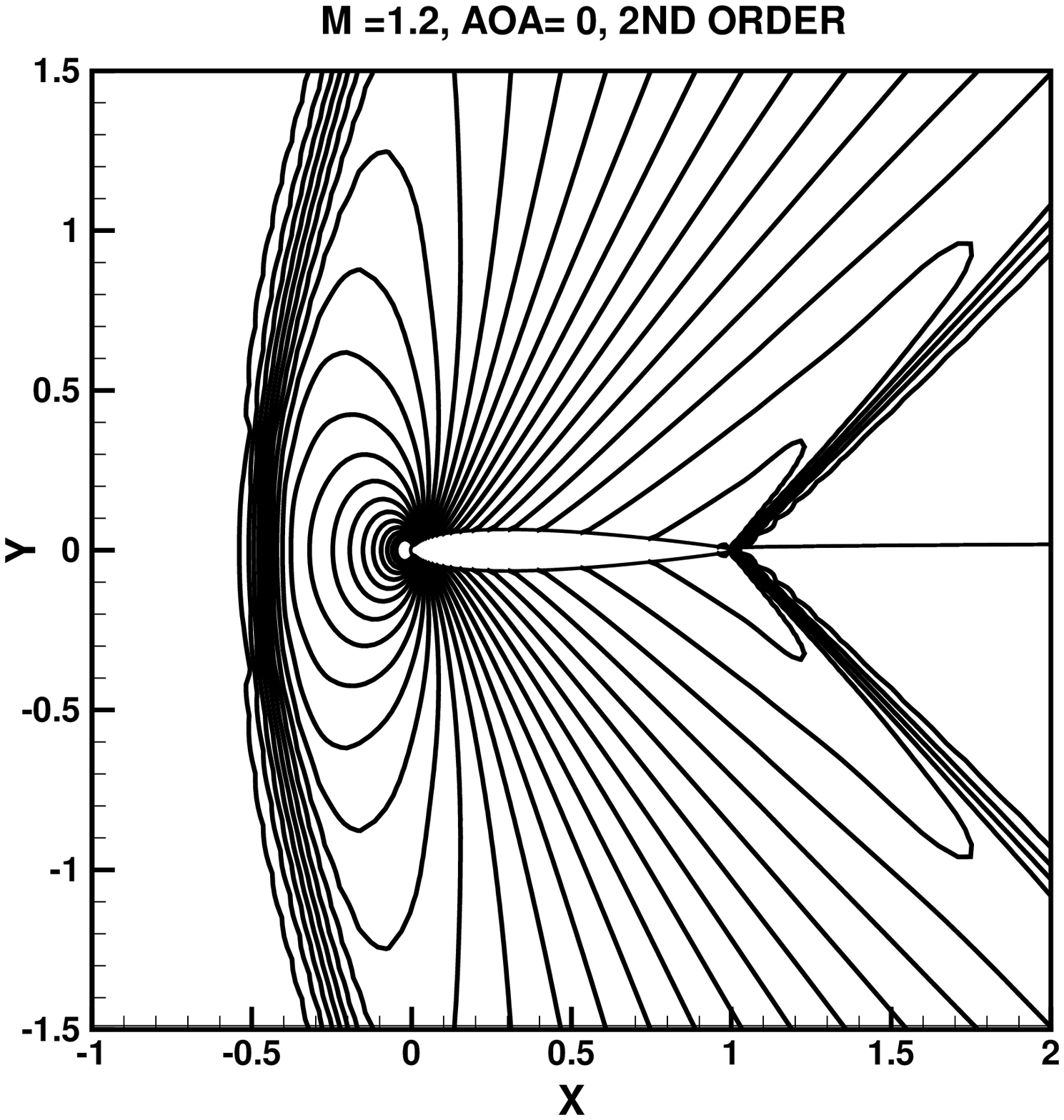}\\
\includegraphics[width=0.25\textwidth]{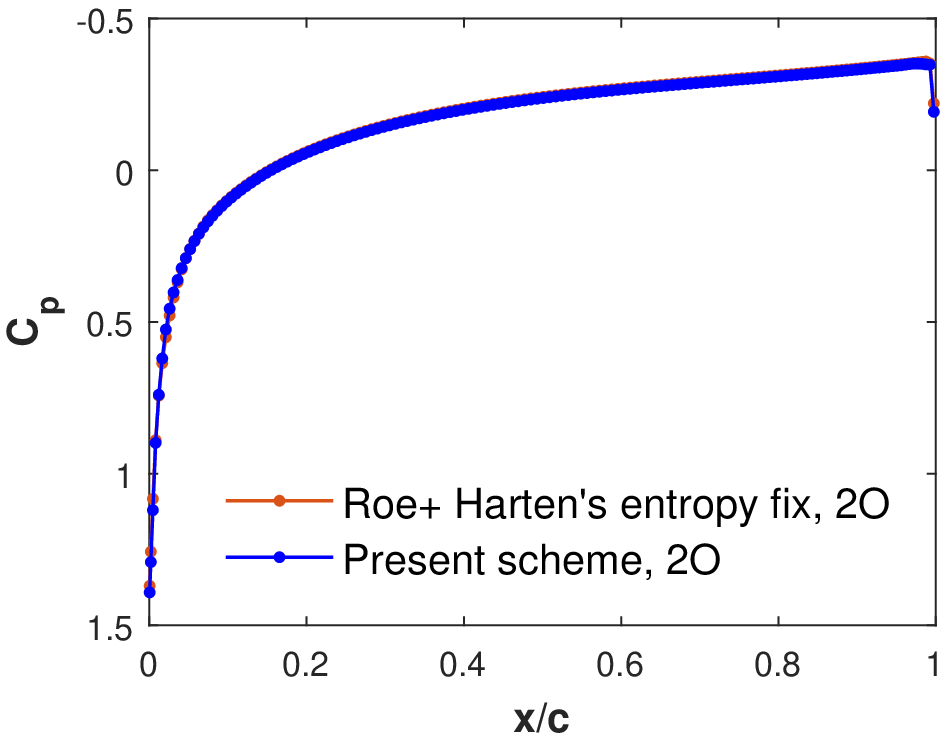} & \includegraphics[width=0.25\textwidth]{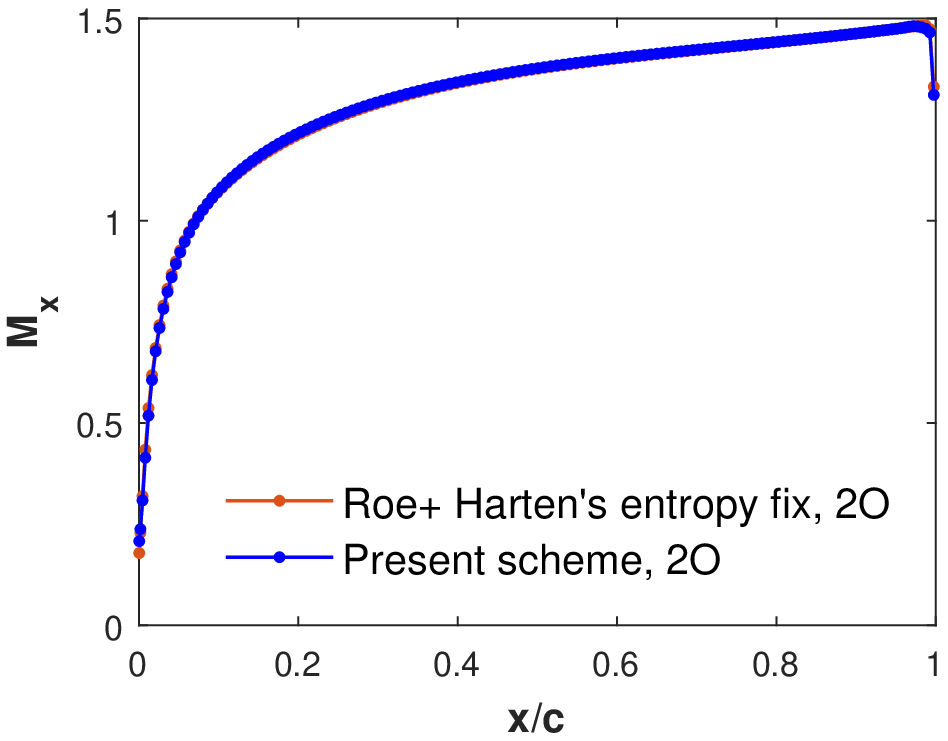} 
\end{tabular}
\caption{NACA0012, $M_{\infty}=$ 1.2, A.O.A$= 0^\circ$: Top) I and II order accurate results on $298 \times 98$ grid; pressure contours (0.4:0.05:2.0), Bottom) plots of $c_{p}$ and $M_{x}$ vs $x/c$ along airfoil surface}
\label{fig:25}
\end{figure}

\begin{figure}[h!] 
\centering
\begin{tabular}{cc}
\includegraphics[width=0.25\textwidth]{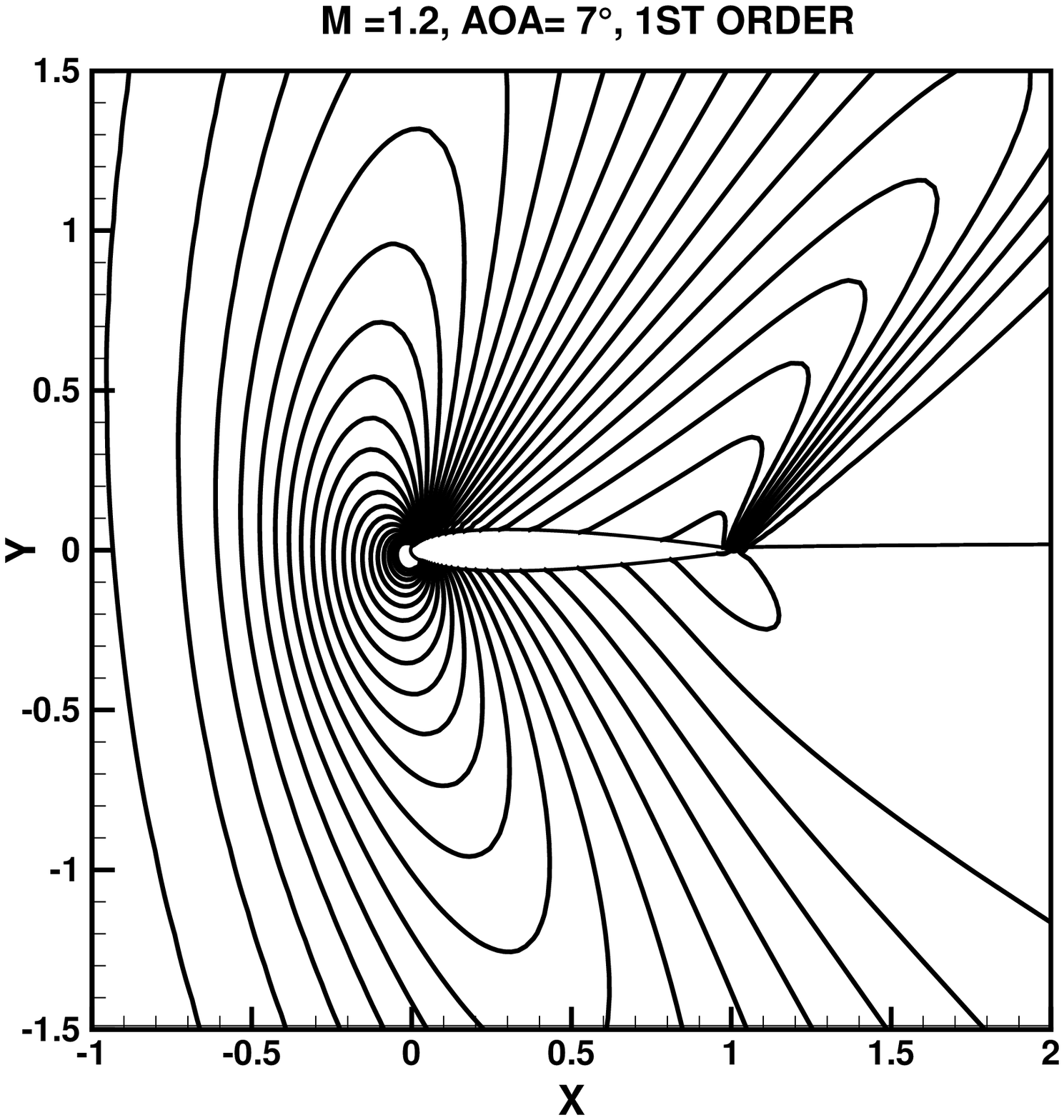} & \includegraphics[width=0.25\textwidth]{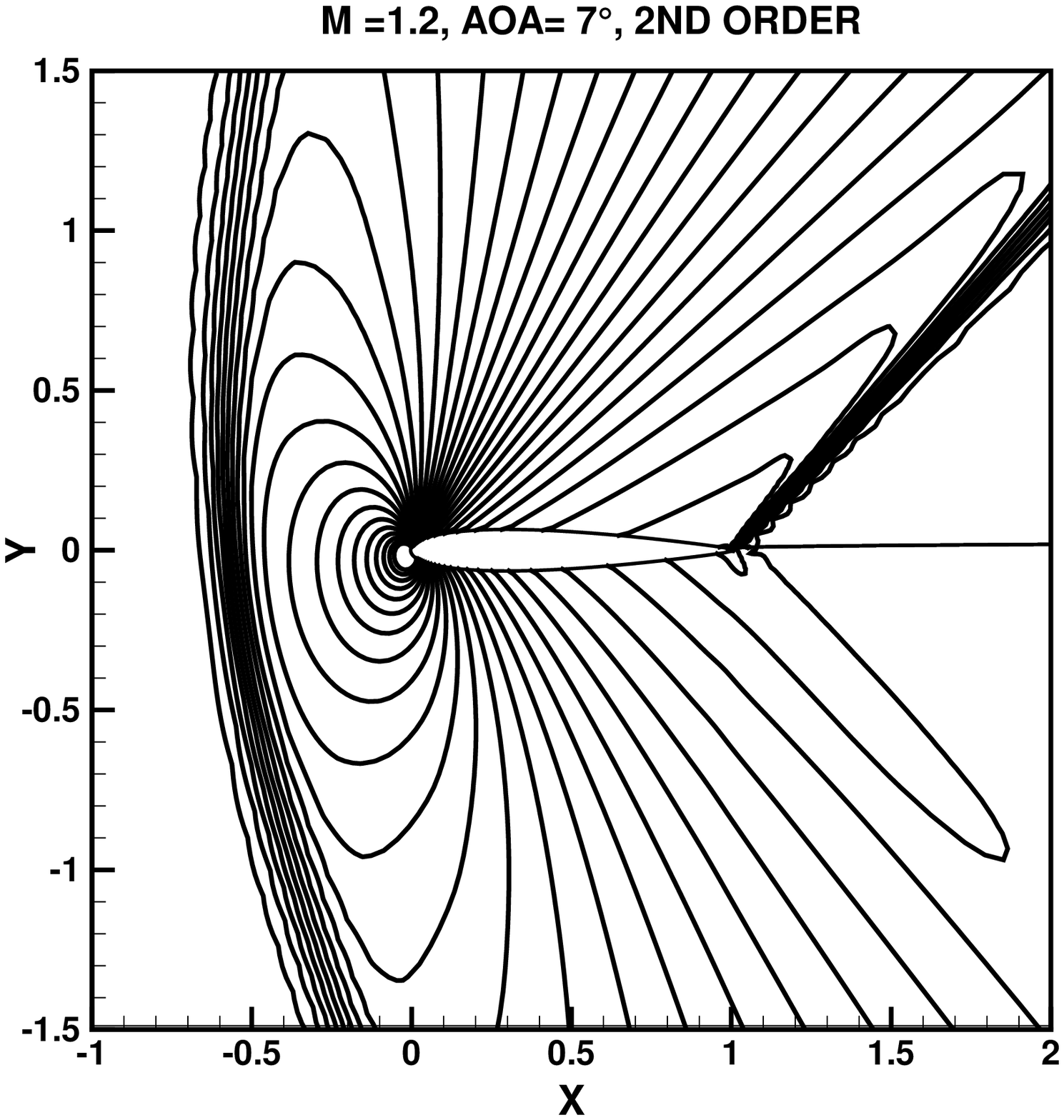}\\
\includegraphics[width=0.25\textwidth]{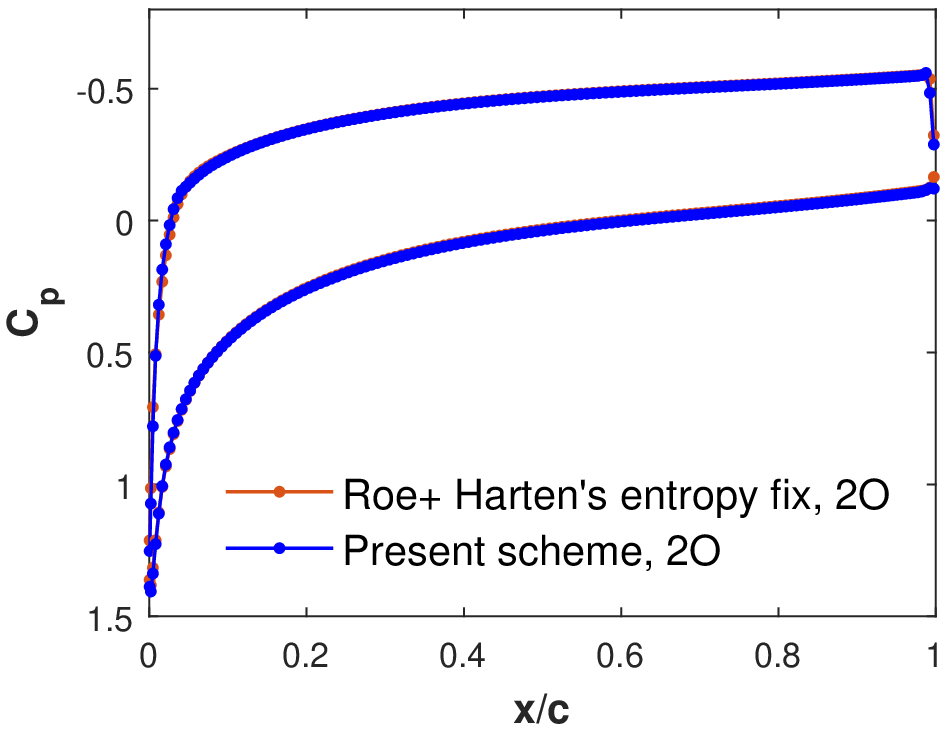} & \includegraphics[width=0.25\textwidth]{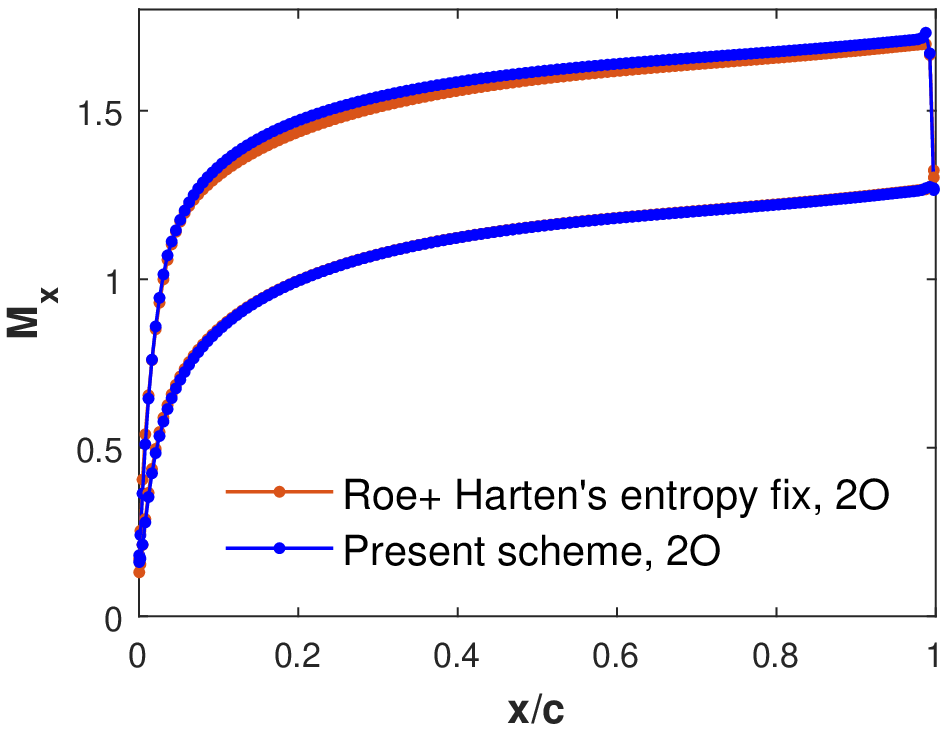} 
\end{tabular}
\caption{NACA0012, $M_{\infty}=$ 1.2, A.O.A$= 7^\circ$: Top) I and II order accurate pressure contours (0.4:0.05:2.0) on $298 \times 98$ grid, Bottom) II order accurate plots of $c_{p}$ and $M_{x}$ vs x/c along airfoil surface}
\label{fig:26}
\end{figure}

\begin{figure}[h!] 
\centering
\begin{tabular}{cc}
\includegraphics[width=0.25\textwidth]{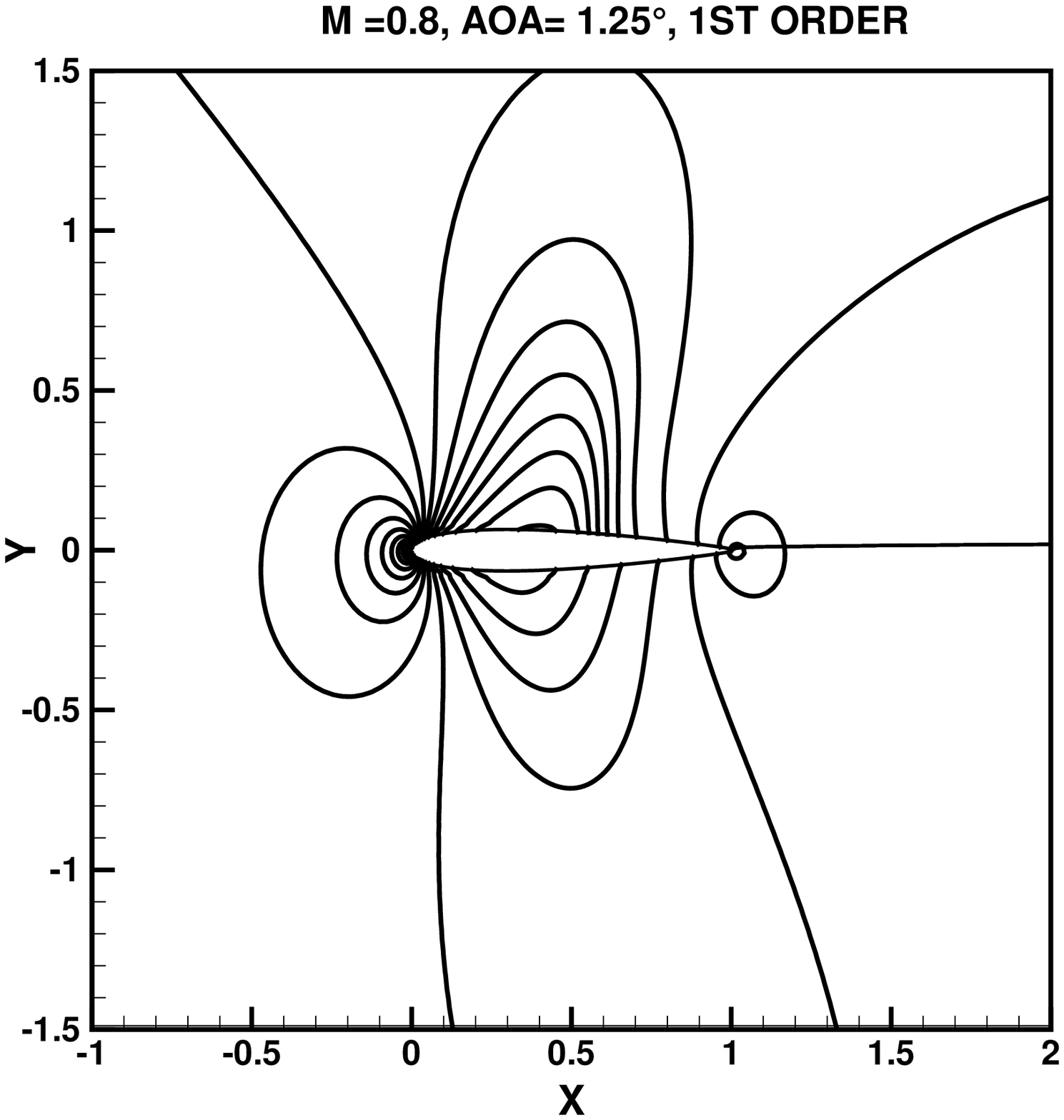} & \includegraphics[width=0.25\textwidth]{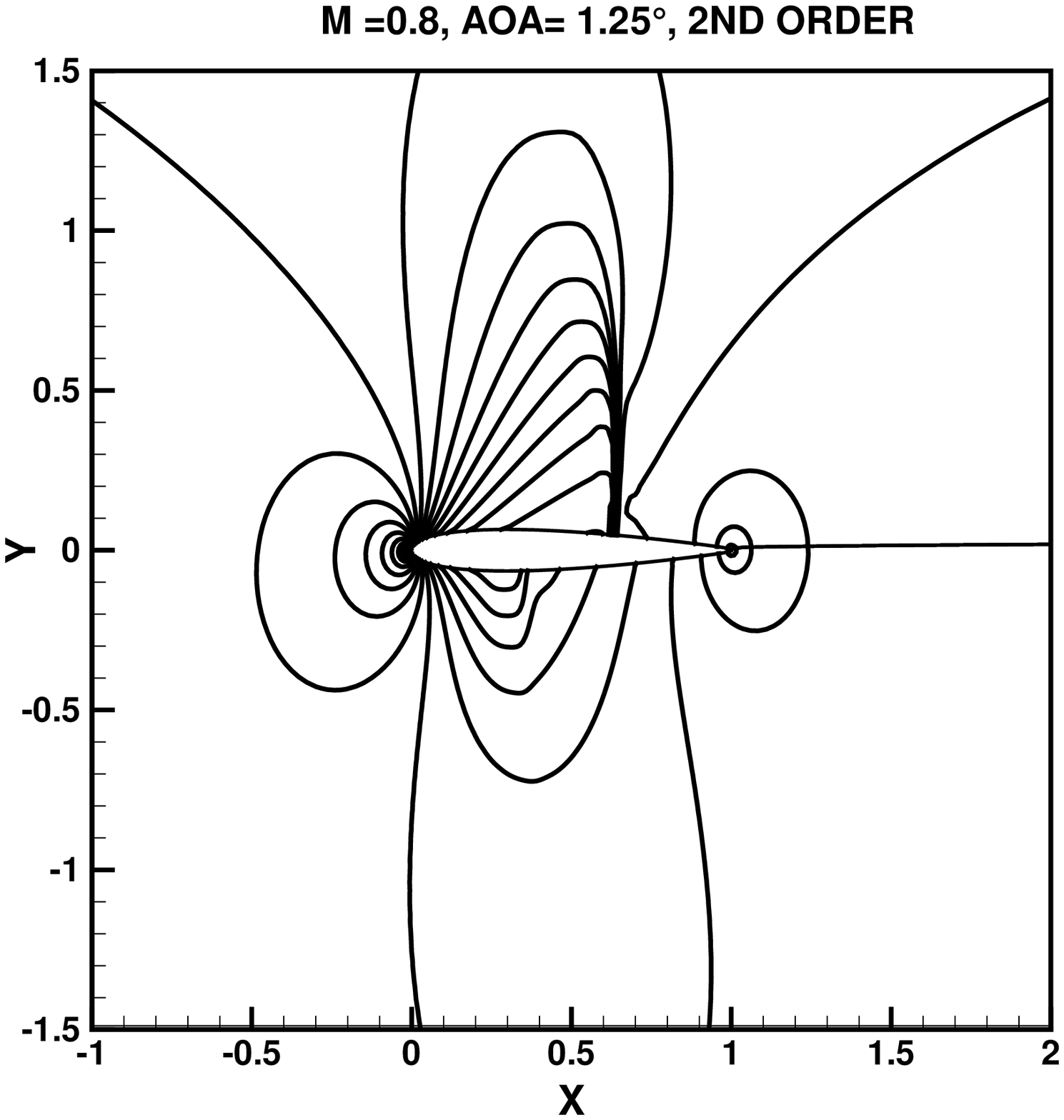}\\
\includegraphics[width=0.25\textwidth]{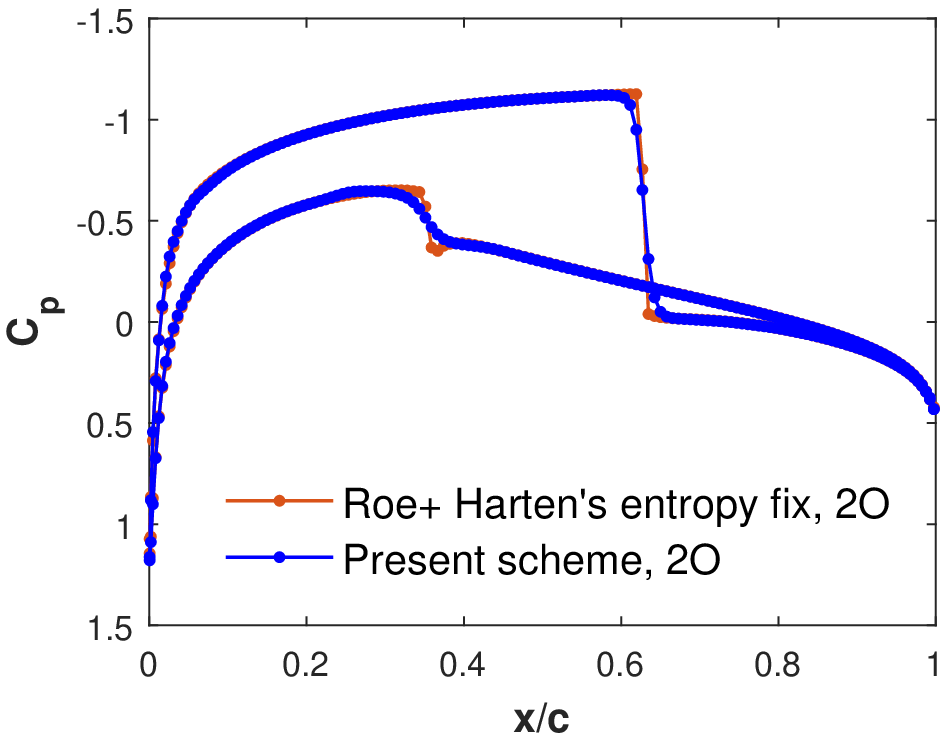} & \includegraphics[width=0.25\textwidth]{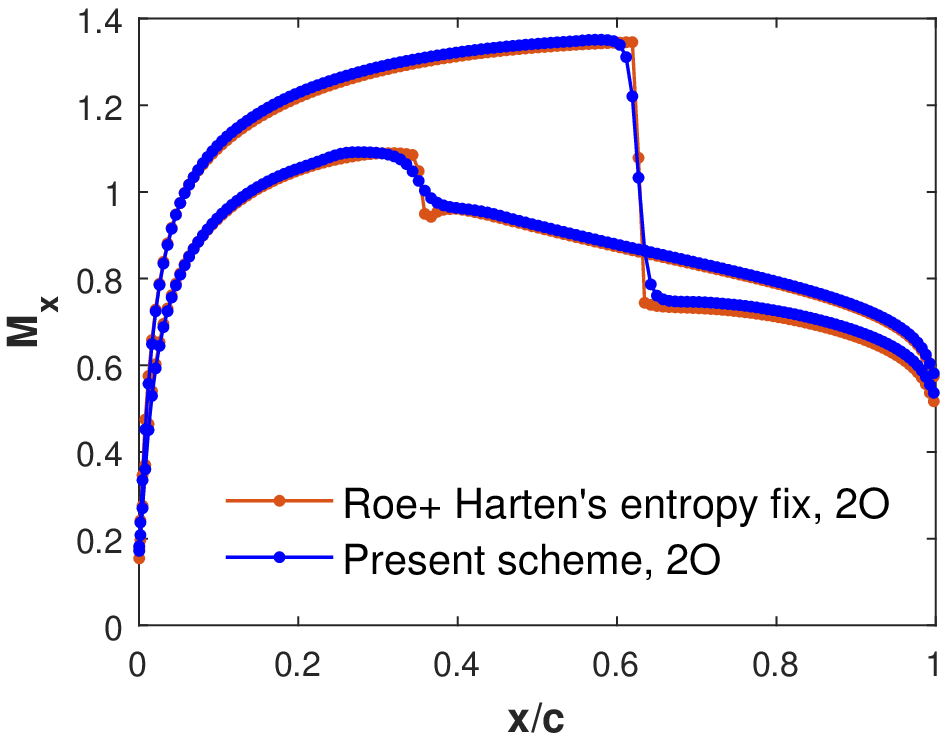} 
\end{tabular}
\caption{NACA0012, $M_{\infty}=$ 0.8, A.O.A$= 1.25^\circ$: Top) I and II order accurate pressure contours (0.4:0.05:2.0) on $298 \times 98$ grid, Bottom) II order accurate plots of $c_{p}$ and $M_{x}$ vs x/c along airfoil surface}
\label{fig:27}
\end{figure}

\begin{figure}[h!] 
\centering
\begin{tabular}{cc}
\includegraphics[width=0.25\textwidth]{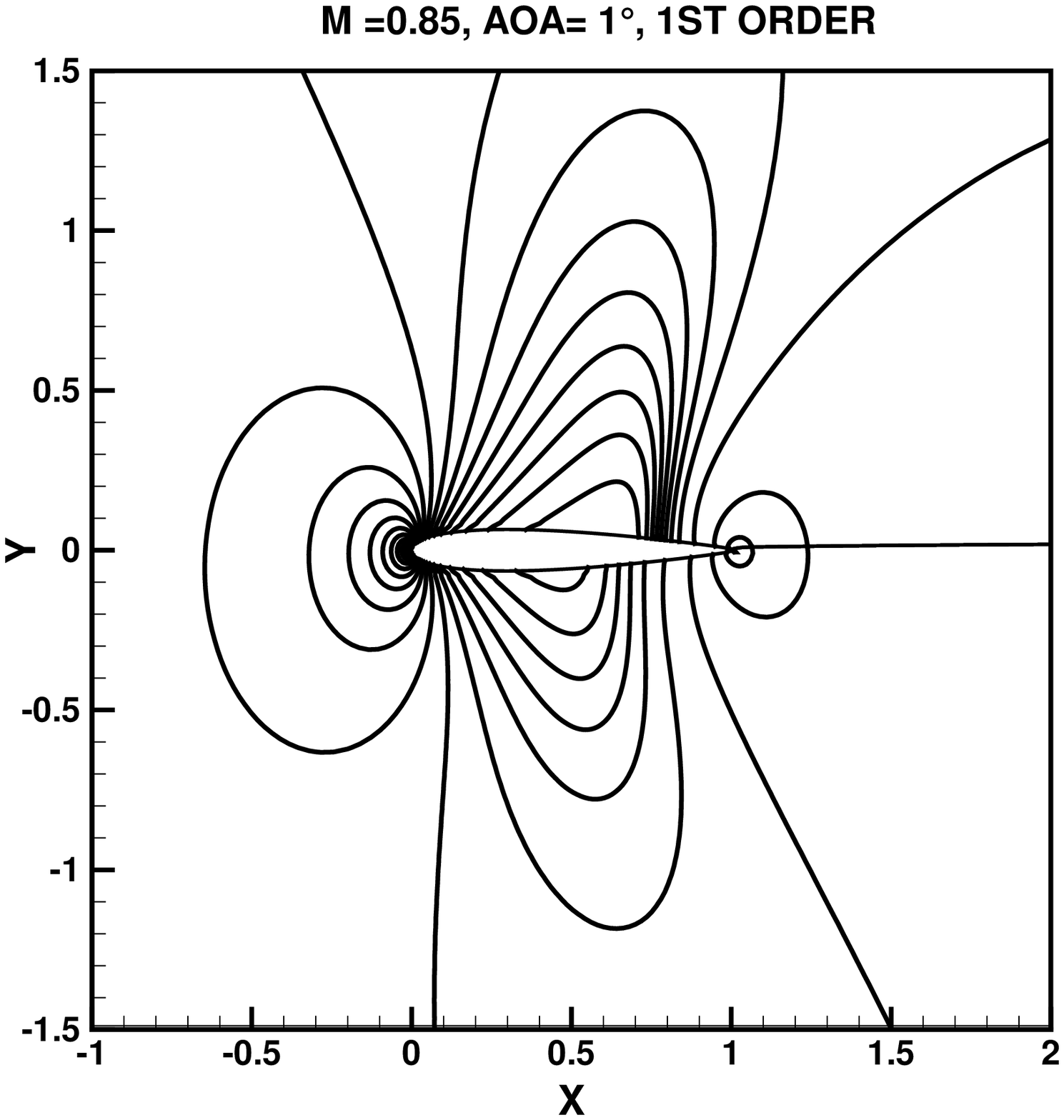} & \includegraphics[width=0.25\textwidth]{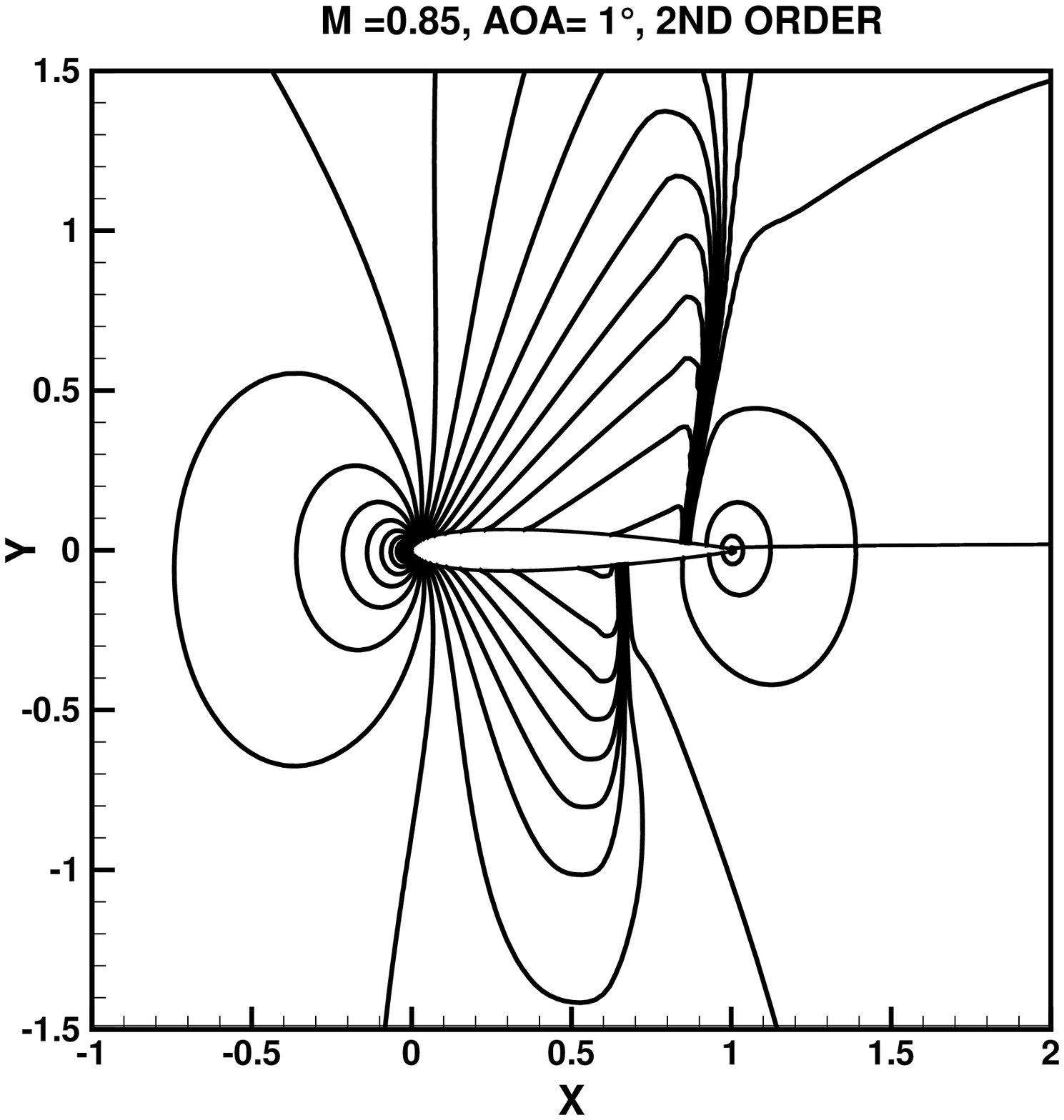}\\
\includegraphics[width=0.25\textwidth]{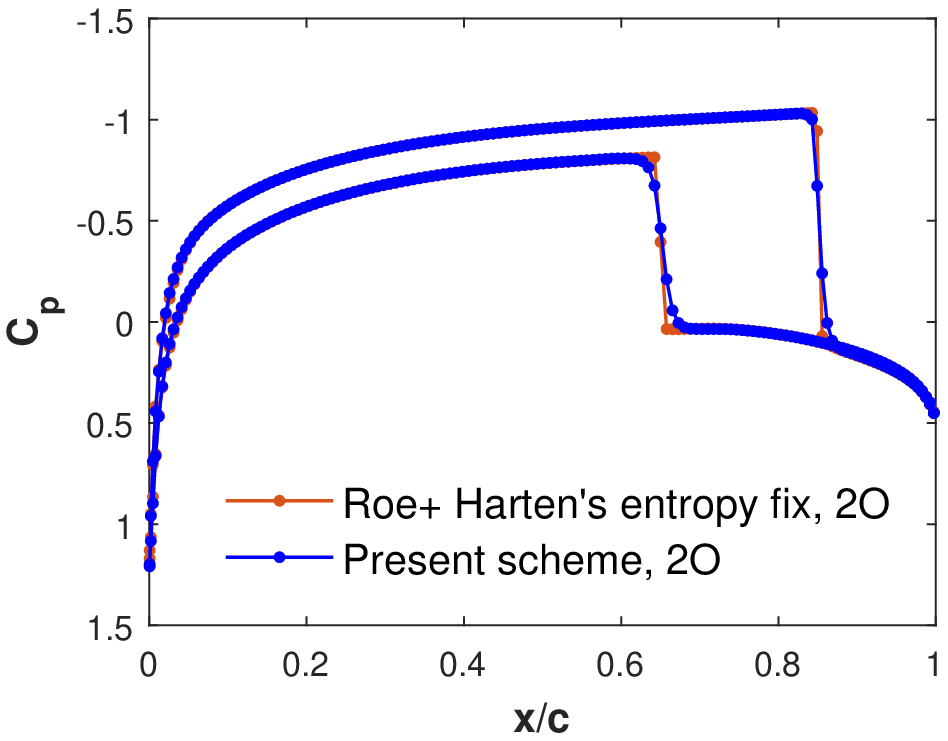} & \includegraphics[width=0.25\textwidth]{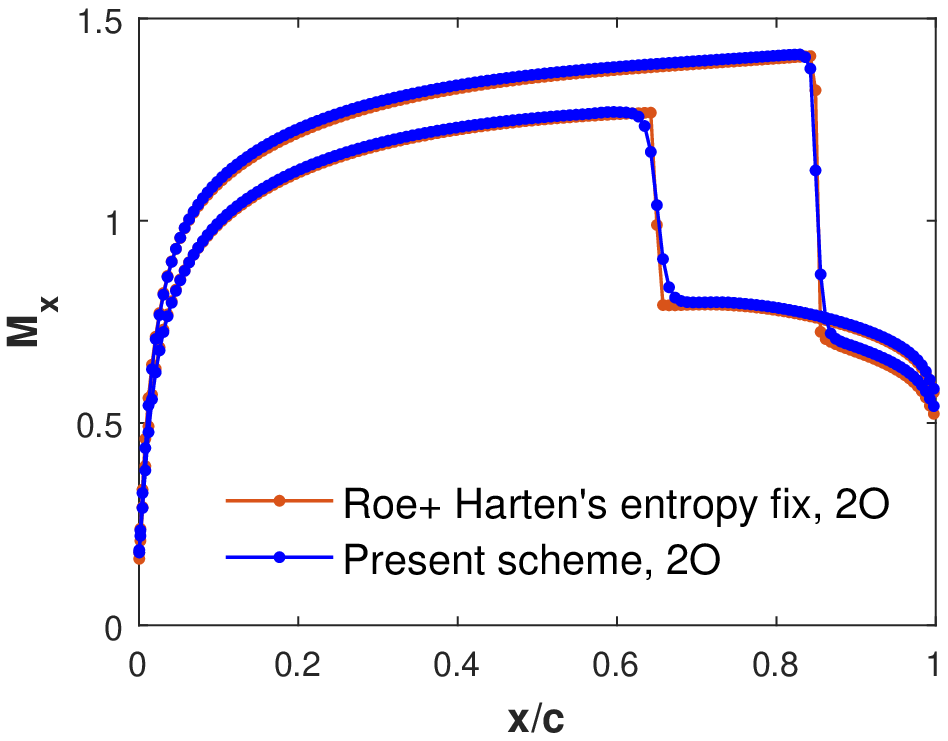} 
\end{tabular}
\caption{NACA0012, $M_{\infty}=$ 0.85, A.O.A$= 1^\circ$: Top) I and II order accurate pressure contours (0.4:0.05:2.0) on $298 \times 98$ grid, Bottom) II order accurate plots of $c_{p}$ and $M_{x}$ vs x/c along airfoil surface}
\label{fig:28}
\end{figure}

\begin{figure}[h!] 
\centering
\begin{tabular}{cc}
\includegraphics[width=0.25\textwidth]{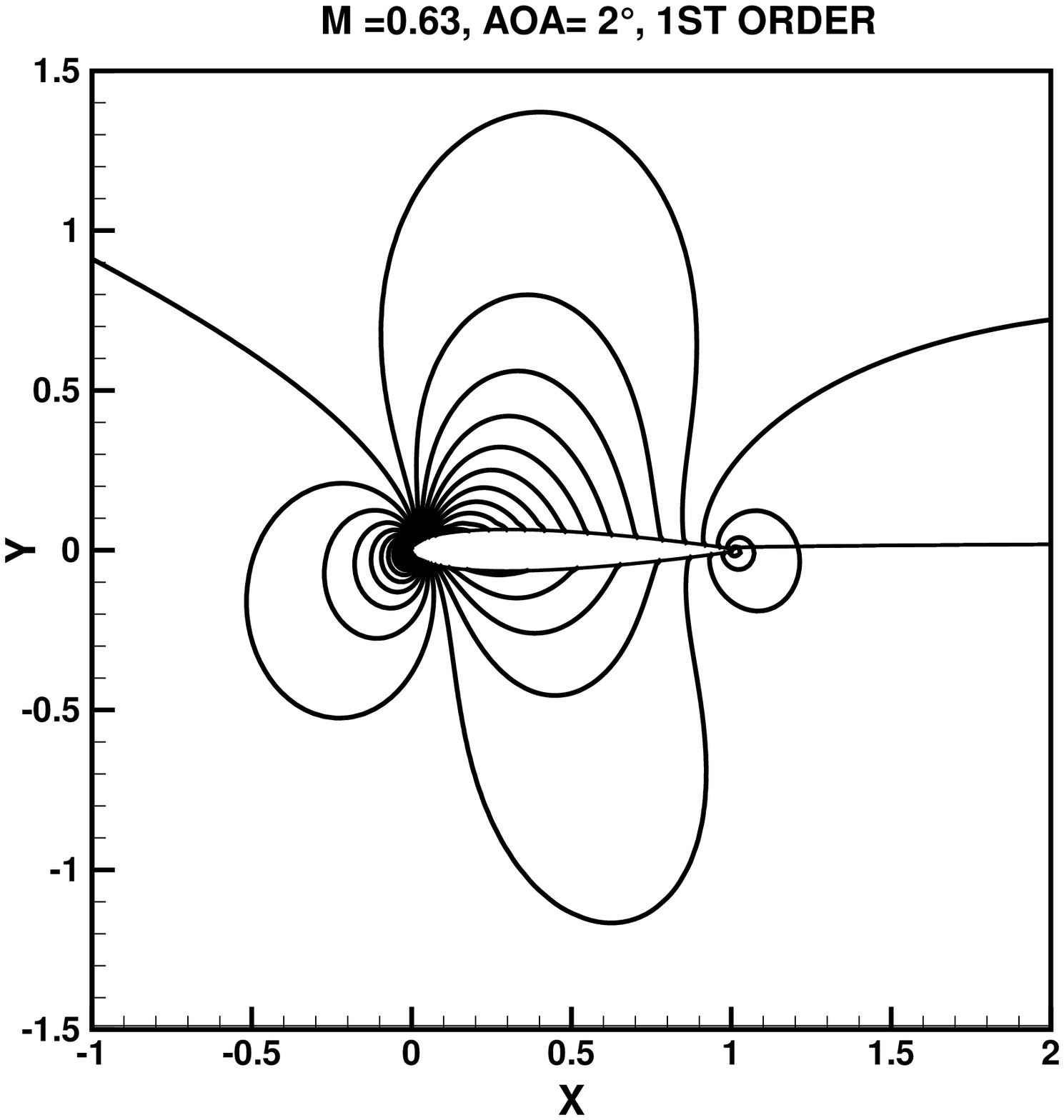} & \includegraphics[width=0.25\textwidth]{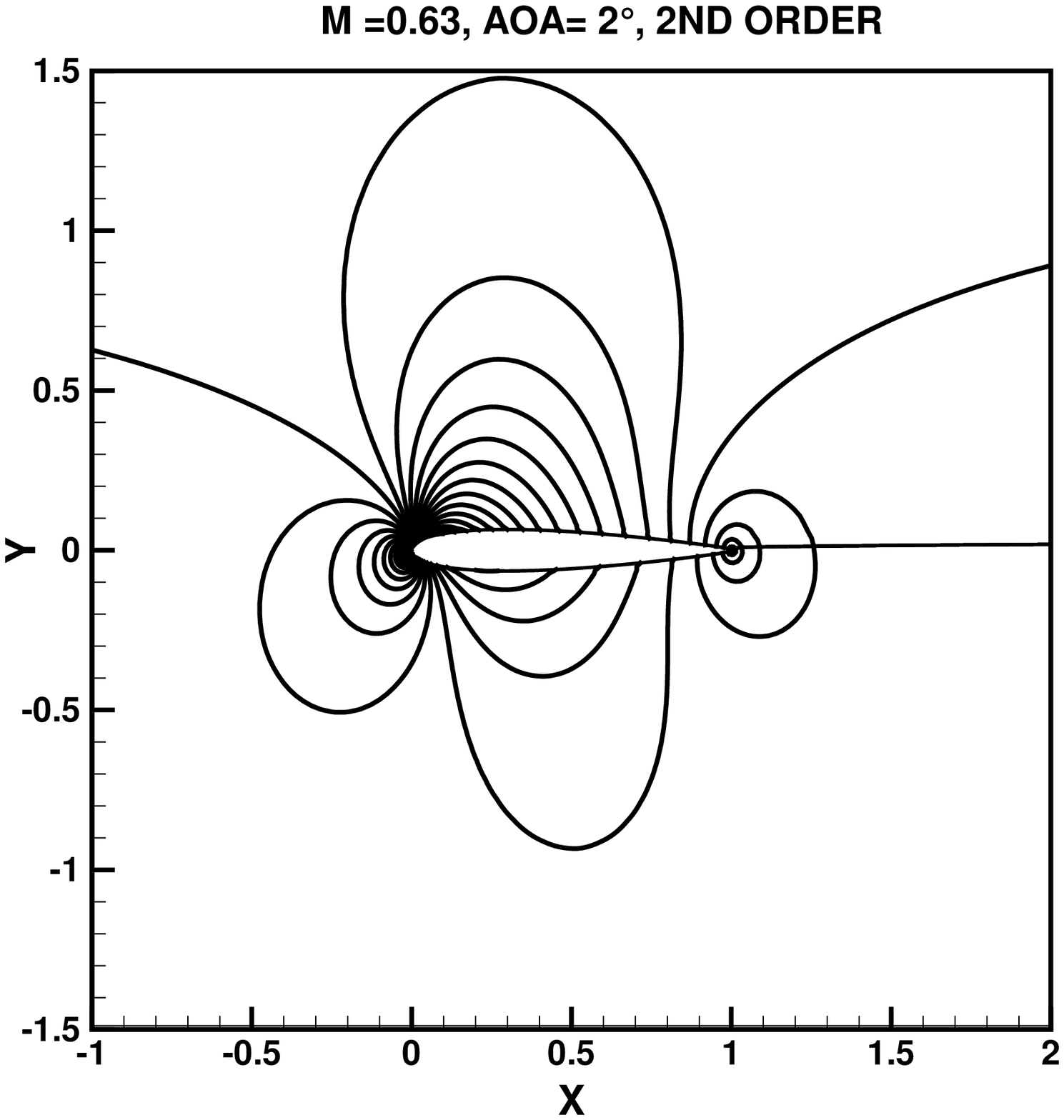}\\
\includegraphics[width=0.25\textwidth]{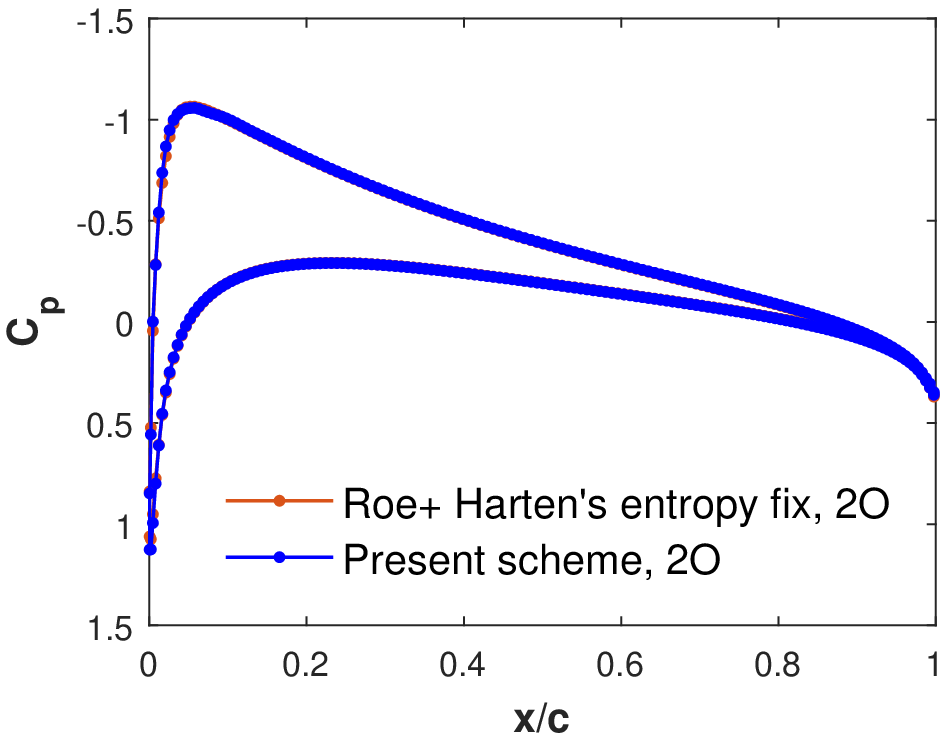} & \includegraphics[width=0.25\textwidth]{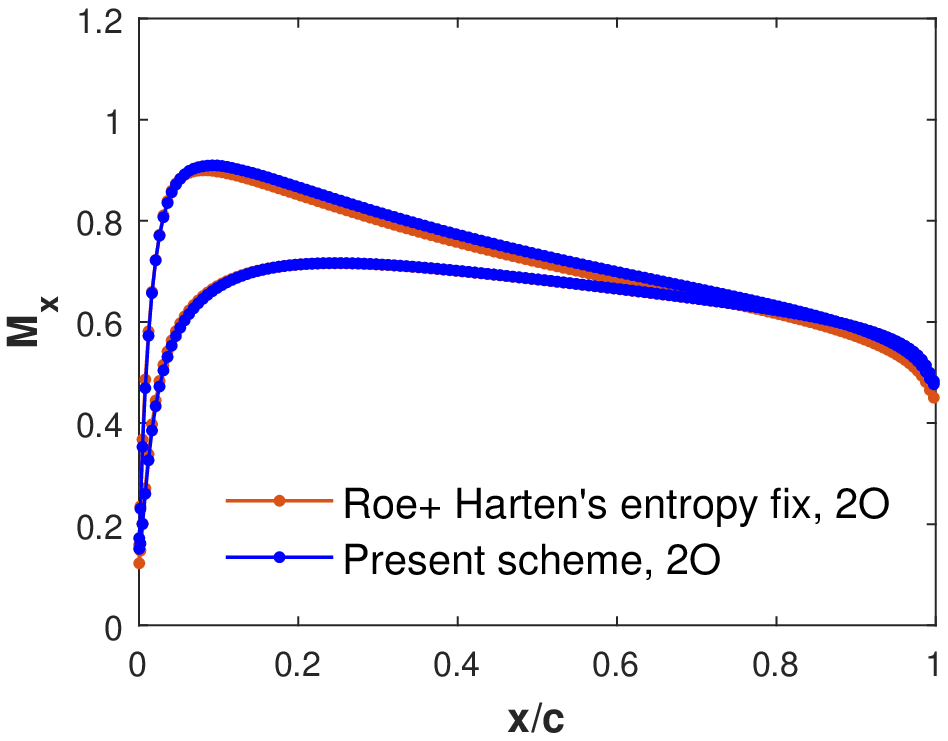} 
\end{tabular}
\caption{NACA0012, $M_{\infty}=$ 0.63, A.O.A$= 2^\circ$: Top) I and II order accurate pressure contours (0.7:0.02:1.4) on $298 \times 98$ grid, Bottom) II order accurate plots of $c_{p}$ and $M_{x}$ vs x/c along airfoil surface}
\label{fig:28_a}
\end{figure}

Some benchmark test cases are performed for the symmetric NACA0012 airfoil \cite{jones1985reference}, \cite{dervieux1989numerical}. An O-type structured mesh of dimensions of 25 times the chord length is used around the airfoil.   The \textit{discrete kinetic farfield} boundary conditions derived in Section 9 are applied at the outer boundary. Whereas \textit{discrete kinetic flow tangency} conditions are applied at the airfoil surface. Periodic conditions are applied along $\eta$ direction where the first and last grid meet. Freestream initial conditions are used, and the steady state solution is sought. Numerical tests are done for the following supersonic, transonic and subsonic test cases:
\begin{enumerate}
	\item $M_{\infty}=$ 1.2, A.O.A. (Angle of attack) = $0^\circ$
	\item $M_{\infty}=$ 1.2, A.O.A. = $7^\circ$
	\item $M_{\infty}=$ 0.8, A.O.A. = $1.25^\circ$
	\item $M_{\infty}=$ 0.85, A.O.A. = $1^\circ$
	\item $M_{\infty}=$ 0.63, A.O.A. = $2^\circ$
\end{enumerate}
The results for the above test cases are shown in Figures \ref{fig:25}-\ref{fig:28_a}. 

The pressure contours around the airfoil for first and second order accuracy are shown. Further, the second order results for pressure coefficient $c_{p}\left(=\frac{p- p_{\infty}}{\frac{1}{2}\rho_{\infty}u_{\infty}^{2}}\right)$ and  $M_{x}\left(= \frac{u_{1}}{a}\right)$ along the top and bottom airfoil surfaces are plotted vs. $x/c$ for our scheme as well as Roe's scheme with Harten's entropy fix on the same grid.

\subsection{2D Viscous tests}

These test cases test the ability of the scheme to solve viscous flow equations and resolve viscous flow features. The viscous terms at the interface are computed using auxiliary volume method. For the inviscid fluxes, we do not use any entropy fix to provide additional numerical diffusion, since physical viscous terms are proved to be sufficient.

\subsubsection{ Shock boundary layer interaction} 
In this test case, an oblique shock wave with freestream Mach number $M_{\infty}~=~2.15$ and shock angle of $30.8^\circ$ strikes a flat plate at the bottom on which a laminar boundary layer is evolving \cite{degrez1987interaction}. The shock impinging on the boundary layer causes the flow to locally separate and then reattach to the surface. The reflected waves comprise of  compression waves converging into a shock, expansion fans, followed again by compression waves. For this test, the computational domain taken is $[-0.2, 1.8] \times [0, 1]$. Freestream conditions are applied at the left boundary for  $y \leq 0.765$, whereas post-shock conditions are applied for $y > 0.765$ at the left end as well as the top boundary. At the bottom, flow symmetry conditions are applied for $x \leq -0.2$, whereas no slip conditions are applied for $x > 0.2$. Supersonic outflow conditions are applied at the right boundary. The freestream Reynolds no, Re= $10^{5}$ and Prandtl no Pr= 0.72 are prescribed. The domain is discretized into $140 \times 120$ cells with constant grid size along $x-$ direction, whereas along $y-$ direction the grid is geometrically stretched with a regular $4.5 \%$ increment in grid size.  
\begin{figure}[h!] 
\centering
\begin{tabular}{cc}
\includegraphics[width=0.45\textwidth]{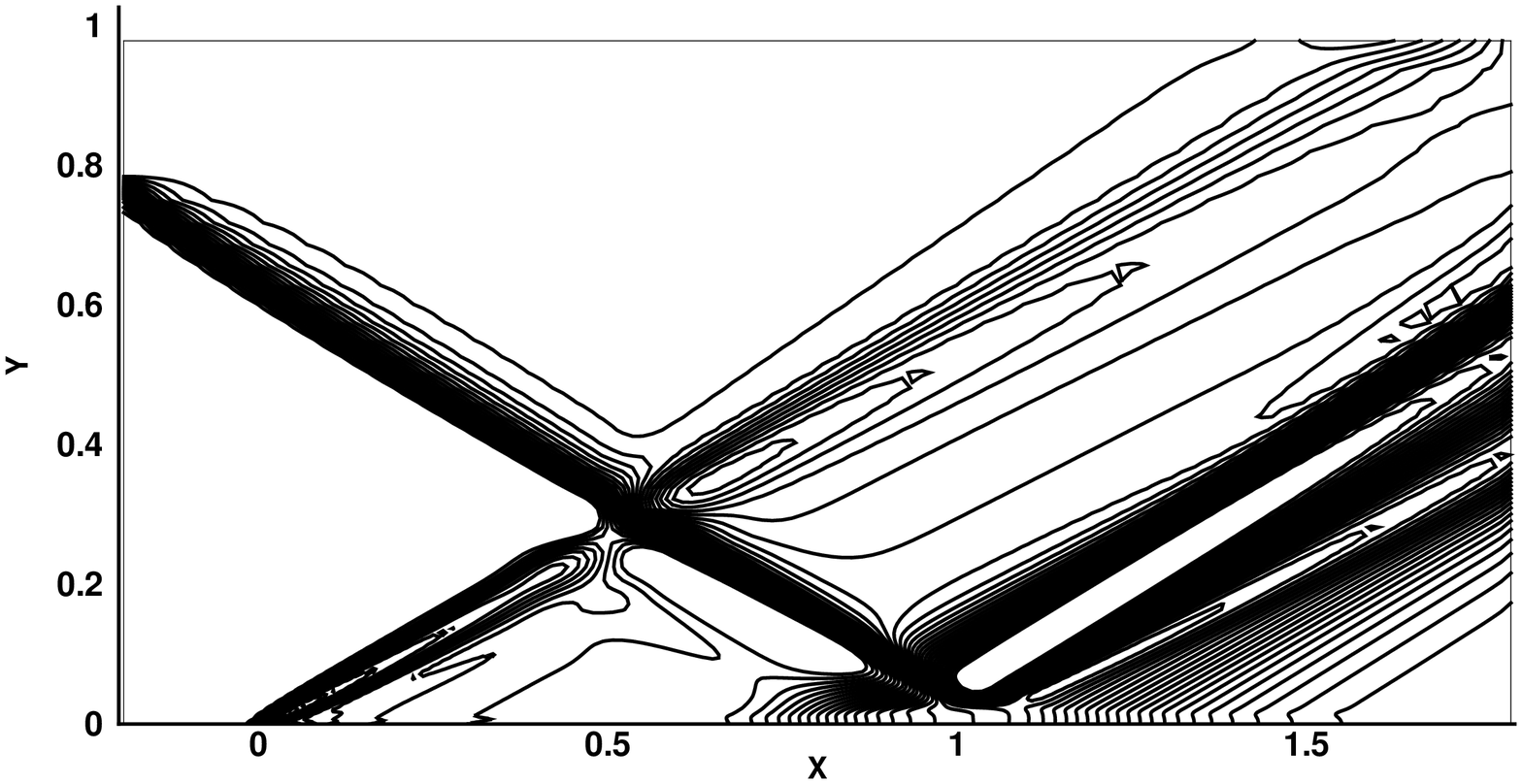} & \includegraphics[width=0.45\textwidth]{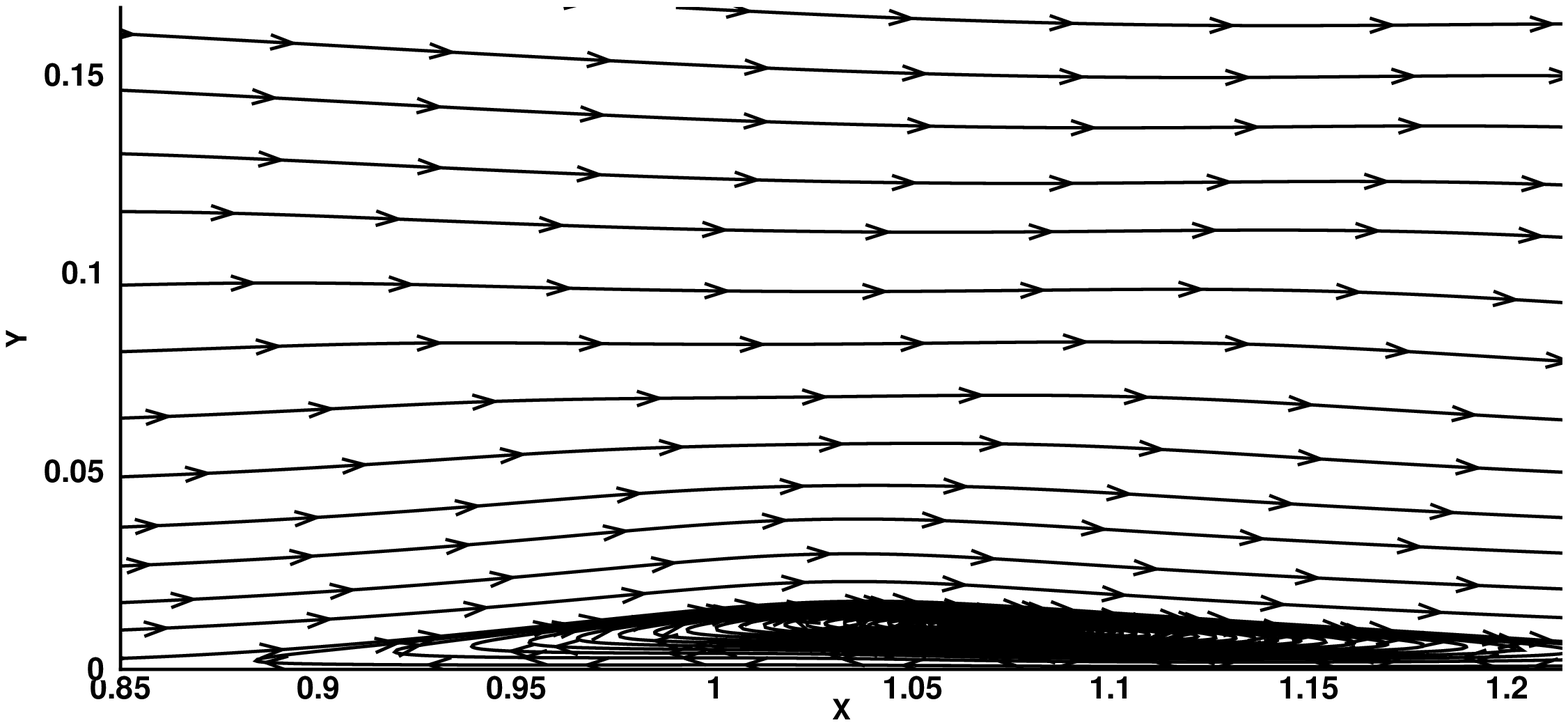}
\end{tabular}
\caption{Test case: Shock wave - boundary layer interaction ($140 \times 120$), a) II order accurate Pressure contours, b) Streamlines showing the recirculation zone}
\label{fig:29}
\end{figure}

\begin{figure}[h!] 
\centering
\begin{tabular}{cc}
\includegraphics[width=0.45\textwidth]{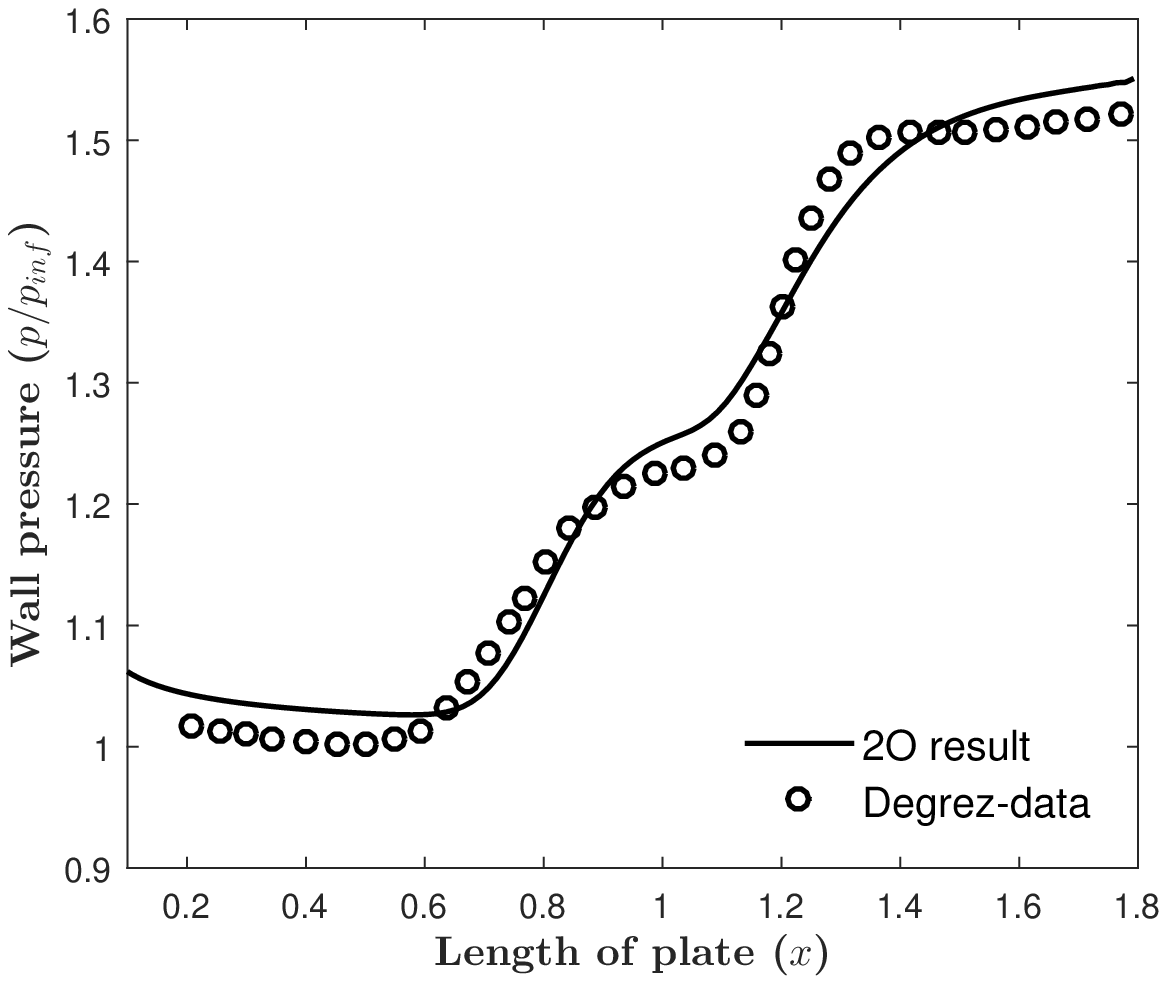} & \includegraphics[width=0.45\textwidth]{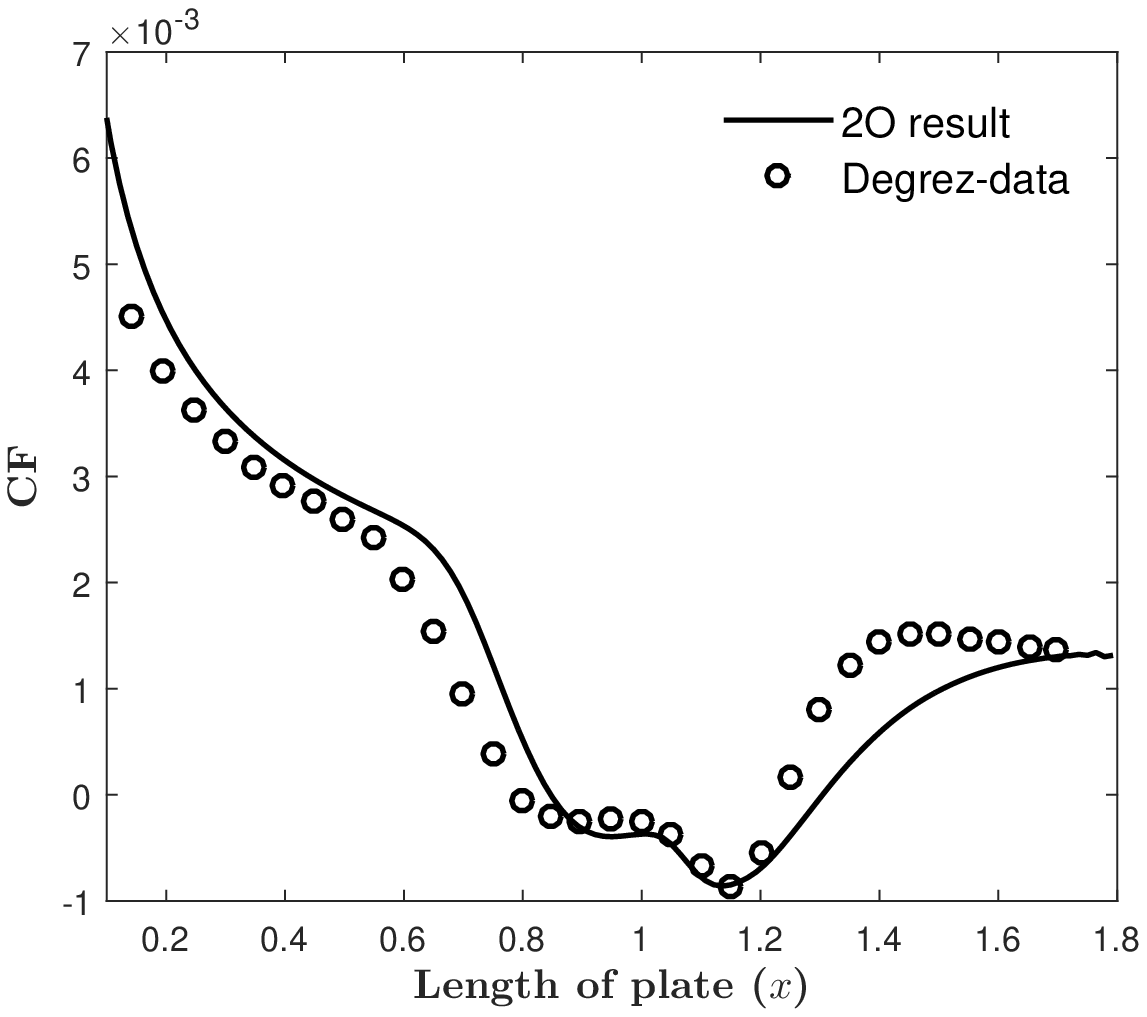}
\end{tabular}
\caption{Test case: Shock wave - boundary layer interaction ($140 \times 120$), a) Wall pressure, b) Skin friction coefficient along plate length}
\label{fig:30}
\end{figure}

The pressure contours for second order accurate results are shown in Figure \ref{fig:29}. As can be observed, flow features like reflected compression and expansion waves and recirculated flow in separated flow region are properly captured. The wall pressure $\frac{p_{w}}{p_{\infty}}$ and skin friction coefficient $c_{f}$ along the length of the plate are plotted in Figure \ref{fig:30}. Our results match well with the data from Degrez \cite{degrez1987interaction}.   

\subsubsection{Supersonic flow over a bump}  
This test case comprises of a Mach 1.4 flow across a circular bump in a channel \cite{parthasarathy1995directional}. The computational domain is $[-1,2] \times [0,1]$ with a $4 \%$ cylindrical bump at the bottom from $x=0$ to $x=1$. Freestream Mach no, $M_{\infty}= 1.4$, Reynolds no $Re = 8000$ and Prandtl no $Pr= 0.72$ are prescribed. Supersonic inflow conditions are used at the left boundary. Flow tangency conditions are used at the top. At the bottom, symmetry conditions are used from $x= -1$ to $x=0$, whereas for $x > 0$, no slip boundary conditions are used. Supersonic outflow conditions are used at the right boundary. The domain is discretized into $240 \times 80$ cells with constant grid size along $x-$ direction, whereas along $y-$ direction the grid is geometrically stretched with a $4.5 \%$ increase in grid size. 
\begin{figure}[h!] 
\centering
\includegraphics[width=0.8\textwidth]{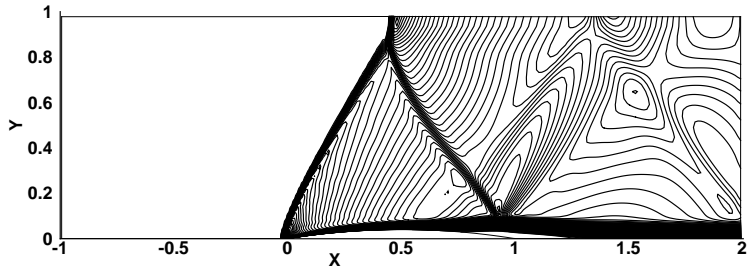}%
\caption{Test case: Supersonic flow over a bump ($240 \times 80$), II order $u_{1}$ contours}%
\label{fig:31}%
\end{figure}
\begin{figure}[h!] 
\centering
\includegraphics[width=0.4\textwidth]{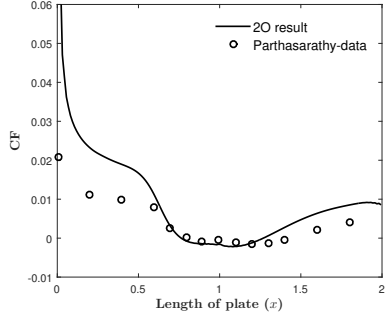}%
\caption{Test case: Supersonic flow over a bump ($240 \times 80$), Skin friction coefficient along wall}%
\label{fig:32}%
\end{figure}
$u_{1}$ contours for the second order scheme are shown in Figure \ref{fig:31}. An oblique shock forms at the leading edge of the bump, which then reflects from the top. This reflected shock then interacts with separated flow at the end of the bump and reflects from it. The skin friction coefficient $c_{f}$ is plotted along the length of the plate and compared with data from  \cite{parthasarathy1995directional} (see Figure \ref{fig:32}).  

\section{Conclusions}
A new kinetic scheme with a set of compactly supported equilibrium distribution functions is presented. The distribution function has flexible average velocities as well as ranges of velocities. The average velocities are used to satisfy the Rankine-Hugoniot jump conditions at discontinuities, leading to exact capture of grid-aligned steady discontinuities. The variable range of velocities is used to provide additional diffusion in expansion and smooth flow regions, preventing formation of any entropy violating solutions. A novel formulation for relative entropy is introduced in the discrete velocity framework. This relative entropy, along with an additional criterion, is used to identify expansions and smooth flow regions. Flow tangency and farfield boundary conditions are formulated for the described kinetic model and used wherever applicable.  Various benchmark 1D and 2D inviscid tests as well as 2D viscous tests are solved for an extensive validation of the proposed scheme.  

\section*{CRediT author statement}
\textbf{Shashi Shekhar Roy}: Methodology, Conceptualization, Investigation, Software, Validation, Formal analysis, Writing- Original draft.\\

\textbf{S. V. Raghurama Rao}: Conceptualization, Investigation, Formal analysis, Supervision, Resources,  Writing- Review \& Editing

\section*{Declaration of competing interest}
The authors declare that they have no known financial interests or personal relationships with any other people or organizations that could influence the work presented here.

\section*{Data availability}
No data was used for research described in this article.

\section*{Acknowledgments}
This research did not receive any specific grant from funding agencies in the public, commercial, or not-for-profit sectors.

\appendix
\section{Chapman Enskog type expansion in 1D}
\label{appendix:a1}
We consider 1D Boltzmann-BGK equations. For a zeroth order approximation to $\textbf{f}$, {\em i.e.}, substituting $\textbf{f}= \textbf{f}^{eq}$ into the Boltzmann equations and taking their moments (which leads to macroscopic inviscid Euler equations), the following moment relations are needed. 
\begin{subequations}
\begin{equation}
\left\langle \textbf{f}^{eq} \right\rangle= \textbf{U}
\label{eq:a1_1}
\end{equation}
\begin{equation}
\left\langle v\textbf{f}^{eq} \right\rangle= \textbf{G}
\label{eq:a1_2}
\end{equation}
\end{subequations}
where $\textbf{U}$ and $\textbf{G}$ are conserved variable vector and inviscid flux vector, respectively. For a first order approximation, we take $\textbf{f}= \textbf{f}^{eq}+ \epsilon \textbf{f}^{\epsilon}$. Taking moment of $\textbf{f}$, we get
\begin{equation}
\left\langle \textbf{f} \right\rangle= \left\langle \textbf{f}^{eq}+ \epsilon \textbf{f}^{\epsilon} \right\rangle= \textbf{U} \Rightarrow \left\langle \textbf{f}^{\epsilon} \right\rangle= 0 
\label{eq:a1_3}
\end{equation}
Further, let
\begin{equation}
\left\langle v\textbf{f} \right\rangle= \left\langle v(\textbf{f}^{eq}+ \epsilon \textbf{f}^{\epsilon}) \right\rangle= \textbf{W} \Rightarrow \left\langle v\textbf{f}^{\epsilon} \right\rangle= \frac{\textbf{W}- \textbf{G}}{\epsilon} 
\label{eq:a1_4}
\end{equation}
Now, substituting $\textbf{f}= \textbf{f}^{eq}+ \epsilon \textbf{f}^{\epsilon}$ into the Boltzmann equations and taking moments, we get
\begin{equation}
\frac{\partial \textbf{U}}{\partial t}+ \frac{\partial \textbf{W}}{\partial x}= 0
\label{eq:a1_5}
\end{equation}
Further, substituting $\textbf{f}= \textbf{f}^{eq}+ \epsilon \textbf{f}^{\epsilon}$ into the Boltzmann equations multiplied by 
$v$ and taking moments, we get 
\begin{equation}
\frac{\partial \textbf{W}}{\partial t}+ \frac{\partial}{\partial x}\left\langle v^{2}\textbf{f} \right\rangle= \frac{\textbf{G}- \textbf{W}}{\epsilon}
\label{eq:a1_6}
\end{equation}
or
\begin{equation}
\textbf{W}= \textbf{G}+ O(\epsilon)
\label{eq:a1_7}
\end{equation}
Taking $\frac{\partial}{\partial t}$ \eqref{eq:a1_7}, 
\begin{eqnarray}
\frac{\partial \textbf{W}}{\partial t} &&= \frac{\partial \textbf{G}}{\partial \textbf{U}}\frac{\partial \textbf{U}}{\partial t}+ O(\epsilon)\nonumber \\ &&= - \frac{\partial \textbf{G}}{\partial \textbf{U}}\frac{\partial \textbf{W}}{\partial x}+ O(\epsilon) = - \frac{\partial \textbf{G}}{\partial \textbf{U}}\frac{\partial \textbf{G}}{\partial x}+ O(\epsilon) \nonumber\\ &&= - \left(\frac{\partial \textbf{G}}{\partial \textbf{U}}\right)^{2}\frac{\partial \textbf{U}}{\partial x}+ O(\epsilon)
\label{eq:a1_8}
\end{eqnarray}
Now,
\begin{eqnarray}
\textbf{W}&&= \textbf{G} -\epsilon \left[\frac{\partial \textbf{W}}{\partial t}+ \frac{\partial}{\partial x}\left\langle v^{2}\textbf{f} \right\rangle \right] \nonumber\\
&&= \textbf{G} -\epsilon \left[ - \left(\frac{\partial \textbf{G}}{\partial \textbf{U}}\right)^{2}\frac{\partial \textbf{U}}{\partial x}+ \frac{\partial}{\partial x}\left\langle v^{2}\textbf{f}^{eq} \right\rangle \right]+ O(\epsilon^{2}) \nonumber\\
&&= \textbf{G} -\epsilon \left[ \left\{ \frac{\partial}{\partial \textbf{U}}\left\langle v^{2}\textbf{f}^{eq} \right\rangle - \left(\frac{\partial \textbf{G}}{\partial \textbf{U}}\right)^{2} \right\} \frac{\partial \textbf{U}}{\partial x} \right] + O(\epsilon^{2})
\label{eq:a1_9}
\end{eqnarray}
Finally, substituting \eqref{eq:a1_9} into \eqref{eq:a1_5}, we get
\begin{equation}
\frac{\partial \textbf{U}}{\partial t}+ \frac{\partial \textbf{G}}{\partial x}= \epsilon \frac{\partial}{\partial x}\left[ \left\{ \frac{\partial}{\partial \textbf{U}}\left\langle v^{2}\textbf{f}^{eq} \right\rangle - \left(\frac{\partial \textbf{G}}{\partial \textbf{U}}\right)^{2} \right\} \frac{\partial \textbf{U}}{\partial x} \right] + O(\epsilon^{2})
\label{eq:a1_10}
\end{equation}
The term in the RHS acts as viscous term. Thus, the $\left\langle v^{2}\textbf{f}^{eq} \right\rangle$ moment adds to viscosity for a first order approximation. For our kinetic model, computing the second moment and substituting in \eqref{eq:a1_10}, we get
\begin{equation}
\frac{\partial \textbf{U}}{\partial t}+ \frac{\partial \textbf{G}}{\partial x}= \epsilon \frac{\partial}{\partial x}\left[ \left\{ \lambda^{2}+ \frac{\delta \lambda^{2}}{3} - \left(\frac{\partial \textbf{G}}{\partial \textbf{U}}\right)^{2} \right\} \frac{\partial \textbf{U}}{\partial x} \right] + O(\epsilon^{2})
\label{eq:a1_11}
\end{equation}
Thus $\delta \lambda$ adds diffusion to the system. We also note that in the limit $\delta \lambda \rightarrow 0$, the above system simplifies to the relaxation model of Jin and Xin \cite{jin1995relaxation}.

\section{Relative entropy in 2D}
\label{appendix:a2}

Our equilibrium distribution for the 2D Flexible Velocity Boltzmann Equation \eqref{eq:75} is given by
\begin{eqnarray}
\widetilde{\textbf{f}}^{eq}_{i} &&= \begin{bmatrix} \widetilde{f}^{eq}_{1i} \\ \widetilde{f}^{eq}_{2i} \\ \widetilde{f}^{eq}_{3i} \\ \widetilde{f}^{eq}_{4i} \end{bmatrix}= \frac{1}{4} \begin{bmatrix}U_{i}+ \frac{G_{1i}}{\widetilde{\lambda}_{1}}+ \frac{G_{2i}}{\widetilde{\lambda}_{2}} \\ U_{i}- \frac{G_{1i}}{\widetilde{\lambda}_{1}}+ \frac{G_{2i}}{\widetilde{\lambda}_{2}} \\ U_{i}- \frac{G_{1i}}{\widetilde{\lambda}_{1}}- \frac{G_{2i}}{\widetilde{\lambda}_{2}} \\ U_{i}+ \frac{G_{1i}}{\widetilde{\lambda}_{1}}- \frac{G_{2i}}{\widetilde{\lambda}_{2}} \end{bmatrix} \nonumber \\
 &&= \frac{1}{4} \begin{bmatrix} 1\\ 1\\ 1\\ 1 \end{bmatrix} U_{i}+ \frac{1}{4 \widetilde{\lambda}_{1} } \begin{bmatrix} 1\\ -1\\ -1\\ 1 \end{bmatrix} G_{1i}+ \frac{1}{4 \widetilde{\lambda}_{2}} \begin{bmatrix} 1\\ 1\\ -1\\ -1 \end{bmatrix} G_{2i} \nonumber \\
&&= \textbf{B}_{0}U_{i}+ \textbf{B}_{1}G_{1i}+ \textbf{B}_{2}G_{2i}
\label{eq:a2_1}
\end{eqnarray}
or 
\begin{eqnarray}
\widetilde{\textbf{f}}^{eq}=&& \begin{bmatrix} \textbf{B}_{0} & 0 & 0 & 0\\ 0 & \textbf{B}_{0} & 0 & 0\\ 0 & 0 & \textbf{B}_{0} & 0\\ 0 & 0 & 0 & \textbf{B}_{0} \end{bmatrix}\textbf{U}+ \begin{bmatrix} \textbf{B}_{1} & 0 & 0 & 0\\ 0 & \textbf{B}_{1} & 0 & 0\\ 0 & 0 & \textbf{B}_{1} & 0\\ 0 & 0 & 0 & \textbf{B}_{1} \end{bmatrix}\textbf{G}_{1}+ \begin{bmatrix} \textbf{B}_{2} & 0 & 0 & 0\\ 0 & \textbf{B}_{2} & 0 & 0\\ 0 & 0 & \textbf{B}_{2} & 0\\ 0 & 0 & 0 & \textbf{B}_{2} \end{bmatrix}\textbf{G}_{2} \nonumber \\
=&& \boldsymbol{\alpha}_{0}\textbf{U}+ \boldsymbol{\alpha}_{1}\textbf{G}_{1}+ \boldsymbol{\alpha}_{2}\textbf{G}_{2}
\label{eq:a2_2}
\end{eqnarray}

Here, $\boldsymbol{\alpha}_{0}$, $\boldsymbol{\alpha}_{1}$ and $\boldsymbol{\alpha}_{2}$ are 16x4 matrices. The kinetic entropy $\hat{H}$ is then given by
\begin{equation}
\hat{H} = \boldsymbol{\alpha}_{0}\eta+ \boldsymbol{\alpha}_{1} \psi_{1}+ \boldsymbol{\alpha}_{2} \psi_{2}
\label{eq:a2_3}
\end{equation}
where $\eta$, $\psi_{1}$ and $\psi_{2}$ are the macroscopic entropy and entropy fluxes along $x-$ and $y-$ directions. For Euler equations, $\eta= \rho s$, $\psi_{1}= \rho u_{1} s$ and $\psi_{2}= \rho u_{2} s$. Now,
\begin{subequations}
\label{eq:a2_4}
\begin{equation}
\left\langle \widetilde{\textbf{f}}^{eq} \right\rangle= \textbf{P}\widetilde{\textbf{f}}^{eq}=\textbf{U} \Rightarrow \textbf{P}\boldsymbol{\alpha}_{0}= I, \textbf{P}\boldsymbol{\alpha}_{1}= 0, \textbf{P}\boldsymbol{\alpha}_{2}= 0
\end{equation}
\begin{equation}
\left\langle \widetilde{\Lambda}_{1}\widetilde{\textbf{f}}^{eq} \right\rangle= \textbf{P}\widetilde{\Lambda}_{1}\widetilde{\textbf{f}}^{eq}=\textbf{G}_{1} \Rightarrow \textbf{P}\widetilde{\Lambda}_{1}\boldsymbol{\alpha}_{0}= 0, \textbf{P}\widetilde{\Lambda}_{1}\boldsymbol{\alpha}_{1}= I, \textbf{P}\widetilde{\Lambda}_{1}\boldsymbol{\alpha}_{2}= 0
\end{equation}
\begin{equation}
\left\langle \widetilde{\Lambda}_{2}\widetilde{\textbf{f}}^{eq} \right\rangle= \textbf{P}\widetilde{\Lambda}_{2}\widetilde{\textbf{f}}^{eq}=\textbf{G}_{2} \Rightarrow \textbf{P}\widetilde{\Lambda}_{2}\boldsymbol{\alpha}_{0}= 0, \textbf{P}\widetilde{\Lambda}_{2}\boldsymbol{\alpha}_{1}= 0, \textbf{P}\widetilde{\Lambda}_{2}\boldsymbol{\alpha}_{2}= I
\end{equation}
\end{subequations}
Therefore,
\begin{subequations}
\label{eq:a2_5}
\begin{equation}
\left\langle \hat{H} \right\rangle= \textbf{P}(\boldsymbol{\alpha}_{0}\eta+ \boldsymbol{\alpha}_{1} \psi_{1}+ \boldsymbol{\alpha}_{2} \psi_{2})= \eta
\end{equation}
\begin{equation}
\left\langle \widetilde{\Lambda}_{1} \hat{H} \right\rangle= \textbf{P}\widetilde{\Lambda}_{1}(\boldsymbol{\alpha}_{0} \eta+ \boldsymbol{\alpha}_{1} \psi_{1}+ + \boldsymbol{\alpha}_{2} \psi_{2})= \psi_{1}
\end{equation}
\begin{equation}
\left\langle \widetilde{\Lambda}_{2} \hat{H} \right\rangle= \textbf{P}\widetilde{\Lambda}_{2}(\boldsymbol{\alpha}_{0} \eta+ \boldsymbol{\alpha}_{1} \psi_{1}+ + \boldsymbol{\alpha}_{2} \psi_{2})= \psi_{2}
\end{equation}
\end{subequations}
Now the relative entropy is 
\begin{eqnarray}
d^{2}=&& \langle \Delta \omega \cdot \Delta \widetilde{\textbf{f}}^{eq} \rangle = \left\langle \Delta \left\{ \frac{\partial \hat{H} (\widetilde{\textbf{f}}^{eq})}{\partial \widetilde{\textbf{f}}^{eq}} \right\} \cdot \Delta \widetilde{\textbf{f}}^{eq}\right\rangle \nonumber \\
=&& \left\langle \Delta \left\{ \boldsymbol{\alpha}_{0} \frac{\partial \eta}{\partial \widetilde{\textbf{f}}^{eq}}+ \boldsymbol{\alpha}_{1} \frac{\partial \psi_{1}}{\partial \widetilde{\textbf{f}}^{eq}}+ \boldsymbol{\alpha}_{2} \frac{\partial \psi_{2}}{\partial \widetilde{\textbf{f}}^{eq}} \right\} \cdot  \Delta \widetilde{\textbf{f}}^{eq} \right\rangle \nonumber\\ 
=&& \left\langle \left\{ \boldsymbol{\alpha}_{0} \Delta \left( \frac{\partial \eta}{\partial \widetilde{\textbf{f}}^{eq}} \right) + \boldsymbol{\alpha}_{1} \Delta \left( \frac{\partial \psi_{1}}{\partial \widetilde{\textbf{f}}^{eq}} \right)+ \boldsymbol{\alpha}_{2} \Delta \left( \frac{\partial \psi_{2}}{\partial \widetilde{\textbf{f}}^{eq}} \right) \right\} \cdot \Delta\widetilde{\textbf{f}}^{eq} \right\rangle \nonumber\\
=&&\textbf{P}\boldsymbol{\alpha}_{0} \Delta \left( \frac{\partial \eta}{\partial \widetilde{\textbf{f}}^{eq}} \right) \cdot \Delta\widetilde{\textbf{f}}^{eq}+ \textbf{P}\boldsymbol{\alpha}_{1} \Delta \left( \frac{\partial \psi_{1}}{\partial \widetilde{\textbf{f}}^{eq}} \right) \cdot \Delta\widetilde{\textbf{f}}^{eq}+ \textbf{P}\boldsymbol{\alpha}_{2} \Delta \left( \frac{\partial \psi_{2}}{\partial \widetilde{\textbf{f}}^{eq}} \right) \cdot \Delta\widetilde{\textbf{f}}^{eq} \nonumber\\
=&&\Delta \left( \frac{\partial \eta}{\partial \widetilde{\textbf{f}}^{eq}} \right) \cdot \Delta\widetilde{\textbf{f}}^{eq} \nonumber\\
=&& \left(\Delta \frac{\partial \eta}{\partial \textbf{U}}\right)^{T}\cdot \Delta \textbf{U} \text{ (on simplifying)}\nonumber\\
=&&\Delta \left(\frac{\gamma -s}{\gamma -1} - \frac{\rho u_{1}^{2}}{2p} - \frac{\rho u_{2}^{2}}{2p}\right) \Delta (\rho)+ \Delta \left(\frac{\rho u_{1}}{p}\right) \Delta (\rho u_{1})+ \nonumber \\ && \Delta \left(\frac{\rho u_{2}}{p}\right) \Delta (\rho u_{2}) + \Delta (-\frac{\rho}{p}) \Delta (\rho E)
\end{eqnarray}

%
%
%
%
%


\end{document}